\documentclass[12pt]{article}
\pdfoutput=1

\newlength{\abstractwidth}
\usepackage[nolabel, final]{showlabels} 

\usepackage{epsf}
\usepackage{color}
\usepackage{graphicx}
\usepackage{tikz}
\usepackage{hyperref}

\usepackage{amsmath}
\usepackage{amssymb}
\usepackage{latexsym}
\usepackage{showlabels}
\numberwithin{equation}{section}
\usepackage{mathtools}

\usepackage{tikz}
\usetikzlibrary{trees}
\usetikzlibrary{decorations.pathmorphing}
\usetikzlibrary{decorations.markings}

\usetikzlibrary{arrows}

\tikzset{
    photon/.style={decorate, decoration={snake}, draw=red},
    electron/.style={draw=blue, postaction={decorate},
        decoration={markings,mark=at position .55 with {\arrow[draw=blue]{>}}}},
    gluon/.style={decorate, draw=magenta,
        decoration={coil,amplitude=4pt, segment length=5pt}} 
}

\makeatletter
\renewcommand\section{\@startsection {section}{1}{\z@}%
                                   {-3.5ex \@plus -1ex \@minus -.2ex}
                                   {2.3ex \@plus.2ex}%
                                   {\normalfont\large\bfseries}}
\renewcommand\subsection{\@startsection{subsection}{2}{\z@}%
                                     {-3.25ex\@plus -1ex \@minus -.2ex}%
                                     {1.5ex \@plus .2ex}%
                                     {\normalfont\bfseries}}
\makeatother

\renewcommand{\thefootnote}{\fnsymbol{footnote}}
\renewcommand{\thanks}[1]{\footnote{#1}}
\newcommand{\starttext}{
\setcounter{footnote}{0}
\renewcommand{\thefootnote}{\arabic{footnote}}}

\newcommand{\bea}{\begin{eqnarray}}
\newcommand{\eea}{\end{eqnarray}}
\newcommand{\be}{\begin{eqnarray}}
\newcommand{\ee}{\end{eqnarray}}

\def\cD{{\cal D}}
\def\cE{{\cal E}}
\def\cF{{\cal F}}

\def\cO{{\cal O}}
\def\cP{{\cal P}}

\def\cR{{\cal R}}

\def\nn{\nonumber}

\def\CC{{\mathbb C}}
\def\ZZ{{\mathbb Z}}
\def\RR{{\mathbb R}}

\def\CC{{\mathbb C}}

\def\GG{{\mathbb G}}

\def\KK{{\mathbb K}}

\def\det{{\rm det \,}}

\def\half{{\scriptstyle {1 \over 2}}}

\def\threeh{{\scriptstyle {3 \over 2}}}
\def\fiveh{{\scriptstyle {5 \over 2}}}

\def\sm{\smallskip}

\def\DD{{D_\theta}}

\def\F{{F}}
\def\Z{{Z}}
\def\H{H}

\def\D{D}
\def\EE{\cE^{(3)}}

\begin{document}
\starttext
\setcounter{footnote}{0}

\begin{flushright}
QMUL-PH-19-07\\
DAMTP-2019-15
\end{flushright}

\vskip 0.3in

\begin{center}

{\Large \bf Modular Forms and $SL(2, {\mathbb Z})$-covariance of type IIB superstring theory}

\vskip 0.2in

{\large  Michael B. Green$^{(a,b)}$ and Congkao Wen$^{(a)}$}

\vskip 0.15in

{\tt \small  M.B.Green@damtp.cam.ac.uk,  c.wen@qmul.ac.uk}

\vskip 0.15in

{ \sl (a) Centre for Research in String Theory, School of Physics and Astronomy, }\\
{\sl Queen Mary University of London, Mile End Road, London, E1 4NS, UK}

\vskip 0.1in

{ \sl (b) Department of Applied Mathematics and Theoretical Physics }\\
{\sl Wilberforce Road, Cambridge CB3 0WA, UK}

\vskip 0.2in

\begin{abstract}
\vskip 0.1in

The local higher-derivative interactions that enter into the low-energy expansion of  the effective action of  type IIB superstring theory with constant complex modulus generally violate the $U(1)$ R-symmetry of IIB supergravity by $q_U$ units. These interactions  have coefficients that transform as non-holomorphic modular forms under  $SL(2, {\mathbb Z})$ transformations  with holomorphic and anti-holomorphic weights $(w,-w)$, where $q_U=-2w$. 

In this paper $SL(2, {\mathbb Z})$-covariance and supersymmetry are used to determine first-order differential equations on moduli space that relate the modular form coefficients of  classes of BPS-protected  maximal $U(1)$-violating interactions that arise at low orders in the low-energy expansion.  These are the moduli-dependent  coefficients of BPS interactions of the form $d^{2p} \mathcal{P}_n$ in linearised approximation, where $\mathcal{P}_n$ is the product of $n$ fields that has dimension $=8$ with $q_U=8-2n$, and $p=0$, $2$ or $3$.
These first-order equations imply that the coefficients satisfy $SL(2, {\mathbb Z})$-covariant Laplace eigenvalue equations on moduli space with solutions that contain information concerning perturbative and non-perturbative contributions to superstring amplitudes. For $p=3$  and $n\ge 6$  there are two independent modular forms, one of which has a vanishing tree-level contribution.

The analysis of super-amplitudes for  $U(1)$-violating processes  involving arbitrary numbers of external fluctuations of the complex modulus  leads to a diagrammatic derivation of the first-order differential relations and Laplace equations  satisfied by the coefficient modular forms.  Combining this with a $SL(2, {\mathbb Z})$-covariant soft axio-dilaton limit  that relates amplitudes with different values of $n$ determines most of the modular invariant coefficients, leaving  a single undetermined constant.
\end{abstract}
\end{center}

\newpage
 
\setcounter{tocdepth}{2} 
\tableofcontents
\newpage

\baselineskip=15pt
\setcounter{equation}{0}
\setcounter{footnote}{0}

\section{Overview and outline of the paper}
\label{overview}
\setcounter{equation}{0}

At low energy, or small curvature, closed string theory reduces to a version of Einstein's theory that may be described in terms of the Einstein--Hilbert action coupled to a variety of massless fields.  The low-energy expansion of the effective string theory action is  a power series in $p\,\ell_s$, where $p$ is the energy-momentum scale and $\ell_s=\sqrt{\alpha'}$ is the string length scale. Successive terms in this expansion may be expressed in terms of higher-derivative interactions that generalise the Einstein--Hilbert action.  Such interactions have a rich dependence on the moduli fields associated with the geometry of the target space.  In the case of superstring theory, the dependence on the moduli is highly  constrained by perturbative and non-perturbative dualities.

\subsection{Overview}

The focus of this paper is on the structure of the coefficients of higher-derivative interactions that arise in the low-energy expansion of scattering amplitudes  in ten-dimensional type IIB superstring theory, which is the simplest example of a theory with a non-trivial S-duality group, namely,  $SL(2,\ZZ)$.   It contains a single complex scalar field, or modulus, $\tau=\tau_1+i\tau_2$, which parameterises the coset $SL(2,\RR)/U(1)$, where $U(1)\sim SO(2)$ is the R-symmetry of classical IIB supergravity.\footnote{Our conventions for  parameterising the embedding of the $U(1)$ R-symmetry in the supergravity coset are summarised in  appendix~\ref{app:SUGRA}.}

\sm

The continuous $SL(2, \RR)$ symmetry of  classical supergravity does not survive in the quantum theory since it is not preserved in the string extension of type IIB supergravity. Indeed it is well known  that the classical superstring is not invariant under the $U(1)$ subgroup of $SL(2, \RR)$ that rotates the two supercharges into each other since the two supercharges move in opposite directions on the world-sheet.  The theory is only invariant under a discrete $Z_4$ subgroup of this $U(1)$ which interchanges the two supercharges and reverses the parameter $\sigma$ that labels points along the string.
This  $Z_4$  is the intersection of $U(1)$ with $SL(2,\ZZ)$.
As a result, the  modulus field is subject to discrete identifications that restrict it to a single fundamental domain of moduli space $SL(2,\ZZ)\backslash SL(2,\RR)/U(1)$, leading to the arithmetic S-duality group $SL(2,\ZZ)\subset SL(2,\RR)$.\footnote{This pattern of the  breaking of  $SL(2,\RR)$ is also indicated by the presence of a one-loop chiral anomaly in ten-dimensional type IIB supergravity  \cite{Gaberdiel:1998ui}.} 

\sm

A consequence is that in a fixed background, $\tau=\tau^0$,  $n$-particle amplitudes for the scattering of  massless states (i.e. supergravity states) generally violate the continuous $U(1)$ symmetry that is conserved in perturbative type IIB supergravity. However,  there is a particular pattern to the non-conservation of the $U(1)$ charge.   As will be reviewed in section~\ref{effact}, an interaction that violates the $U(1)$ charge by $q_U= - 2w$ units  contributes to $n$-particle amplitudes with $n\ge |w|+4$ and  its coefficients are given by non-holomorphic modular forms that transform with holomorphic and anti-holomorphic weights $(w,-w)$.\footnote{ More generally,  a non-holomorphic modular form has independent weights, $(w,w')$,  but when $w'=-w$  it transforms by a phase under the action of $SL(2,\ZZ)$, as described in section~\ref{nonhol}, }   In this paper we will consider $w\ge 0$,  i.e. $q_U\le 0$.  The $w<0$ ($q_U>0$) cases  are complex conjugates of the $w>0$ cases. 

\sm

These modular forms  are highly constrained, and in some cases precisely determined, by the requirement that the effective action should be invariant under $SL(2,\ZZ)$ as well as maximal supersymmetry.  This is true for 
the terms that arise up to dimension $14$, which preserve a fraction of the supersymmetry\footnote{The dimension of the Einstein--Hilbert action is 2. }  -- in that sense they are `$F$-terms' that can be expressed as integrals over a subspace of the full 32-component space of Grassmann on-shell  superspace coordinates.

\sm

For example, the leading higher-derivative terms are associated with $1/2$-BPS interactions of  the same dimension as $(\alpha')^{-1} R^4$ (where the four Riemann tensors are contracted with a well-known sixteen-index tensor).   This interaction arises in the expansion of the tree-level four-graviton amplitude \cite{Gross:1986iv,Grisaru:1986px}
and its  exact non-perturbative structure  \cite{Green:1997tv, Green:1997as,Green:1998by} is encoded in its coefficient, $E(\threeh,\tau)$, which is a non-holomorphic Eisenstein series.  This is a $SL(2,\ZZ)$-invariant function of  the complex scalar field $\tau=\tau_1+i\tau_2$, where $\tau_1=C^{(0)}$ is the Ramond-Ramond zero form (or ``axion'') and $\tau_2= e^{-\varphi}$, where  $\varphi$ is the dilaton field.  In our consideration of scattering amplitudes the background scalar field is the  complex coupling constant $\tau^0:=  \chi +i/g_s$.  The expansion of $E(\threeh,\tau^0)$ for weak coupling ($\tau^0_2=1/g_s\to \infty$) yields two power behaved terms of order $g_s^{-3/2}$ and $g_s^{1/2}$, which are identified with tree-level and one-loop terms in string perturbation theory.  It also gets contributions of order $e^{-2\pi/g_s}$ from D-instantons.  Properties of modular forms and Eisenstein series of relevance to this paper are reviewed in appendix~\ref{holosol}.

\sm

Many other $1/2$-BPS interaction terms arise in the low-energy expansion of the type IIB string action at order $(\alpha')^{-1}$
(dimension 8) in a constant background $\tau=\tau^0$,  and these generally violate the $U(1)$ R-symmetry.  There is a bound on the $U(1)$ violation of an $n$-particle interaction, $|q_U| \leq |8-2n|$ (where $n\ge 4$), which is saturated by the `maximal $U(1)$-violating' interactions \cite{Boels:2012zr}. The lowest-dimension interactions that saturate the bound  can be expressed in linearised approximation as  monomials $\cP_{n}(\{\Phi\})$, which are products of $n$ on-shell fields and field strengths  of type IIB supergravity  with total  $U(1)$ charge violation equal to $q_U=-2w=8-2n$ and with dimension $8$ (as will be discussed in appendix~\ref{summact}).
Interactions of the form $d^{2p} \, \cP_{n}(\{\Phi\})$ have dimension $8+2p$  -- with $p=2$ they are $1/4$--BPS interactions and with $p=3$ they are  $1/8$-BPS.\footnote{Interactions with  $p=1$  are  absent.}  For now we will not specify how the derivatives act, but this will be clarified in an economical manner by the kinematic structures involving Mandelstam invariants in scattering amplitudes later in this paper.   These fractional BPS interactions are known to be ``protected''  by supersymmetry from receiving perturbative contributions.    When $p\ge 4$ (dimension $\ge 16$) the interactions are non-BPS (at least, in any conventional sense) and their coefficients are not constrained by  supersymmetry in any obvious manner.

\sm

We will be concerned with contributions of  the ``protected''  maximal $U(1)$-violating interactions to the effective action.    We may write terms of this type involving $n$ fields  in the form 
 \bea
S^{(p)}_{n\, i} &=& (\alpha')^{p-1}\int d^{10}x\, e\, \tau_2^{\frac{1-p}{2}}\, \F_{w\, i}^{(p)}(\tau)\,  d_{(i)}^{2p}\, \cP_{n}(\{\Phi\})\nn\\
&=& \kappa^{\frac{p-1}{2}}\int d^{10}x\, e\, \F_{w\,i}^{(p)}(\tau)\,  d_{(i)}^{2p}\,\cP_{n}   (\{\Phi\}) \, ,
 \label{chargeq}
 \eea
where $n=w+4$ and $e$ is the determinant of the zehnbein.   In the second expression, in which the  fields have been transformed to the Einstein frame, the gravitational coupling, $\kappa$, is related to $\alpha'$  by $\kappa =  (\alpha')^2\, g_s$. 
The interaction $\cP_{n}  (\{\Phi\})$  carries a charge $q_U=-2w=-2n+8$ and so transforms by a phase under  $U(1)$ transformations embedded in $SL(2,\RR)$,  and therefore invariance of the low-energy  action under $SL(2,\ZZ)$  implies that $\F_{w,i}^{(0)}(\tau)$  must be a $(w,-w)$ modular form that transforms by a compensating phase. 

\sm

The  subscript $i$ on the symbol $ d_{(i)}^{2p}\, \cP_{n}$ labels the independent invariants made out of the $2p$ derivatives -- the independent ways in which the derivatives acting on the fields are contracted into each other, up to terms  which vanish on-shell.  These correspond to the independent symmetric polynomials of degree $p$ in the Mandelstam variables for the $n$-particle amplitude.
This degeneracy of the kinematic invariants  is correlated with the number of independent moduli-dependent coefficients, $\F_{w\, i}^{(p)}(\tau)$, and plays an important r\^ole in the following discussion. A two-fold degeneracy of these invariants first arises for  $n=4$ when $p= 6$, for $n=5$  when $p = 4$, and  for $n\ge 6$ when $p\ge 3$.  The BPS interactions have $p\le 3$ and therefore degeneracy only arises with $p=3$ and $n\ge 6$.   In other cases the ``degeneracy'' index $i$ is redundant so we will generally use the notation $\F^{(p)}_{w}(\tau)$  unless the index $i$ is needed. Each coefficient $\F_{w\, i}^{(p)}(\tau)$ is a $(w,-w)$ modular form, which transforms with a   $U(1)$ charge $q=2w=2n-8$  under $SL(2,\ZZ)$, so its transformation compensates for that of 
$\cP_n(\{\Phi\})$ and the action  \eqref{chargeq} is invariant.  
 
 \sm

 All of the known examples of $F^{(p)}_{w}  (\tau)$, which will be reviewed in section~\ref{effact}, are non-degenerate.
In the $1/2$-BPS case ($p=0$)  the coefficients $\F_{w}^{(0)}(\tau)$  in \eqref{chargeq} are non-holomorphic modular forms of weight $(w,-w)$  that are known to be generalisations of non-holomorphic Eisenstein series that  transform non-trivially under $SL(2,\ZZ)$.   
A coefficient $F^{(p)}_{w}  (\tau)$ is related to $F^{(p)}_{w+1}  (\tau)$ by first-order differential equations implied by supersymmetry, which, in turn, lead to Laplace eigenvalue equations in the upper half plane \cite{Green:1997as, Green:1998by}. These have unique solutions proportional to generalisations of non-holomorphic Eisenstein series with modular weights $(w,-w)$ that are reviewed in appendix~\ref{holosol}. 
 The complete list of dimension-8  ($1/2$-BPS) linearised interactions $\cP_{n}(\{\Phi\})$ can be obtained from supersymmetry considerations making use of a linearised on-shell scalar superfield introduced in  \cite{Howe:1983sra},  as will be reviewed in appendix~\ref{summact}. Some examples of these dimension-8 interaction polynomials are:
 \bea
 \label{examps}
  R^4\ \  (w=0), \qquad G^2 R^3\ \  (w=1), \qquad G^4 R^2\ \  (w=2),\ \   \dots, \ \ \Lambda^{16}\  \  (w=12)\,,
  \eea
  where $R$ is the linearised curvature, $G$ is the complex third-rank field strength and $\Lambda$ is the complex dilatino. 

\sm

Similar comments apply to the $1/4$--BPS interactions \cite{Green:1998by,Sinha:2002zr}, which have $p=2$.         In the cases with $p=0$ and $p=2$ (the $1/2$-BPS and $1/4$--BPS cases) there is a complete understanding of  $F^{(p)}_{w}(\tau)$  for all $U(1)$ charges, $0\le q_U\le 24$ ($q_U=2w=2n-8$). 
Less is known in  the $p=3$  cases  (which are $1/8$--BPS interactions), for which  only the $w=0$ coefficient has been determined  (the coefficient of $d^6 R^4$) \cite{Green:2005ba}.  This satisfies an inhomogeneous Laplace eigenvalue equation on the upper-half plane and has many fascinating features.
 The homogeneous Laplace equations for $R^4, d^4 R^4$ and the inhomogeneous equation for $d^6 R^4$ can also be motivated by supersymmetric Ward identities that are implied by the structure of super-amplitudes \cite{Wang:2015jna}.\footnote{Other interesting applications of such super-amplitude constraints on effective actions can be found e.g. in \cite{Elvang:2010jv,Chen:2015hpa, Wang:2015aua, Lin:2015dsa, Bianchi:2016viy}}  The coefficient of $d^6R^4$ has a weak coupling expansion that reproduces results of explicit superstring perturbation theory from genus-zero to genus-three
\cite{Green:1997tv, Green:1999pu, Green:2005ba, DHoker:2005jhf, DHoker:2013fcx, Gomez:2013sla}.
 
\subsection{Outline of results}
 
 The generalisation of the $p=3$, $w=0$ Laplace equation to cases with $w>0$ will be considered in section \ref{eighthbps}.  We will show that  requiring consistency with the $w=0$ case leads to first-order differential equations relating coefficients of $\F^{(3)}_w(\tau)$ with different values of $w$.  This determines  a novel inhomogeneous Laplace eigenvalue equation for the $p=3$, $w=1$  modular form $\F_1^{(3)} (\tau)$, which is the coefficient of  five-particle maximal $U(1)$-violating interactions.  In the $p=3$, $w=2$ case there are two distinct forms $\F^{(3)}_{2,i}(\tau)$ that are related to $\F_1^{(3)} (\tau)$, and  which satisfy  distinct inhomogeneous Laplace eigenvalue equations.  Importantly, it is also known that there are two symmetric cubic invariants for the six-particle amplitude with  $p=3$.    The modular form $\F^{(3)}_{2,1}(\tau)$ is related to the known modular function, $\F^{(3)}_0(\tau)$, by the relation $\F^{(3)}_{2,1}(\tau) \sim \cD_1 \cD_0 \F^{(3)}_0(\tau)$ (where $\cD_w$ is a modular covariant derivative that will be defined later), whereas   $\F^{(3)}_{2,2}(\tau)$ is qualitatively distinct.   In particular, its weak coupling expansion does not contain a tree-level contribution, but starts with the genus-one term of order $\tau$.
  
\sm

The preceding pattern of effective interactions has an interesting interpretation in terms of type IIB superstring scattering amplitudes in the low-energy expansion,  which is the subject of sections~\ref{uoneamp},~\ref{sec:softB} and~\ref{constraints}. Such amplitudes describe the scattering of fluctuations of the massless fields around a fixed background, which will be taken to be flat ten-dimensional Minkowski space with a constant value of the complex coupling constant, $\tau=\tau^0$.  We will  be particularly concerned with maximal $U(1)$-violating $n$-particle amplitudes   with  $n=4+w$,   which  violate the $U(1)$ charge by $2w$ units and  have no massless intermediate poles.

\sm

A subset of these are amplitudes in which the external fluctuating states correspond to the fields in $d^{2p}_{(i)}\cP_{n}$.    The dependence of their low-energy expansion on the coupling constant is given by $\F_{w,i}^{(p)} (\tau^0)$.  The amplitudes transform covariantly under $SL(2,\ZZ)$  in the sense  that the coefficient  $\F_{w,i}^{(p)}(\tau^0)$ transforms as  a $(w,-w)$ modular form under  a  $SL(2,\ZZ)$ transformation of the background in a manner that compensates for the  transformation of the external states that correspond to the fields in $\cP_{n}(\{\Phi\})$.

\sm

More generally there are maximal $U(1)$-violating amplitudes obtained by adding $m$ axio-dilaton fluctuations, $\delta \tau=\tau-\tau^0$,  to the $n$-particle states associated with the $n$ fluctuating fields in the $d^{2p}_{(i)}\,\cP_{n}(\{\Phi\})$ contact terms.     Such amplitudes are obtained by an $m$th order Taylor expansion of $\F_{w,i}^{(p)}(\tau)$.   But $\delta\tau$ does not respect the $U(1)$ symmetry  of the coset and such an expansion generates $(n+m)$-particle amplitudes for which the  $SL(2,\ZZ)$ duality is not manifest.  Following the usual normal coordinate expansion  for nonlinear sigma models on coset spaces,  invariance will be restored by a suitable reparameterisation that replaces $\delta \tau$ by  the  complex field $\Z $
 \bea
 \delta\tau \to \Z :=\frac{\tau-\tau^0}{\tau-\bar \tau^0}\,.
 \label{bdefone}
 \eea
This is a $SL(2,\CC)$ transformation that maps the upper-half $\tau$ plane to the unit disk in the $\Z $ plane.\footnote{The field, $\Z $,  was introduced  (but called $B$) as the modulus field in the $SU(1,1)$ formulation of  type IIB supergravity in \cite{Schwarz:1983qr}.}   The field $\Z $ transforms with a phase appropriate to a charge-1 field under  $SL(2,\ZZ)$.   Consequently, the $m$th order term in the expansion of  $\F_{n-4,i}^{(p)}(\tau)$ in powers of $\Z $ is proportional to the  $m$th order modular-covariant derivative of  $\F_{n-4,i}^{(p)}(\tau)$.  This guarantees that the scattering amplitude with $m$ $\Z $ fields and $n$ fields from $\cP_{n}$ transforms covariantly under $SL(2,\ZZ)$ acting on the fluctuations and the constant modulus, $\tau^0$.  For example, the term of dimension  $8+2p$ in the  low  energy expansion of the amplitude with $m$ complex scalars and four gravitons, is proportional to 
 \bea
  \F_{m,1}^{(p)}(\tau^0)\, \langle g_1\,g_2\,g_3\,g_4\, \Z _1\,\Z _2\dots \Z _m\rangle|_{8+2p}\,,
 \label{4gmt}
 \eea
 where $\langle  \dots \rangle|_{8+2p}$  indicates the term with dimension $(8+2p)$  in the low-energy expansion  of  order $d^{2p} R^4$ and 
 \bea
 \F_{m,1}^{(p)}(\tau^0)=\left.  2^m \cD_{m-1}\, \cD_{m-2} \dots \cD_0\, \F^{(p)}_0(\tau)\right|_{\tau=\tau^0}\,.
 \label{covdf}
 \eea
The subscript $1$ is redundant except for the cases with $p=3$ and $w=m\ge 2$.  In these cases there is a separate contribution proportional to $F^{(3)}_{m, 2} (\tau^0)$, which corresponds to the coefficient of another independent interaction term at this order. This interaction has a vanishing tree-level contribution and the lowest-order term in its perturbative  expansion is the one-loop term.  

\sm

In section~\ref{uoneamp} we will also briefly  review  the ten-dimensional helicity-spinor formalism, which is an efficient  framework for constructing supersymmetric amplitudes.    We  will  describe a general soft axio-dilaton limit, which is confirmed by explicit type IIB superstring tree amplitudes with $n\le 6$ external states.
\sm 

 Soft axio-dilaton limits will be further  considered in section~\ref{sec:softB}. In particular,  for the low-dimension  terms in the low-energy expansion considered in this paper the soft limits determine the expansion coefficients of  higher-point amplitudes completely in terms of the lower-point ones. The soft limits relate the coefficient modular functions of different weights in the manner of \eqref{covdf}.

\sm

Supersymmetry constraints on the BPS terms are investigated in section~\ref{constraints}   by using an extension of the ideas in \cite{Wang:2015jna}, where it was shown that the well-known Laplace equations satisfied by $w=0$ functions $F^{(p)}_{0}(\tau)$ can also be understood from the constraints imposed by on-shell super-amplitudes. Here we will generalise this approach  to obtain first-order differential equations, including the equations for the coefficient modular functions $F^{(3)}_{2, 1} (\tau^0)$ and $F^{(3)}_{2, 2} (\tau^0)$. This confirms the results obtained in section~\ref{eighthbps} that were based on somewhat different considerations.  
 
\sm

The conclusions of this paper will be summarised in section~\ref{discussion},  where we will also discuss some of their implications.
     
\section{Effective interactions in the low-energy expansion}
\label{effact}

The effective interactions that arise in the first few orders in the low-energy expansion of the ten-dimensional type IIB superstring effective action  have been determined by imposing the requirements of maximal supersymmetry together with $SL(2,\ZZ)$ S-duality.  The  coefficients of these higher-derivative interactions are functions of the complex scalar field $\tau$ that transform covariantly under the action of $SL(2,\ZZ)$.   

\sm

To establish our conventions recall that the classical supergravity action has the form  
\bea
\label{einsteinH}
S^{EH}= \frac{1}{(\alpha')^4} \int d^{10} x \, e \, \tau_2^2  \,  R + \cdots
= \frac{1}{\kappa^2} \int d^{10} x \, e\,  R + \cdots \,,
\eea
where the ellipsis denotes the presence of many other terms of the same dimension that complete the supersymmetric action of type IIB supergravity\footnote{The problematic issue of writing an action for the self-dual five-form  field strength in the type IIB theory  is not relevant here since our focus will be on the on-shell scattering amplitudes.}.  Note that when the dilaton is constant,  $(\tau_2^0)^{-1} = g_s$ is the string coupling constant, and therefore $\kappa^2 = (\alpha')^4 \, g_s^2$.

\sm

The classical type IIB equations of motion were determined in component form in \cite{Schwarz:1983qr} (up to terms quadratic in fermion fields) and in terms of on-shell  superfields in  \cite{Howe:1983sra}.  The expression \eqref{einsteinH} is invariant under $SL(2,\ZZ)$ as is obvious when expressed in the Einstein frame.
  In order to proceed further we will review properties of higher order terms in the low-energy expansion of the effective action, for which $SL(2,\ZZ)$ invariance is more subtle since the moduli-dependent coefficients transform as non-holomorphic modular forms.
 
 \subsection{Non-holomorphic modular forms}
 \label{nonhol}

 Recall that $SL(2,\ZZ)$ acts on the scalar field $\tau$ as
\bea
\label{tautrans}
\tau\to \frac{a\tau+ b}{c\tau+d}\,,
\eea
 with $a,b,c,d\in \ZZ$ and $ad-bc=1$.   A non-holomorphic modular form $f^{(w,w')}(\tau)$ has holomorphic and anti-holomorphic modular weights $(w,w')$ and its transformation under $SL(2,\ZZ)$ is given by
\bea
f^{(w,w')} (\tau) \to (c\tau+d)^w  \, (c\bar \tau +d)^{w'}\,  f^{(w,w')} (\tau) \,.
\label{modtrans}
\eea
Modular covariant derivatives can be defined by
\bea
 \cD_w= i\left(\tau_2 \frac{\partial}{\partial \tau} - i\,\frac{w}{2} \right),
\qquad  \bar{\cD}_{w'} = -i\left(\tau_2
\frac{\partial}{\partial \bar\tau} +i \frac{w'}{2}\right)\, , 
\label{covderdef}
\eea
where $\cD_w$ transforms $(w,w') \to (w+1,w'-1)$ and $\bar \cD_{w'}$ transforms $(w,w')\to (w-1,w'+1)$.  In other words,
\bea
\label{faction}
\cD_w\,  f^{(w,w')} := f^{(w+1,w'-1)}\,,\qquad\quad \bar \cD_{w'}\, f^{(w,w')} := f^{(w-1,w'+1)} \, .
\eea
Non-holomorphic forms for which $w'=-w$ transform by a phase characterised by a $U(1)$ charge, $q=2w$, as is evident from \eqref{modtrans}.

\sm

For future reference we note that since  $\tau_2=(\tau-\bar\tau)/(2i)$,  the action of $\cD_w$ on a power of $\tau^0_2=1/g_s$ is given by 
\bea
  \cD_w \tau_2^\alpha \bigg |_{\tau_2=\tau^0_2} \sim   \frac{1}{2}\left( \tau_2 \frac{\partial}{\partial \tau_2} +w\right)\tau_2^\alpha \bigg|_{\tau_2=\tau^0_2} 
=  \frac{1}{2} \left(-g_s \frac{\partial}{\partial g_s} +w\right)  g_s^{-\alpha}\,.
\label{tauind}
 \eea
In the next sub-section we will describe the coefficients of the first few terms in the low-energy expansion, which are known to be modular forms that  are related by first-order differential equations of the form \eqref{faction}. These imply that they satisfy various kinds of Laplace equations.  The simplest examples of such equations are Laplace eigenvalue equations that have solutions parameterised by $s\in \CC$ and have the form 
\bea
\label{laplaceone}
\Delta_{(-) w} \, f_s^{(w,-w)}(\tau) = \left(s(s-1)-w(w-1)\right)\, f_s^{(w,-w)}(\tau)\,.
\eea
where $\Delta_{(-)w}:=4  \cD_{w-1} \bar \cD_{-w}$   is a laplacian acting on a weight $(w,-w)$ non-holomorphic modular form.  
This equation has a unique solution, subject to the physically required boundary condition that  it has moderate growth (power behaviour) in the large-$\tau_2$ limit (the weak-coupling limit).  The solution is given in terms of Eisenstein series  as reviewed in appendix~\ref{holosol}.

\subsection{Some coefficients of low-order terms}
\label{coeffrev}

In order to motivate our subsequent discussion, we will now summarise the known coefficients of terms in the low-energy expansion. In each case the interaction takes its simplest form in the Einstein frame, in which S-duality is manifest.  As  is seen in  \eqref{chargeq}, this is related to the form of the interaction in the string frame by a rescaling of the metric by a dilaton-dependent factor.  Since the dilaton is constant in the backgrounds of relevance to this paper, this simply introduces a power of the coupling constant that depends on the dimension of the interaction (the order in $\alpha'$).

\sm

\subsubsection{Terms at order $(\alpha')^{-1}$}
In our conventions the classical supergravity action \eqref{einsteinH} is of order  $(\alpha')^{-4}$ and has dimension 2. The first non-leading terms arise from a super-multiplet of $1/2$-BPS  interactions at order $(\alpha')^{-1}$   (dimension 8), with distinct $U(1)$ charges ranging from $0$ to $24$.\footnote{As will be discussed later, expanding the coefficient function $\F^{(p)}_{w,i}(\tau)$ in \eqref{chargeq} in fluctuations of $\tau$ leads to amplitudes with higher $U(1)$-charge violation.}  The fully supersymmetric nonlinear effective action has not been determined,\footnote{However, the fully nonlinear action at $O((\alpha')^{-1})$  has been determined in the special case in which  $G= \partial_\mu \tau= 0$, where $G$ is the complex  three-form and $\tau$ is the complex scalar field~\cite{Green:2003an, Rajaraman:2005ag}} but it is relatively straightforward to determine it in linearised approximation.  In this description the supergravity fields are the components of a  linearised scalar superfield that is a function of  a $16$-component chiral $SO(9,1)$ spinor, $\theta$,  as described in appendix~\ref{summact}. The   component interactions described by $\cP_{n}(\{\Phi\})$ in \eqref{chargeq} result from integrating a function of this  superfield over $\theta$.

\sm

The coefficient of any such interaction with $U(1)$  charge $q=-2w$ is a  $(w,-w)$  modular form $\F_{w}^{(0)}(\tau)$, where we recall that $n=4+w$ indicates the number of factors in the  product of $n$ fields, $\cP_{n}(\{\Phi\})$.
These component interactions  include the $R^4$ interaction, which has charge $q_{R^4}= 0$  and a coefficient that is a weight-$(0,0)$ form (a modular function),   as well as many other dimension-8  interactions with non-zero $q_U$. The  $O((\alpha')^{-1})$  interaction with the greatest value of $q$ is the  sixteen-dilatino interaction, $\Lambda^{16}$, which has $q_{\Lambda^{16}} =-24$ and its coefficient is a modular form $\F_{12}^{(0)}(\tau)$.   

\sm

\sm

The coefficient of the effective term $R^4$ is the solution of \eqref{laplaceone} with $w=0$ and $s=3/2$, which has the form 
\bea
S^{(0)}_4 =\frac{1}{\alpha'}\int d^{10}x  e\, \tau_2^\half\,E(\threeh,\tau) R^4=  \kappa^{-\half} \int d^{10}x  e\, E(\threeh,\tau) R^4\,.
\label{rfournew}
\eea
The weak-coupling ($\tau_2\to \infty$) expansion of $E(\threeh,\tau)$  that arises in \eqref{rfournew}  has the form 
\bea
E(\threeh,\tau)=2\zeta(3) {\tau_2^{\threeh}} + 4 \zeta(2) \, \tau_2^{-\half}+ 4 \pi \, \sum_{n\ne 0} {\sigma_{-2}(|n|) } {|n|^{\frac{1}{2}}}\, e^{2\pi (in\tau_1- |n| \tau_2)} \left(1+O(\tau_2^{-1})\right)\, , 
\label{eisenform}
\eea 
where $\sigma_{-2}(|n|)$ is the divisor sum defined in \eqref{eq:divisor-sum}. After including the factor of $\tau_2^{1/2}$ in \eqref{rfournew}, which translates the expression into the string frame, the power-behaved terms in \eqref{eisenform} correspond to tree-level and one-loop terms in string perturbation theory. Whereas the sum of exponential terms is interpreted as the contribution of D-instantons, each of which has an infinite series of perturbative corrections in powers of $\tau_2 = g^{-1}_s$. 

\sm

More generally, the coefficients of dimension-8 ($p=0$) interactions with $0\le w \le 12$  are proportional to the modified Eisenstein series', $E_{w}(s,\tau)$,  which are $(w,-w)$ modular  forms  defined in  appendix~\ref{holosol}, so that 
\bea
F_{w}^{(0)}(\tau) = c^{(0)}_w\, E_{w}(\threeh,\tau)\,.
\label{fwr4}
\eea
The normalisation constants $c^{(0)}_w$ may be determined by comparison of the tree-level term in $F_{w}^{(0)}(\tau) $ (the term of order $\tau_2^{3/2}$) with tree-level superstring perturbation theory (and we have chosen $c^{(0)}_0=1$ for later convenience).
The expression $E_{w}(\threeh,\tau)$, in \eqref{fwr4} is obtained by setting $s=3/2$ in \eqref{zwdef} and \eqref{zswdef},
\bea
\label{ewtext}
E_{w}(\threeh,\tau)= {2^{w-1} \sqrt{\pi}\over\Gamma ( \frac{3}{2} +w)}\, \cD_{w-1}\dots \cD_0\, E(\threeh,\tau)\
 =\sum_{(m,n)\ne (0,0)} \left({m+n\bar\tau\over m+n\tau}\right)^w \, {\tau_2^\threeh \over |m + n \tau|^{3}}  \, .
\ee
This  has a weak-coupling expansion (given in \eqref{expadef}) that  has two terms that are power behaved in $\tau_2=g_s^{-1}$ and an infinite sequence of  exponentially suppressed D-instanton and anti D-instanton contributions (where the D-instantons dominate by a factor of $\tau_2^{2w}$).  The normalisation in \eqref{ewtext} has been chosen so that $E_{w}(\threeh,\tau) \underset{\tau_2\to \infty}{\to} 2\zeta(3)\tau_2^{3/2} + O(\tau_2^{-1/2})$ for all $w$.  
The first order differential equations \eqref{dzsn} and \eqref{dbarzsn} satisfied by $E_{w}(s,\tau)$ imply Laplace eigenvalue equations given in \eqref{laplaceplusn} and  \eqref{laplaceminusn}.

\subsubsection{Terms at order $O\left(\alpha' \right)$}

The next order in the low-energy expansion has dimension 12.  The subset of the interactions that we are considering are those that are given by integrals of $\F_{w}^{(2)} (\tau)\, d^4\, \cP_{n}(\{\Phi\})$.  The  notation does not indicate which of the  $n=4+w$ fields in  $\cP_{n}(\{\Phi\})$ the four derivatives act on, or how they are contracted.  This is specified precisely by  the form of the scattering amplitudes, where  derivatives are replaced by momenta and $d^4$ becomes a quadratic monomial in Mandelstam invariants. The modular forms are given in this case by Eisenstein series with $s=5/2$,
\bea
\label{eisfive}
\F_{w}^{(2)}(\tau)  =c^{(2)}_w\, E_{w}(\fiveh,\tau)\,,
\eea
where the modified Eisenstein series, $E_{w}(\fiveh,\tau)$, is defined by setting $s=5/2$ in \eqref{zwdef} and \eqref{zswdef},
and $c^{(2)}_w$ are normalisation constants.  These may be fixed by comparison of the coefficient of the $(\tau_2)^{5/2}$  term  in $\F_{w}^{(2)}$ with the term of order $\alpha'$  in the expansion of $n$-point  tree-level superstring amplitude, where $n=4+w$.   The coefficient of the $w=0$ term is $c^{(2)}_0=1/2$.

\subsubsection{Terms at order $O\left((\alpha')^2 \right)$}

Up to now the only interaction  of order $O\left((\alpha')^2 \right)$ that has been fully analysed in the literature is the $U(1)$-conserving  interaction $\F_{0}^{(3)}(\tau) \, d^6R^4$ \cite{Green:2005ba}.  This is a $w=0$ component of the series of maximal $U(1)$ charge-violating interactions defined by the integrals  
\bea
S^{(3)}_{n,i}=\int d^{10}x\, e\, \F_{w,i}^{(3)} (\tau)\, d^6_{(i)}  \cP_{n}(\{\Phi\})\,,
\label{sixinter}
\eea
which will be discussed in the following sections.  The operator $d^6$ gives rise to a single kinematic invariant in the $w=0$ case, namely,  the symmetric monomial in Mandelstam invariants, $s^3+t^3+u^3$, in the expansion of the four-graviton amplitude.  However, as we will discuss, for general $w=n-4$, in particular for those with $w \ge 2$, there is a two-fold degeneracy in the kinematic invariants that is indicated by the index $i=1,2$ in the operator $ d^6_{(i)} $ in \eqref{sixinter}.

\sm

Whereas the coefficients of the $1/2$-BPS and $1/4$--BPS interactions satisfy Laplace  eigenvalue equations of the form \eqref{laplaceone},  the equation satisfied by $\F_{0}^{(3)}(\tau)$ is the inhomogeneous Laplace equation  \cite{Green:2005ba}, 
\bea
\left(\Delta - 12\right) \, \F_{0}^{(3)}(\tau)= -  E(\threeh,\tau)^2\,.
\label{inlap}
\eea
The zero Fourier mode of $ \F_{0}^{(3)}(\tau)$  with respect to $\tau_1$ of the solution to this equation contains four power-behaved terms, which correspond to genus-zero to genus-three contributions  in string perturbation theory, as well as the contribution of D-instanton/anti D-instanton pairs,
\bea
 \F_{0}^{(3)}(\tau) = \frac{2}{3} \zeta(3)^2 \tau _2^3 +\frac{4}{3}  \zeta(2)\zeta(3) \,\tau_2  + {8\over 5}\zeta(2)^2  \tau _2^{-1} +{4\over 27}\zeta(6) \tau _2^{-3} + \cO(e^{-4\pi \tau _2})\, .
\label{ethreehthreeh}
\eea  
The first three of the power-behaved terms are easily obtained by equating the coefficients in the expansion of the left-hand and right-hand sides of \eqref{ethreehthreeh} in powers of $\tau_2$ using the expansion \eqref{eisenform} for $E(\threeh, \tau)$.  However, the $\tau_2^{-3}$ term is in the kernel of the operator on the left-hand side and does not arise in the expansion of $E(\threeh, \tau)^2$  on the right-hand side.  The determination of its coefficient is therefore a little subtle, and originates from the presence of an infinite series of instantonic contributions, as was described in \cite{Green:2005ba} (and amplified in \cite{Green:2014yxa}).
These coefficients agree with the explicit string perturbation theory calculations.   The complete large-$\tau_2$ expansion of $F_{0}^{(3)}(\tau)$, including its rich assortment of instanton contributions, was determined in \cite{Green:2014yxa}. One of the challenges that we address in this paper is the extension of this equation to cases in which $w>0$.

\section{Consistency constraints on $1/8$-BPS modular coefficients.}
\label{eighthbps}
 
As described in section~\ref{coeffrev}  the $1/2$-BPS and $1/4$-BPS interactions $\F^{(p)}_{w}\, \cP_{n}$ ($p=0,2$)  in \eqref{chargeq} are related by first order differential equations that are implied by supersymmetry.  The Laplace eigenvalue equations that follow by iterating these equations determine the $\tau$-dependent coefficients to be modular forms that generalise the standard non-holomorphic Eisenstein series at $s=(3+p)/2$.

\sm

We will here generalise the $p=3$, $w=0$ case to coefficients $\F_{w, i}^{(3)}$  with $w>0$ by requiring consistency with the $w=0$ case.   We will consider the cases $w=1$ and $w=2$ explicitly although the procedure generalises in an obvious manner to all $w$. The cases of $w>2$ will be further studied in section~\ref{sec:softB} using soft limits. As we will see in section~\ref{w2case}, for $w=2$ a two-fold degeneracy arises ($i=1,2$), which is connected with the presence of two possible symmetric polynomials in the Mandelstam variables for massless six-particle scattering.  One particular combination arises in the tree-level expansion as we will see later.  This  implies that there is a second $w=2$ modular form that is unrelated to the $w=0$ case (in the manner of \eqref{covdf}) that has no tree-level (genus-zero) term in its zero Fourier mode.

\subsection*{A comment on notation}
 
\noindent  
The coefficients of the $O\left((\alpha')^2\right)$ $1/8$-BPS interactions are proportional to modular forms denoted $\EE_{w,i}(\tau)$
\bea
F^{(3)}_{w,i}(\tau)= c^{(3)}_{w,i}\, \EE_{w,i}(\tau)\,,
\label{p6def}
\eea
where the index $i$  again allows for a possible degeneracy, which will arise in cases with $w\ge 2$.
We will suppress this index for the cases with $w=0$ and $w=1$, where there is no degeneracy.  

\sm

Furthermore, we will choose a normalisation in which $c^{(3)}_{0}=1$ so that \eqref{inlap}  may be rewritten in the form
\bea
\Delta \EE_0 = 4  \bar \cD \cD \EE_0  = 12 \EE_0  - \left(E_0(\threeh)\right)^2\,,
\label{d6eq}
\eea 
where 
we are suppressing the arguments $\tau$, and we will often drop the labelling on covariant derivatives since it is implied by the context.\footnote{For example, an expression such as $\cD f^{(w,-w)}$  is identified with $\cD_w f^{(w,-w)}$.}

\subsection{The $p=3$, $w=1$ case}
\label{w1case}

This is the first example of a $U(1)$-violating interaction that is related  by supersymmetry to the $d^6 R^4$ interaction.  It involves fields in $\cP_{5}$ and contributes to amplitudes with $n=5$ external particles.  

\sm

We begin by  {\it defining}
\bea
 \EE_1 := a \cD \EE_0 \,,
\label{onedef}
\eea
where $a$ is a constant.  The coefficient $\EE_1$ is proportional to the $q_U=2$ modular form that is the coefficient of maximal $U(1)$-violating five-point interactions such as $d^6 G^2 R^3$.\footnote{This definition can be modified, for example, by defining $ \EE_1 := a \cD \EE_0+ b E_0 E_1$, but such a shift is mathematically trivial and  \eqref{onedef} coincides with the definition of physical interest as we discuss in the next section.}

\sm

With the definition \eqref{onedef}  (and noting \eqref{tauind})  the leading term at large $\tau_2$ is given by 
$\EE_1\underset{\tau_2\to \infty}{\to}  a\, \zeta(3)^2 \tau_2^3$. Applying the covariant derivative $\cD$ to the Laplace equation  \eqref{d6eq} leads to the inhomogeneous Laplace equation for $ \EE_1$,
\bea
\Delta_{(-)} \EE_1 =  4 \cD \bar \cD \EE_1  = 12 \EE_1 - \frac{3}{2}\,a\, E_1(\threeh )\, E_0(\threeh)\,,
\label{onederiv}
\eea
where we have used the properties of the Eisenstein series given in \eqref{dzsn} and \eqref{dbarzsn}. 
We will now check the consistency of this equation by applying $\bar\cD$ to it, which gives 
\bea
\Delta(\bar \cD \EE_1  ) =4 \bar \cD \cD(\bar \cD \EE_1  ) =  12\,  \bar \cD \EE_1   - \frac{3}{2}\,a\, \bar \cD \left( E_1(\threeh )\, E_0(\threeh) \right)\, .
\label{lapcon}
\eea
Using the relation
\bea
E_1(\threeh )\, E_0(\threeh) ={2\over 3} \cD \, (E_0(\threeh ))^2  \, ,
\label{eissquare}
\eea
we then obtain
\bea
\Delta(\bar \cD \EE_1  ) =  12  \, \bar \cD \EE_1  - \frac{a}{4}\Delta \left( (E_0(\threeh))^2 \right) \,.
\label{lapcorny}
\eea
The above equation can be recast as
\bea
\Delta\left(\bar \cD \EE_1 +\frac{a}{4} (E_0(\threeh))^2 \right) =  12  \left(\bar \cD \EE_1 +\frac{a}{4} (E_0(\threeh))^2 \right)  - 3\,a\,  (E_0(\threeh))^2 \, . 
\label{intstep}
\eea
This reproduces \eqref{d6eq}  if we make  the  identification 
\bea
 \bar{\cD} \EE_1  =3 a\, \EE_0 -\frac{a}{4} (E_0(\threeh))^2\,,
\label{barddef}
\eea
in which case \eqref{intstep} reduces to the $w=0$ Laplace equation, \eqref{d6eq}.  

\sm

The value of $a$ is arbitrary, but for later convenience we we will make the choice $a=2$. With this choice, the first order differential relations for $\EE_1$ become
\bea \label{eq:DEE0}
 \EE_1 &=& 2 \cD \EE_0\, , \\
  \bar \cD \EE_1  &=& 6\, \EE_0 -\frac{1}{2} (E_0(\threeh))^2 \, ,
\label{eq:DEE1}
\eea
and the inhomogeneous Laplace equation \eqref{onederiv} becomes
\bea \label{onednew}
\Delta_{(-)} \EE_1  = 12 \EE_1 - 3\, E_1(\threeh )\, E_0(\threeh)\,.
\eea

\subsection{The $p=3$, $w=2$ cases}
\label{w2case}
We may anticipate that there is a two-fold degeneracy in $w=2$, $p=3$ modular forms, labelled by an index $i$ on $\EE_{2,i}$, where $i=1,2$.  This expectation is based on the following analysis, coupled with known facts about the low-energy expansion of superstring six-particle scattering amplitudes.     
We will consider the $i=1$ case in some detail and follow that with a somewhat more conjectural discussion of the $i=2$ case, which will be justified from the consideration of scattering amplitudes.

\subsubsection*{The modular form $\EE_{2,1}$}   \label{sec: EE21}

The $\EE_1$ Laplace equation \eqref{onednew}  may be rewritten using the identification $\Delta_{(-)}= \Delta_{(+)} +2$ (see  \eqref{difflap}), giving
\bea
\Delta_{(+)}\EE_1 =  4 \bar \cD \cD \EE_1  = 10 \EE_1 - 3\, E_1(\threeh )\, E_0(\threeh)\, .
\label{onenewderiv}
\eea
Applying the covariant derivative, $\cD$, to this equation gives
\bea
\Delta_{(-)} (\cD \EE_1)=  4 \cD \bar\cD (\cD \EE_1)  = 10 \cD \EE_1  - 3\, \cD \left( E_1(\threeh )\, E_0(\threeh)  \right) \,.
\label{sixdriv}
\eea
To proceed, we will define
\bea
\EE_{2,1} := b\, \cD \EE_1  \,.
\label{newsix}
\eea
Substituting \eqref{newsix} in \eqref{sixdriv} gives 
\bea
\Delta_{(-)} \EE_{2,1} =  4 \cD \bar\cD \EE_{2,1}  = 10 \EE_{2,1} - 3b\, \cD \left( E_1(\threeh )\, E_0(\threeh)  \right) \,.
\label{sixnew}
\eea
Applying  $\bar \cD$ to both sides of  the above equation  leads to
\begin{align}
 \Delta_{(+)}(\bar\cD \EE_{2,1}) = 4 \bar \cD \cD (\bar\cD \EE_{2,1})  &= 10  (\bar \cD \EE_{2,1}) -3b\,\bar \cD \cD \left( E_1(\threeh )\, E_0(\threeh)  \right)  \nn\\
 &= 10  (\bar \cD \EE_{2,1}) -\frac{3b}{4} \Delta_{(+)}( E_1(\threeh )\, E_0(\threeh))  \,. 
\label{sixderivs}
\end{align}
Now we can identify 
\bea
\bar\cD \EE_{2,1} = {5 b\over 2} \, \EE_1 - \frac{3b}{4} E_1(\threeh )\, E_0(\threeh)  \,,
\label{fonerev}
\eea
in which case \eqref{sixderivs} reduces to \eqref{onenewderiv}. Again, $b$ is arbitrary and is correlated with the normalisation constant $c^{(3)}_{2,1}$. It is again convenient to make the choice $b=2$ so that the first-order  differential  equations for  $\EE_1 $ become
\bea \label{sixderivs-2}
\EE_{2,1} &=& 2 \cD \EE_1  \, , \\
 \qquad \bar\cD \EE_{2,1} &=& 5 \, \EE_1 - \frac{3}{2} E_1(\threeh )\, E_0(\threeh)  \, .
 \label{eq:DbE21}
\eea
The inhomogeneous Laplace equation that follows by combining these equations is 
\bea  \label{eq:LapE21}
\Delta_{(-)} \EE_{2,1} =  4 \cD \bar\cD \EE_{2,1}  = 10 \EE_{2,1} - {15 \over 2}\,  \left( E_0(\threeh ) E_2(\threeh) + {3\over 5} E_1(\threeh) E_1(\threeh ) \right) \, ,
\eea
where we have used the relation $\cD \left( E_1(\threeh )\, E_0(\threeh) \right)=5/4 E_0(\threeh ) E_2(\threeh) + 3/4 E_1(\threeh) E_1(\threeh )$. 

\sm

Since $\EE_{2,1}(\tau)=2\cD \EE_1(\tau)=4 \cD \cD \, \EE_0(\tau)$, it  is  straightforward to deduce its large-$\tau_2$ expansion from \eqref{ethreehthreeh}, which gives 
\bea
 \EE_{2,1} (\tau) =8 \zeta(3)^2 \tau _2^3 +\frac{8}{3}  \zeta(2)\zeta(3) \,\tau_2   + \frac{8}{9}\zeta(6)  \tau _2^{-3} + \cO(e^{-2\pi \tau _2})\, .
\label{e21}
\eea  
Note that the genus-two term proportional to $\tau_2^{-1}$ is absent.    This is an  example  of  a general point concerning  perturbative terms in the  $p=3$ coefficient  modular forms with $w\ge2$, which are positive or negative integer powers of $\tau_2$.
The action of successive covariant derivatives on a particular power $\tau_2^x$  is
\bea
\cD_{w-1}\dots \cD_{u}\, \tau_2^x=  2^{u-w}\prod_{j=u}^{w-1} (x+j)\, \tau_2^x\, ,
\label{multideriv}
\eea
which is killed by a sufficient number of derivatives if $-x>j$ and $x\in \ZZ$.
The $p=0$ and $p=2$ perturbative  interactions have half-integer powers of  $\tau_2$ in the Einstein frame and so the above argument does not apply.

\sm

 \subsubsection{Comments concerning six-particle amplitudes}
 
Before discussing the other $p=3$, $w=2$ modular form, $\EE_{2,2}$, we will  discuss the $w=2$ contribution to six-particle scattering amplitudes with $q_U=-4$. 
As mentioned earlier, the  structure of the six-particle superstring amplitude suggests that there should be two distinct modular forms that contribute to $\EE_{2,i}\,d^6_{(i)}\, \cP_{6}$.   We have seen that one of these,  $\EE_{2,1}$, has a large-$\tau_2$ (weak coupling) expansion that contains a component proportional to $\tau_2^3$ that corresponds to a tree-level contribution (in the Einstein frame)  to the $d^6_{(1)}\, \cP_{6}$ interaction.   This matches the expectation based on the explicit tree-level superstring calculations to be described in section~\ref{uoneamp}, where we will see that the derivative factor $d^6_{(1)}$ translates into a particular cubic polynomial in the Mandelstam invariants of the six-particle amplitude,\footnote{In the following expressions the Mandelstam invariants are defined by $s_{ij}=-(p_i+p_j)^2$ and $s_{ijk}= -(p_i+p_j+p_k)^2$, where $p_i$ is the momentum for the $i$th external massless particle, which satisfies $p_i^2=0$ and $\sum_{i=1}^6 p_i=0$.} 
\bea
d^6_{(1)}\to \cO^{(3)}_{6,1} :={1\over 32} \left( 10  \sum_{1 \leq i<j \leq 6} s_{ij}^3 + 3 \sum_{1 \leq i<j<k \leq 6} s_{ijk}^3 \right)\,.
\label{sixinv1}
\eea  
This invariant has been chosen to coincide with the combination that arises in the tree-level calculation of the six-particle amplitude as will be shown in section~\ref{uoneamp}. Furthermore, in the soft limit, it reduces to the unique kinematic  invariant of the five-particle amplitude, 
\bea
 \label{eq:soft-1}
\cO^{(3)}_{6,1}\bigl|_{p_6\rightarrow 0} \rightarrow  \cO^{(3)}_{5} \, ,
\eea
 where we have defined $\cO^{(3)}_{5} := 1/2\, \sum_{ 1\le i<j \le 5} s_{ij}^3$. 

\sm 

The above soft behaviour of the six-particle kinematic invariant   $\cO^{(3)}_{6,1}$ is consistent with the fact that  a six-particle  $U(1)$-violating amplitude with a number of external $\Z $ states  reduces to a five-particle amplitude  with one less $\Z $ when one of the  $\Z $s becomes soft.  As we will  see in greater detail in section~\ref{softsec} this is related to the fact that the  coefficient of the  six-particle  interaction  $\F_{2,1}^{(3)}(\tau)$ is a covariant derivative of  $\F_1^{(3)}(\tau)$.

\sm

 The other independent kinematic structure translates into
\bea
d^6_{(2)}\to \cO^{(3)}_{6,2} := 2 \sum_{1 \leq i<j \leq 6} s_{ij}^3 - \! \sum_{1 \leq i<j<k\leq 6} s_{ijk}^3 = {1\over 8} \sum_{\rm  permutation} s_{12}s_{34}s_{56}  \, ,
\label{sixinv2}
\eea  
where  the  sum is over $6!$ permutations. Up to an overall constant this is the unique symmetric polynomial  of degree 3 in the six-particle Mandelstam invariants that vanishes  in the single soft limit, $p_i\to 0$ for any $i$.   This soft behaviour implies that in any maximal $U(1)$-violating  amplitude with external axio-dilaton states there is at least one derivative on each $\Z $ or $\bar \Z $.  This is important since it shows that the coefficient of $\cO^{(3)}_{6,2}$,  which is $\F_{2,2}^{(3)}(\tau)$, does not come from the expansion of the coefficient of a $n=5$ amplitude.  If it did it would contain at least one ``naked" $\Z $ or $\bar \Z $ factor (a factor with no derivative  acting on it), as we will see in section~\ref{uoneamp}.

\subsubsection*{The modular form $\EE_{2,2}$} \label{sec:E22}

The coefficient of $d^6_{(2)}\, \cP_{6}$ is given by $F^{(3)}_{2,2}(\tau)= c^{(3)}_{2,2}\, \EE_{2,2}(\tau)$, which is proportional to the second $w=2$, $p=3$ modular form. 
The following  discussion of $\EE_{2,2}$  is based  on the following inputs.
\begin{itemize}
\item[1]
We will assume that $\EE_{2,2}(\tau)$ satisfies a $SL(2,\ZZ)$-covariant first-order differential equation analogous to \eqref{eq:DbE21}.\footnote{This assumption will be justified later in section \ref{constraints}.}
\item[2]
Since $\cO^{(3)}_{6,2}$ does not contribute to the  tree-level  $p=3$ and $w=2$ interaction, so the leading term in  $\EE_{2,2}(\tau)$ in the large-$\tau_2$ limit is the genus-one term of order $\tau_2$.
\end{itemize}

\sm

Item $1$ implies that the inhomogeneous term in the first-order differential equation of $\bar\cD \EE_{2,2}$ in terms of a linear combination of $\EE_1$ and $E_0(\threeh) E_1(\threeh)$, namely 
\bea 
\bar\cD \EE_{2,2} = c_1 \EE_1 + c_2 E_0(\threeh )\, E_1(\threeh) \, .
\eea
Item $2$ determines the relative coefficients $c_1$ and $c_2$ of these two terms that is required for the $\tau_2^3$  contribution to cancel.  Making use of the  perturbative  expansions
\bea
\EE_1 = 2 \cD \EE_0 = 2 \zeta(3)^2 \tau _2^3 +\frac{4}{3}  \zeta(2)\zeta(3) \,\tau_2 - {8\over 5}\zeta(2)^2  \tau _2^{-1} - {4\over 9}\zeta(6) \tau _2^{-3} + \cO(e^{-2\pi \tau _2}) \, ,
\eea
and 
\bea
E_0(\threeh )\, E_1(\threeh) = 4 \zeta(3)^2 \tau _2^3 +\frac{16}{3}  \zeta(2)\zeta(3) \,\tau_2 - \frac{16}{3} \zeta(2)^2  \tau _2^{-1} + \cO(e^{-2\pi \tau _2}) \, ,
\eea
we see that the cancellation of the tree-level term (proportional to $\tau_2^3$) requires  $c_2=-1/2 \, c_1$, and therefore
\bea \label{eq:DbE22}
\bar\cD \EE_{2,2} = c_1 \left( \EE_1 -{1\over 2} E_0(\threeh )\, E_1(\threeh) \right)\, .
\eea
Since the tree-level term  in the large-$\tau_2$ expansion of the right-hand side is designed to be zero, the leading power-behaved term is the genus-one term proportional to $\tau_2$. The value of the constant $c_1$ may therefore  be determined by an explicit evaluation of the six-point maximal $U(1)$-violating one-loop amplitude.

\sm

Comparing (\ref{eq:DbE21}) and  (\ref{eq:DbE22}), and using $E_0(\threeh)E_1(\threeh)=2\bar \cD  (E_1(\threeh)^2)$, we find 
\bea \label{eq:E22asE21}
\EE_{2,2}  ={ c_1 \over 5} \left( \EE_{2,1} - 2 E_1(\threeh)E_1(\threeh)\right)\, .
\eea
The Laplace equation satisfied by $\EE_{2,2}$ follows by applying $\Delta_{(-)}$ to  the above equation and using \eqref{eq:LapE21}, 
\bea
\label{laplace22}
\Delta_{(-)} \EE_{2,2} = 10 \EE_{2,2}  - {5c_1 \over 2} \left(E_0(\threeh) E_2(\threeh) - E_1(\threeh) E_1(\threeh) \right) \, ,
\eea
which has the same eigenvalue as the Laplace equation for $\EE_{2,1}$ but with a different inhomogeneous term such that the tree-level contribution vanishes. 

\sm

By knowing $\EE_{2,1}=4\cD \cD \EE_0$, from \eqref{eq:E22asE21} we may obtain $\EE_{2,2}$ up to an unknown overall constant $c_1$. Expanding near the cusp,  $\tau_2\to \infty$,  and  dropping the constant factor $c_1$ gives the following weak coupling expansion, (in the Einstein frame, as usual) 
\bea \label{eq:E22-expansion}
\EE_{2,2}(\tau)  =   \zeta(2)  \zeta(3) \tau_2 - 
{4 \over 15} \zeta(2)^2 \tau^{-1}_2  +  {1 \over 15}  \zeta(6) \tau_2^{-3}  + \cO(e^{-2\pi \tau _2}) \, ,
\eea
which contains  contributions corresponding to genus-one, genus-two  and genus-three superstring loop amplitudes, but with  no tree contribution.

\subsection{Comments on coefficients with  $w>2$ and on $p>3$}

As was commented on earlier,  the modular  form coefficients with weights $w>2$ are related to the lower-weight ones by applying  covariant derivatives.  Thus, the modular form $\EE_{w,2}(\tau)$ ($w>2$)  accompanying the $n$-particle $p=3$ kinematic invariant,  $\cO^{(3)}_{n-4,2}$  can be obtained by acting with covariant derivates on $\EE_{2,2}(\tau)$. This corresponds the expansion of $\EE_{2,2}(\tau)$ in powers of $\Z $ fields around a fixed background, as will be discuss  further in section \ref{uoneamp}. 

\sm

Interactions with $p>3$, which have dimension $> 14$, are non-BPS terms.  Therefore in general they are expected to receive all-loop perturbative contributions, in addition to non-perturbative D-instanton contributions.  The first such interaction is $d^8R^4$, which has a unique kinematic invariant $s^4+t^4+u^4$ and a coefficient  $\mathcal{E}_0^{(4)}(\tau)$. Once again,  higher-point terms have a degenerate  set  of kinematic invariants. In fact, already at five points  there are  two independent kinematic invariants, 
\bea
\label{fiveinv}
\mathcal{O}_{5,1}^{(4)} = \sum_{i<j} s_{ij}^4 + {1\over 12}  (\sum_{i<j} s_{ij}^2)^2 \, , \qquad
\mathcal{O}_{5,2}^{(4)} = \sum_{i<j} s_{ij}^4 - {1\over 4}  (\sum_{i<j} s_{ij}^2)^2 \, ,
\eea 
where $\mathcal{O}_{5,1}^{(4)}$ is the invariant arising in the five-point $U(1)$-violating tree-level string amplitude, which has a single-particle soft limit that results in the unique  four-particle kinematic factor.  
As in the case  of  $\mathcal{O}_{6,1}^{(3)}$,  the coefficient  of $\mathcal{O}_{5,1}^{(4)}$ can be obtained by acting with a covariant derivative on  the coefficient of $d^8R^4$, so  its coefficient is given by $\cD \mathcal{E}_0^{(4)}(\tau)$, as  was discussed  in  \cite{Green:2013bza}.

The second $p=4$ five-particle kinematic invariant  in  \eqref{fiveinv},  $\mathcal{O}_{5,2}^{(4)}$, is determined by requiring it to vanish in the soft limit. Therefore, $\mathcal{O}_{5,2}^{(4)}$  does not appear at tree level and first appears at one loop (and its form precisely agrees with the expression obtained from the  matrix $M_7'$ in equation (5.5) of \cite{Green:2013bza}). Clearly, the same analysis applies to the interactions with more general $w$'s and $p$'s \footnote{We have checked that the kinematic invariants obtained from the matrices $M_8'$ (i.e. $w=1, p=5$) and $M_9', M_9''$ (i.e. $w=1, p=6$) in equation (5.5) of \cite{Green:2013bza} also vanish in the soft limit. This is expected from our considerations since $M_8', M_9', M_9''$  do not contribute at tree level (as was also the  case with $M_7'$).}.  We expect that it is generally true that interactions can be separated into different sets whose coefficients are related by covariant derivatives (which will become more evident in the next section).  However, as will be shown in section~\ref{constraints}, equations such as \eqref{eq:DbE21} and \eqref{eq:DbE22}, or the Laplace equations discussed in the previous section, are special properties of $F$-terms with $p \leq 3$.


\section{Low-energy expansion of $U(1)$-violating scattering amplitudes}
\label{uoneamp}
Before discussing details of the scattering amplitudes  we will make some important comments about the special features of maximal $U(1)$-violating amplitudes.

\subsection{Preliminary comments concerning maximal $U(1)$-violating amplitudes}

The simplest class of superstring amplitudes are those $n$-particle amplitudes that violate $U(1)$ maximally since these do not have any massless poles in any channel.  At low orders in the low-energy expansion these amplitudes correspond to the contact  interactions in the effective  action that was considered in earlier sections.  
Simply setting the  modulus field equal to its background value, $\tau=\tau^0= \chi + i/g_s$ in the effective action  leads immediately to  expressions for on-shell maximal-violating $n$-particle scattering amplitudes, in which each of the fields in $d_{(i)}^{2p}\, \cP_n(\{\Phi\})$ is associated with an external on-shell state.  The coupling constant dependence is determined by $(\alpha')^{p-1}(\tau_2^0)^{\frac{1-p}{2}}\, \F_{w}^{(p)}(\tau^0)=\kappa^{\frac{p-1}{2}}\, \F_{w}^{(p)}(\tau^0)$ with $w=n-4$.   
For example, the leading correction to the four-graviton amplitude beyond the classical  supergravity amplitude has $p=0$ and is  proportional to $\kappa^{-1/2}\, E(\threeh, \tau^0)\, R^4$, while the sixteen-dilatino amplitude is proportional to $\kappa^{-1/2}\,E_{12}(\threeh, \tau^0)\, \Lambda^{16}$,  and so on.  

\sm

Among the $n$-field terms in $\cP_n(\{\Phi\})$ there are dimension-$8$ terms containing at most two powers of $\bar\tau$ and two powers of $\tau$, which are related by supersymmetry to the $R^4$ interaction.  For example there is a term of the form $d^2 \tau\,d^2 \tau\,d^2 \bar\tau\,d^2\bar\tau$.  But there can be no ``naked'' powers of $\tau$ or $\bar\tau$ in  $\cP_n(\{\Phi\})$  -- i.e., no factors of $\tau$ or $\bar\tau$ that are not acted on by derivatives.  The naked $\tau$ and $\bar\tau$ fields are moduli that enter into the instanton contributions to the coefficients  $\F_{w,i}^{(p)}(\tau)$  in \eqref{chargeq}.

\sm

There are, however,  further maximal $U(1)$-violating scattering  amplitudes that have arbitrary numbers of additional $\tau$ fluctuations  that are  obtained by expanding the modular coefficients in the action in fluctuations of $\tau$.   
 As will be described in the next sub-section, it is important in performing such an expansion to parameterise the fluctuations in a manner that preserves the induced $U(1)$ symmetry.   This is a special case of the general procedure for  expanding  nonlinear sigma models defined on a $\GG/\KK$ coset space, in which the fluctuating fields are defined by a normal coordinate expansion that is covariant with respect to the $\KK$ symmetry. In our case we need to re-parameterise the fluctuations $\delta\tau$ around the background $\tau^0$ in order that the fluctuations transform with a given $U(1)$ charge.  In the next sub-section we will see that the covariant expansion is given in terms of a reparameterisation of the complex scalar $\tau$ of the form of a Cayley map from the upper-half plane to the unit disk,
 \bea
 \tau  \to \Z  = \frac{\tau-\tau^0}{\tau-\bar\tau^{0}}\,.
 \label{Brefdef}
\eea

\begin{figure}[h]
\begin{center}
\begin{tikzpicture}[scale=1]
\begin{scope}[xshift= -4.5  cm,yshift=0.0cm,very thick, every node/.style={sloped,allow upside down}]
\draw [red,ultra thick,dashed, domain=55:125] plot ({cos(\x)}, {sin(\x)});
\draw [red,ultra thick,dashed, domain=-15:-50] plot ({cos(\x)}, {sin(\x)});
  \filldraw [black]  (0,0) ellipse (.3 and .3);
\draw [ultra thick] (-1,1) -- (0,0);
\draw [ultra thick]  (1,1) --  (0,0);
\draw [ultra thick] (-1.0,-1.0) -- (0,0);
\draw [ultra thick]  (-1.4,0) --  (0,0);
\draw [ultra thick, decorate,decoration=snake]  (0.8,-1.3) --  (0,0);
\draw [ultra thick, decorate,decoration=snake]  (1.4,0) --  (0,0);
           \draw  (1.1, 0.3) node{$\Z $};
           \draw  (0.2,-1.1) node{$\Z $};
     \draw      (-1.2,-1.0)  node{$1$} ;
         \draw      (-1.6,0.0)  node{$2$} ;
            \draw      (-1.2,1.0)  node{$3$} ;
               \draw      (1.2,1.0)  node{$n$} ;
                \draw      (2.1,0.0)  node{$n+1$} ;
     \draw      (1.4,-1.6)  node{$n+m$} ;
   \draw  (-3.2,0.0) node{$\cD^m \F^{(p)}_{n-4,i}(\tau^0)$};
             \draw  (0.0,-1.9) node{$(a)$};
          \end{scope}
\begin{scope}[xshift= 4.5 cm,yshift=0.0cm,very thick, every node/.style={sloped,allow upside down}]
\draw [red,ultra thick,dashed, domain=100:170] plot ({cos(\x)}, {sin(\x)});
  \filldraw [black]  (0,0) ellipse (.3 and .3);
\draw [ultra thick, decorate,decoration=snake]  (0.0,1.2) --  (0,0);
\draw [ultra thick, decorate,decoration=snake]  (-1.0,-1.0) -- (0,0);
\draw [ultra thick, decorate,decoration=snake] (-1.4,0) --  (0,0);
\draw [ultra thick, decorate,decoration=snake]  (0.8,-1.3) --  (0,0);
\draw [ultra thick, decorate,decoration=snake]  (1.4,0) --  (0,0);
           \draw  (1.1, 0.4) node{$\bar \Z $};
           \draw  (0.2,-1.1) node{$\bar \Z $};
            \draw  (0.30,1.1) node{$ \Z $};
                 \draw  (-1.3,0.4) node{$ \Z $};
                        \draw  (-0.80,-1.1) node{$ \Z $};
     \draw      (-1.2,-1.0)  node{$3$} ;
         \draw      (-1.6,0.0)  node{$4$} ;
               \draw      (0.0,1.6)  node{$n$} ;
                \draw      (1.7,0.0)  node{$1$} ;
     \draw      (1.0,-1.0)  node{$2$} ;
   \draw  (-3.2,0.0) node{$\cD^{n-4} \F^{(p)}_{0,i}(\tau^0)$};
                \draw  (0.0,-1.9) node{$(b)$};
          \end{scope}
\end{tikzpicture}
\end{center}
\caption{  Maximal $U(1)$-violating amplitude  (a)  With $n$ particles interacting via $d^{2p}_{(i)} \,\cP_n$ and $m$ fluctuations of the complex scalar field $\Z $.  (b) With two $\Z $ particles  and two $\bar \Z $ particles  interacting via $d^{2p}_{(i)} \,\cP_4$ and $n-4$ fluctuations of $\Z $. } 
\label{fig:flucts}
\end{figure}
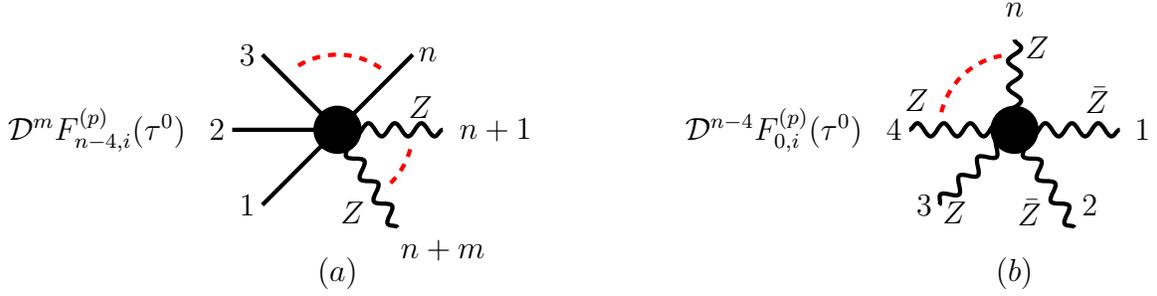

\sm

Figure~\ref{fig:flucts}(a) illustrates a  maximal $U(1)$-violating amplitude with $n$ external  states taken from $\cP_n$ and $m$ scalar particle fluctuations (so $q_U=8-2m-2n$).  This amplitude can be expressed in the form \footnote{In this expression, as well as later expressions for amplitudes, each field  is  to be replaced by its wave function in momentum space (i.e., each field represents an external state of definite momentum).}  
 \bea
 \cD^m \F^{(p)}_{n-4, \, i}(\tau^0)\, \cO_{m+n,i}^{(p)} \,\cP_n(\{\Phi\})\, \Z ^m\,,
 \label{contamp}
 \eea
$\cO_{m+n,i}^{(p)}$ is a monomial in the Mandelstam invariants of the $(m+n)$-particle amplitude of degree $p$. 
\sm
Among many such component amplitudes, there are maximal $U(1)$-violating interactions  in which all $n$ external states are complex scalars.  These arise by choosing the component of $\cP_4(\{\Phi\})$ that has two $\Z $ states and two $\bar  \Z $ states and expanding 
$\F^{(4)}_0(\tau)$ to give $n-4$ $\Z $ fluctuations
which is illustrated in figure~\ref{fig:flucts}(b).  These maximal $U(1)$-violating amplitudes of scalars are simply monomials of Mandelstam invariants given by 
\bea
 \cD^{n-4} \F^{(p)}_{0, \, i}(\tau^0)\, \cO^{(p)}_{m+n,i}\,  \cO^{(4)}_{\Z^2\bar{\Z}^2} \, \Z ^{m+n-2}\, \bar \Z ^2\,, 
 \label{contamp2}
 \eea
 where the Mandelstam invariant $\cO^{(4)}_{\Z^2\bar{\Z}^2}$ represents $8$ derivatives acting on two $\Z $ states and two $\bar  \Z $ states of $\cP_4(\{\Phi\})$. How these derivatives act is determined by maximal supersymmetry. 

\subsection{Expansion in scalar field  fluctuations}
\label{tauexpand}

In order to discuss the structure of the amplitudes in more detail we need to consider the appropriate definition of the fluctuating scalar fields. To illustrate the issue, consider the expansion of any modular form, $\F_w(\tau)$, in powers of normalised small fluctuations around the background, $\delta \tau=\tau-\tau^0$, 
\bea \label{eq:hat-tau}
\hat{\tau}:= {i\over 2} { \tau-\tau^0 \over \tau^0_2 } = {i\over 2} { \delta \tau \over \tau^0_2 } \, , \qquad \qquad \bar{\hat{\tau}}:=- {i\over 2} { \delta \bar{\tau} \over \tau^0_2 }\, . 
\eea 
The quantity $\hat \tau$ does not transform covariantly under $SL(2,\ZZ)$ acting on $\tau$ and $\tau^0$.   Consequently, the  coefficients in the expansion of a $w=0$ modular form, $F_0(\tau)$, in powers of $\hat \tau$, 
\bea
\label{eq:expansion} 
F_0(\tau^0 + \delta \tau) = F_0(\tau^0) +2 i \tau^0_2 \, \partial_{\tau^0}F_0(\tau^0) \hat{\tau}  
 - 2 (\tau^0_2)^2 \, \partial^2_{\tau^0}F_0(\tau^0) \hat{\tau}^2  + \cdots\,, 
\eea
do not transform as modular forms.  
In such a parameterisation the Feynman rules have contact terms that vanish on shell and  the evaluation of covariant  amplitudes is very complicated.

\sm

 The appropriate redefintion of $\tau$  is achieved by the $SL(2,\CC)$ transformation that defines $\Z $ in \eqref{Brefdef} and which has an expansion as an infinite series of powers of $\hat \tau$,
 \bea
 \Z =-(\hat \tau+ \hat \tau^2 + \hat \tau^3+\dots)\,.
 \label{btauh}
 \eea
   As  required, the transformation of  $\Z $  under the action of  $SL(2,\ZZ)$ is the linear $U(1)$ transformation given by 
\bea
\Z  \to \frac{c \bar\tau^0+d}{c \tau^0 + d}\,\Z \,,
\label{slact}
\eea
which means that  $\Z $ is a weight $(-1,1)$ modular form and so carries $U(1)$  charge $q_\Z =-2$,

\sm 

From  the  definition of  $\delta \tau$  we have 
\bea
\delta \tau = 2i \tau_2^0 {\Z  \over 1- \Z }\,,
\eea
and it is straightforward to verify that the Taylor expansion of $F_0(\tau)$ around the background $\tau=\tau^0$ given in (\ref{eq:expansion}) has the required covariant form,
\bea
F_0(\tau^0 + \delta \tau(Z)) =  \sum_{w=0}^\infty \,2^w\, \cD_{w-1}\dots \cD_0\, F_0(\tau)\bigg|_{\tau=\tau^0} \left( \frac{\Z ^w}{w!}  \right) + \cdots \, ,
\label{bexpand}
\eea
where we have indicated  the $\Z $-dependence in the fluctuations of $\tau$ around  its  background value. 
The reparameterisation has converted the derivatives in \eqref{eq:expansion} into covariant derivatives and the coefficients of the powers of $\Z ^w$ are weight $(w,-w)$ modular forms  which  compensates  for  the  charge of  $\Z ^w$.   The systematics of this expansion will be reflected in the expressions for $n$-particle scattering amplitudes  in the following.

\sm

The parameterisation in terms of $\Z $ follows closely the discussion in  \cite{Schwarz:1983qr}, where the type IIB supergravity equations were formulated in a $SU(1,1)$-covariant manner. This is briefly reviewed in appendix~\ref{app:B-field}.   However, in considering the type IIB superstring it is important to  remain in the gauge $\phi=0$ as in \eqref{phigauge}. In that case after setting $\phi=0$ in \eqref{predef} we have
\bea
P_\mu (\tau) := i {\partial_{\mu} \tau \over 2\tau_2} = \frac{\partial_\mu \Z }{1-\bar \Z \Z }\, \left(\frac{1- \bar \Z }{1-\Z }\right) \,.
\label{prenew}
\eea 

\sm

The supergravity action expressed in terms of the fluctuations $\hat\tau$ has interaction terms that vanish on shell.  For example, the expansion of $S_\tau$  in \eqref{scalact}  in powers of $\hat\tau$ and $\bar{\hat\tau}$ leads to interactions, such as $\partial_\mu\hat\tau\, \partial^\mu\bar {\hat\tau} \,\hat\tau$  that violate $U(1)$ but vanish on shell.  
It is an important consequence of the parameterisation of the complex scalar field  in terms of $\Z $ that such on-shell vanishing terms are absent.  Thus, the scalar field action $S_\tau$  in \eqref{scalact} is replaced by 
\bea
S_\Z  = -\frac{2}{\kappa^2} \int d^{10} x \,e\, 
 \frac{\partial_\mu \Z \partial^\mu \bar \Z }{(1-\bar \Z \Z )^2} \,.
\label{Baction}
\eea
All terms in the expansion of this expression in powers of  $\bar \Z \Z $ transform as $U(1)$ singlets and none of them vanish on shell.

\subsubsection*{Other fields}

For consistency it  is important to perform  reparameterisations of other massless fields. For example, consider the Dirac lagrangian density for  the dilatino  of charge $q_\Lambda=- 3/2$,
\bea
\bar \Lambda^a  \gamma^{\mu}(\partial_{\mu} + i q_{\Lambda} Q_{\mu}) \ \Lambda^a \, .
\label{diraclam}
\eea
In a fixed background $\tau=\tau^0$ we need to  use the expression for  $Q_{\mu}$  given in \eqref{newq}.   The resulting Dirac action contains  $U(1)$-violating contact interactions that vanish on shell, the lowest  order being of the form $\bar \Lambda \gamma^\mu \partial_\mu \Z \, \Lambda$.   These again  lead to very complicated  Feynman rules and  are removed by the appropriate  field redefinition, 
         \begin{align}  \label{eq:Lambda}
  \Lambda'_a  =  \Lambda_a \left( { 1 -   \Z  \over 1 -\bar{\Z } }\right)^{q_{\Lambda}/2} \, .
       \end{align}  
It is straightforward to check that the redefined $\Lambda'_a$ transforms linearly by the induced $U(1)$  transformation,
 \begin{align}   
\Lambda'_a \rightarrow \left(  { c \bar{\tau}^0    + d \over c {\tau}^0  + d  }\right)^{q_{\Lambda} /2} \Lambda'_a \, ,
 \end{align}
under $SL(2,\ZZ)$. 
Furthermore, when the interactions are expressed in terms of $\Lambda'_a$, all the $U(1)$-violating (and on-shell vanishing) vertices in  the reparameterised  \eqref{diraclam} are removed.

The  same considerations  apply to the reprameterisation of the gravitino, $\psi$,  which has $q_\psi=-1/2$, as  well as the third-rank field strength $G$ with $q_G=-1$.
After these reparameterisations  the $n$-particle contact interactions in \eqref{chargeq}  are  transformed  into
\bea
\F_{n-4,i}^{(p)}(\tau^0+ \delta \tau(\Z )) \,d^{2p}_{(i)} \cP_n^{(p)}(\{\Phi)\}) \left( { 1 -   \Z  \over 1 -\bar{\Z } }\right)^{n-4} \, ,
\label{transformp}
\eea  
where we have indicated the $\Z $-dependence in the fluctuations $\delta \tau$ as well as the explicit $\Z $-dependence coming  from the transformation of the fields in $\cP_n^{(p)}(\{\Phi\})$.
The expression \eqref{transformp} is appropriate for performing  a covariant expansion in $m$  powers of $\Z $, resulting in an expression for the $(m+n)$-particle amplitude of the form \eqref{contamp}.

\subsection{Supersymmetric scattering amplitudes}
\label{susyamp}

\subsubsection*{Ten-dimensional helicity spinors}

In order to describe the super-amplitudes, we introduce the ten dimensional spinor helicity formalism, following \cite{CaronHuot:2010rj}.  The spinor-helicity formalism expresses the momentum of any massless state in terms of chiral  bosonic spinors $\lambda^A_a$ satisfying the ten-dimensional Dirac equation,
\bea
 (\gamma^{\mu})^{B }_{\  \ A}\, p_\mu \lambda^A_a=0 \,,
\label{dirace}
\eea
where $A=1,\dots,16$  labels  the components of a $SO(9,1)$ chiral spinor under and $a=1,\dots ,8$ labels the components of a $SO(8)$ spinor of the  little group of massless states.  The  ten-dimensional gamma matrices,  $( \gamma^{\mu})^B  _{\  A}$, are projected onto the subspace of 16-dimensional chiral spinors.  The momentum is expressed in terms of $\lambda$ by
\bea
p^{BA}:= (\gamma^{\mu})^{BA}\, p_\mu = \lambda^{Ba} \lambda_a^{A}\,,
\label{momspin}
\eea
and the supercharges satisfying the on-shell super-algebra  
\bea
\{\bar q^B_i,\, q^A_i \} =    \lambda^{Ba}_i \lambda_{i, a}^{A}\,,
\label{supalg}
\eea 
are expressed as \cite{Boels:2012ie}
\bea
q^A_i=\lambda^A_{i,a}\, \eta^a_i \, ,\qquad\quad \bar q^B_i= \lambda^{B,a}_i\, \frac{\partial}{\partial \eta^a_i}\,,
\label{susyc}
\eea
where $\eta^a$ is a Grassman variable satisfying
\bea
\left\{\eta^a,\, \frac{\partial}{\partial \eta^b} \right\}=\delta^a_{\ b}\,.
\label{etacom}
\eea

\sm

Each external single-particle state in a scattering amplitude labelled $i$ is associated with a on-shell super-field that is a function of independent variables  $(p_i, \eta_i)$ and has an expansion in powers of $\eta_i^a$ given by  
\bea
\phi_0(p_i) +\eta_i^a  \phi_a(p_i) + \frac{1}{2!} \eta_i^a\eta_i^b \phi_{ab}(p_i)  +\dots+\frac{1}{8!} (\eta_i)^8 \bar {\phi_0}(p_i)\,.
\label{phiexp}
\eea
The 256 component fields in this expansion correspond to the massless fields of type IIB supergravity that arise in the linearised on-shell superfield in appendix~\ref{summact}.\footnote{This is also very similar to the expansion of the superfield in the light-cone gauge formulation of type IIB supergravity.}  Thus, 
\bea
\phi_0 \sim \Z  \,, \  \qquad  \phi_a \sim \Lambda'_a \, , \,\, \dots ,\qquad \bar {\phi_0}\sim \bar{\Z } \, ,
\label{fieldsamp}
\eea
where we have explicitly used the redefined component fields.  As with the superfield defined in \eqref{expphi} this field has $U(1)$ charge $q_\phi=-2$.  If we assign a $U(1)$ charge $q_\eta= - 1/2$   to $\eta$, a component field with $m$  $SO(8)$ spinor indices  has a charge $q_m= -2+m/2$.  

\sm

There are independent supersymmetry generators, $(q^A_i,\, \bar q^B_i)$,   of the form \eqref{susyc}  on each leg of the diagram with variables $(\eta_i,\lambda_i)$.  The total supercharge for a $n$-particle amplitude are 
\bea
 Q^A_n = \sum_{i=1}^n q^A_i  =   \sum_{i=1}^n  \lambda^A_{i,a}\, \eta^a_i \,,\qquad\quad \bar Q^B_n =\sum_{i=1}^n  \bar q^B_i= \sum_{i=1}^n    \lambda^{B,a}_i\, \frac{\partial}{\partial \eta^a_i} \,,
\label{totq}
\eea 
and the amplitude satisfies the overall supersymmetry conditions,
\bea
Q^A_n\, A_n =0 =\bar Q^A_n \, A_n\,,
\label{susyam}
\eea
 in addition to overall momentum conservation.  This means that an amplitude with $n$ massless external states  
 has the form\footnote{We will suppress the momentum  conservation delta function  from  hereon.}
  \bea
 A_n =\delta^{10}\left(\sum_{i=1}^n p_i\right)\, \delta^{16}(Q_n)\, \hat A_n\,,
 \label{ampcons}
 \eea
 where 
 \bea
 \delta^{16}(Q) = \frac{1}{16!} \epsilon_{A_1\dots A_{16}}Q^{A_1}\dots Q^{A_{16}}\,,
 \label{deldef}
 \eea
 and 
  \bea
  \qquad\quad \bar Q_n^A\hat A_n=0\,.
  \label{ahatprop}
  \eea
 The relation \eqref{deldef} and the condition \eqref{ahatprop} ensure that the amplitude $A_n$ is annihilated by the thirty-two supersymmetries.

\sm

Apart from the three-particle on-shell amplitude, which has degenerate kinematics, these conditions imply that scattering amplitudes vanish unless the total number of $\eta$'s from external states is at least $16$. 
Amplitudes in which there are exactly sixteen $\eta$ variables are those for which $q_U=- 2(n-4)$, (these are called ``maximal R-symmetry violating'' amplitudes in \cite{Boels:2012zr}).    
 In this case the quantity $\hat A$  contains no factors of $\eta$ but it is a function  of the Mandelstam invariants that encodes the $\alpha'$-dependence characteristic of string theory, as well  as the dependence  on the complex coupling constant,  $\tau^0$. 
   
\sm

 In considering the low-energy expansion of amplitudes it is important to take into  account non-analytic features  that come from the effects of higher genus contributions and non-perturbative effects.  Although this is very complicated in  general, 
the  first three terms in the low-energy expansion  of  the ten-dimensional  amplitude, which are protected by supersymmetry, are analytic in the Mandelstam invariants.  For these terms  $\hat A$ has an expansion in a series of  symmetric polynomials of degree $p=0$,  $p=2$ and $p=3$ in the Mandelstam invariants, since maximal $U(1)$-violating amplitudes cannot have poles in momenta. 
 
\sm

This leads to BPS terms  in the low-energy limits of $n$-particle  superstring amplitudes in the form,
 \begin{align}    
 \label{eq:maximal-U(1)}
 A_n^{(p)} = \kappa^{\frac{p-1}{2}}F^{(p)}_{n-4} (\tau^0)\, \delta^{16}(Q_n) \, \hat{A}^{(p)}_n(s_{ij}) \,,
  \end{align}
where the subscript $(n-4)$ indicates the weight, $w$.  In this expression, which includes amplitudes of the form  \eqref{contamp},   the factor  $\hat{A}^{(p)}_n(s_{ij})$ is simply a symmetric homogeneous degree-$p$ polynomial of Mandelstam variables.  Note that in our normalisation the overall power of $\kappa$ for a $n$-particle amplitude is independent of $n$.

\sm

Since these amplitudes have no poles they may be viewed as  on-shell supervertices. For $p\leq 3$ they are BPS $F$-terms, whose coefficients $F^{(p)}_{n-4} (\tau^0)$ are constrained by supersymmetry, as will be shown in later sections. For such amplitudes $\hat{A}^{(p)}_n(s_{ij})$ may contain powers of Mandelstam invariants but these cannot be re-expressed in terms of the other sixteen supercharges $\bar Q^A$ \cite{Wang:2015jna}.

\sm

Terms of higher order in the low-energy expansion -- i.e. of dimension $\ge 16$  (or $p \ge 4$) -- are $D$-terms and they can be written in terms of $32$ supercharges.   For example  if $\hat A_n^{(4)}(s_{ij})$ is a symmetric polynomial in Mandelstam invariants of degree $4$ it can be expressed in the schematic form
\bea
\hat A_n^{(4)}(s_{ij}) \sim \sum_{\rm permutations} (\bar Q)^{16} \eta_i^8  \eta_j^8\, .
\eea
This is simply a consequence of power counting since $(\bar Q)^{16}$ is of order $s_{ij}^4$. This is the on-shell amplitude description of $D$-terms.  Indeed as we will see later such terms are,  unsurprisingly, unconstrained and do not appear to be protected by supersymmetry.

\sm

The coefficient function $F^{(p)}_{n-4} (\tau^0)$  contains the full non-perturbative dependence on the complex type IIB coupling  constant in the Einstein frame.  The leading term  in  the weak  coupling limit is the  tree-level contribution, which is given by $(\tau^0_2)^{\frac{3+p}{2} }$ multiplied by a rational multiple of  a weight-$(3{+}p)$ odd zeta value. 
We will now discuss examples of these BPS terms that emerge explicitly  from the expansion  of tree-level maximal $U(1)$-violating  superstring amplitudes.  

\subsection{Low-energy expansion of tree-level maximal $U(1)$-violating amplitudes}

In recent years various methods have been devised for calculating $n$-particle superstring theory  tree amplitudes \cite{Mafra:2011nv, Mafra:2011nw,Schlotterer:2012ny}.  Closed-string amplitudes are efficiently expressed in the KLT manner by doubling open-string amplitudes, which are stringy extensions of Yang--Mills theory.  This results in expressions for the super closed-string tree amplitudes that are conveniently expressed in  the following manner
\begin{align}
A^n_{closed} = A_{\rm YM\, tree}^{n} \, S_{\rm KLT}(s_{ij}) \, G(\alpha' s_{ij})\, \tilde{A}^{n} _{\rm YM\, tree}
\label{genamp}
\end{align}
where  ${A}^{n} _{\rm YM\, tree}$ and  $ \tilde{A}^{n} _{\rm YM\, tree}$ are $n$-particle colour-stripped super Yang--Mills tree amplitudes with different permutations of the cyclic order, $S_{\rm KLT}(s_{ij})$ is the KLT kernel, and $G(\alpha' s_{ij})$  contains the stringy corrections to the field theory expression so $\lim_{\alpha'\to 0}G(\alpha' s_{ij})=1$.  The low-energy expansion involves expanding $G(\alpha' s_{ij})$ in a power series in $\alpha' s_{ij}$.

\sm 

Such expressions may be efficiently evaluated in  the four-dimensional spinor-helicity formalism in which MHV amplitudes play a distinguished r\^ole.  For instance, the MHV supergravity amplitudes are obtained if the states in both Yang--Mills factors are chosen to be MHV.  However,  $U(1)$-violating amplitudes arise when the helicity assignments in the Yang--Mills factors are distinct.  The maximal $U(1)$-violating closed-string amplitudes result from the choices in which one Yang--Mills factor is MHV and the other is ${\rm \overline{MHV}}$.    This four-dimensional formalism is convenient  for describing the compactification to four-dimensional  $\mathcal{N}=8$ supergravity  but it obscures its origins in ten dimensions.  In particular, it obscures the r\^ole of the ten-dimensional complex axio-dilaton.   

\sm 
A more direct procedure is to consider the KLT construction of \eqref{genamp} in ten dimensions, where the maximally supersymmetric amplitudes have been constructed by use of the pure spinor formalism. The form of the maximal $U(1)$-violating amplitudes is tightly constrained by supersymmetry. One may compute component amplitudes with particular external states, which have the form (\ref{eq:maximal-U(1)}). The results of this analysis give the following low-order terms in the low-energy expansion of $\hat{A}_n(s_{ij}),$\footnote{We are very grateful to Oliver Schlotterer for providing us with the coefficients for the six-point amplitude $\hat{A}_6(s_{ij})$ in the following equations~\cite{SchlottereNew}.}  
   \begin{align} \label{treeamp}
 \hat{A}_4(s_{ij})  &=  2 \,\kappa^{-\half}\, \tau_2^{3\over 2} \,\zeta(3)+  \kappa^\half\, \tau_2^{5\over 2} \,  \zeta(5)\,  \mathcal{O}_4^{(2)} +  {2 \over 3}  \,\kappa\,  \tau_2^3  \, \zeta(3)^2\,  \mathcal{O}^{(3)}_{4}  + \cdots  \nn \\
 \hat{A}_5(s_{ij})  &=   3\,  \kappa^{-\half}\,  \tau_2^{3\over 2} \zeta(3)+ {5\over 2} \kappa^\half \, \tau_2^{5\over 2} \zeta(5) \, \mathcal{O}^{(2)}_{5} + 2\, \kappa\,  \tau_2^3  \zeta(3)^2  \,  \mathcal{O}^{(3)}_{5} + \cdots  \\
  \hat{A}_6(s_{ij})  &=   {15 \over 2}\, \kappa^{-\half}\, \tau_2^{3\over 2} \zeta(3)+ {35 \over 4} \kappa^\half \, \tau_2^{5\over 2} \zeta(5) \, \mathcal{O}^{(2)}_{6}+ 8\, \kappa\,  \tau_2^3 \zeta(3)^2 
 \, \cO^{(3)}_{6,1}+ \cdots  \,, \nonumber
  \end{align}
where we have expressed the amplitudes in the Einstein frame, and 
\bea \label{eq:On2On3}
 \mathcal{O}_n^{(2)} :=  {1\over 2} \sum_{1\leq i<j \leq n} s_{ij}^2\, , \qquad \mathcal{O}_n^{(3)} := {1\over 2} \sum_{1\leq i<j \leq n} s_{ij}^3\, ,
\eea 
and $\cO^{(3)}_{6,1}$ is the kinematics structure defined in \eqref{sixinv1}. Each amplitude has been normalised  to be consistent with the  convention that will be defined in (\ref{eq:softn}).

\sm 
 
Although the  overall normalisations of the tree-level $n$-particle amplitudes depend on conventions, the above equations determine the relative coefficients of the $p=2$ and $p=3$ terms in the low-energy expansion (the dimension-12 and dimension-14 terms)  in terms of the $p=0$ coefficients. This relates the values of constants   $c_{w,i}^{(p)}$ that arose in sections~\ref{effact} and \ref{eighthbps} as follows 
\bea
c_{0}^{(2)} &=&\frac{1}{2} c_{0}^{(0)} \,, \qquad c_{0}^{(3)} =  c_{0}^{(0)}  \nn \\
 c_{1}^{(2)}&=& \frac{5} {6} c_{1}^{(0)} \,, \qquad c_{1}^{(3)} =\frac{2} {3} c_{1}^{(0)}  \nn \\
c_{2}^{(2)} &=& \frac{7}{6}c_{2}^{(0)} \,, \qquad c_{2,1}^{(3)} ={4 \over 15} c_{2}^{(0)}  \, .
\label{cdefs}
\eea 
  Note that the choice of overall normalisations of the amplitudes given  in  \eqref{treeamp}  translates into the choices   $c_{0}^{(0)}=1$, $c_{1}^{(0)}=3/2$,  $c_{2}^{(0)}=15/4$.  This gives the values for the $p=3$ coefficients, $c_{0}^{(3)}=c_1^{(3)}=1$ and $c_{2,1}^{(3)}=1$,  which is consistent with the choice of normalisation to be made in \eqref{eq:softn2} (based on consideration of the soft  $Z$ limit).
\sm 

These tree-level amplitudes are the lowest order terms in the expansion of $SL(2,\ZZ)$-covariant amplitudes so the coefficients in the expansions of amplitudes with different values of $n$  (and hence of $w$)  must be related to each other by $SL(2,\ZZ)$. Since the amplitudes with $(n+1)$ external particles and with $n$ external particles have different kinematics we cannot simply compare the coefficients.  However, a  $(n+1)$-particle  amplitude is expected to reduce to a $n$-particle amplitude in the soft axio-dilaton limit, as we will now discuss. 

\section{The soft  $\Z $ limit and covariant derivatives} \label{sec:softB}

As discussed previously, the general $(n{+}m)$-particle maximal $U(1)$-violating amplitude for $n$ particles in $\cP_n(\{\Phi\})$ together with  $m$ axio-dilaton particles,  ${\Z ^m}$,  is  obtained by  expanding the expression in \eqref{transformp},  giving interactions $\cP_n(\{\Phi\})\Z ^m/m!$ with  a coefficient modular form
\bea \label{eq:R4-expansion}
F^{(p)}_{m+n-4}  (\tau^0) =2^m \,\cD_{m+n-5}\dots \cD_{n-4}\, F^{(p)}_{n-4}  (\tau) \bigg|_{\tau=\tau^0}\, . 
\eea
Importantly the ${\Z }$ fields are trivially attached to the lower-point vertex $\cP_n(\{\Phi\})$, therefore we see that not only are the coefficients related by covariant derivatives as in \eqref{eq:R4-expansion}, but also the kinematic factors in the amplitudes are related by the soft limit on the momentum of a $\Z $ field. 

\sm

Indeed, this is the general property of scalars of a coset space. It is well-known that the amplitudes vanish in the soft scalar limit for the classical theory where the duality symmetry is unbroken \cite{ArkaniHamed:2008gz}. The soft behaviour reflects the fact that the scalars parameterising the coset space are Goldstone bosons.  However, for the case of interest in this paper, the continuous symmetries are in general broken, and correspondingly the amplitudes are non-vanishing in the soft scalar limit. In fact, as we indicated above, the soft $\Z $ limit relates a $n$-point amplitude to a  $(n-1)$-point amplitude with the soft particle $\Z_n $ removed \cite{Wang:2015aua},\footnote{Analogous soft scalar limits for $U(1)$-violating amplitudes in the four-dimensional $\mathcal{N}=4$ supergravity were studied in \cite{Carrasco:2013ypa, Huang:2015sla}}  
\bea \label{eq:softn}
A_{n}(X, \Z _n)\big{|}_{p_n \rightarrow 0}  =  2\, \cD A_{n-1}(X)\, ,
\eea
where $X$ represents the hard particles. More precisely, both $A_{n-1}(X)$ and $A_{n}(X, \Z _n)$ are products of modular forms and kinematic factors, where the modular forms of $A_{n}(X, \Z _n)$ are related to those of $A_{n-1}(X)$ by a covariant derivative $\cD$, whereas the kinematic parts of $A_{n}(X, \Z _n)$ reduce to those of  $A_{n-1}(X)$, so that \eqref{eq:softn} takes the form 
\bea \label{eq:softn2}
\F_{w}^{(p)}(\tau^0)\, \cO_{n,i} ^{(p)} \big |_{p_n\to 0} = 2\, \cD \F_{w-1}^{(p)}(\tau) \big |_{\tau=\tau^0}  \cO_{n-1,i} ^{(p)} \, ,
\eea
where the subscripts $w$ and $w{-}1$ indicate the $U(1)$ weights of the modular form coefficients of the interaction terms before and after the soft limit. 

\sm

The soft limit \eqref{eq:softn} has also been explicitly checked against the known results such as the tree-level amplitudes given in (\ref{treeamp}) (as well as higher-order terms up to order $\tau_2^4$ which we did not  exhibit). On the other hand, the soft limits (\ref{eq:softn}) impose highly non-trivial constraints on the amplitudes, and may be utilised to determine higher-point interactions from lower-point ones as will be analysed in the following section. 

\sm

\subsubsection*{Note on connection with the standard soft dilaton limit}

There is a well-studied soft limit that involves only the real part of $\Z $ field, namely the dilaton, which states that\footnote{In our normalisation of the amplitudes, the following soft factors have no overall factor of $\kappa$.}
\bea \label{eq:soft-dilaton}
A_{n}(X, \varphi_n)\big{|}_{p_n \rightarrow 0}  =  \left( \alpha'  {\partial \over \partial  \alpha' } -2 g_s  {\partial \over \partial  g_s }   \right) A_{n-1}(X)\, ,
\eea
where $\varphi_n$ is the dilaton fluctuation corresponding to the particle with momentum $p_n$. This soft-dilaton limit has been known since the $70$'s \cite{Ademollo:1975pf, Shapiro:1975cz}, and has been revisited recently (see for example, \cite{DiVecchia:2015jaq, DiVecchia:2016szw, DiVecchia:2018dob}).  In order to compare with the soft-$\Z $ limit   \eqref{eq:softn}, we will transform  \eqref{eq:soft-dilaton} to  the  Einstein  frame. To do so, we express amplitudes in terms of $\kappa= (\alpha' )^2 g_s$ and $g_s = \tau^{-1}_2$, such as those in \eqref{treeamp}. We further use the fact that the differential operator in \eqref{eq:soft-dilaton} annihilates $\kappa$, then \eqref{eq:soft-dilaton} translates into 
\bea
 \label{eq:soft-dilaton-2}
A_{n}(X, \varphi_n)\big{|}_{p_n \rightarrow 0}  =   2  \tau_2  {\partial \over \partial  \tau_2 }  \, A_{n-1}(X)\, .
\eea
  Since $\tau_2 = e^{-\varphi}$, to lowest order  the dilaton is related to $\Z $ by
\bea
 \varphi\sim  \frac{ \tau_2-\tau_2^0}{\tau_2^0}=  \left( \frac{\Z }{1-\Z } + \frac{\bar \Z }{1-\bar \Z }\right)\sim \Z +\bar \Z \,.
 \label{dildef}
 \eea
It follows that  \eqref{eq:soft-dilaton-2} is a consequence of the sum of  \eqref{eq:softn} and its conjugate equation, $A_{n}(X,\bar \Z _n)\big{|}_{p_n \rightarrow 0}  =  2\,\bar \cD A_{n-1}(X)$,
\bea
 \label{eq:soft-dilaton-3} 
A_{n}(X, \Z _n{+}\bar{\Z }_n)\big{|}_{p_n \rightarrow 0}  =  2 \left(\cD_w+\bar{\cD}_{-w} \right) A_{n-1}(X)\, .
\eea
Upon using \eqref{tauind}, \eqref{eq:soft-dilaton-3} indeed reduces to \eqref{eq:soft-dilaton-2}. Of course, taking the soft limit on $\varphi$ is unnatural in the content of $SL(2, \ZZ)$ symmetry.  In particular the right-hand side  of \eqref{eq:soft-dilaton-3} is a sum of modular functions of different $SL(2, \ZZ)$ weights.

\sm
  
 \subsection{Applications of the soft $\Z $ limit}
 \label{softsec}
 
Here we consider the consequences of  soft $\Z $ limits for the maximal $U(1)$-violating amplitudes. As discussed in the previous sections, the amplitudes take the form 
\begin{align}    
 A_n^{(p)} =\kappa^{\frac{p-1}{2}}\,F^{(p)}_{n-4} (\tau^0)\, \delta^{16}(Q_n) \, \hat{A}^{(p)}_n(s_{ij}) \, ,
  \end{align}
where  $\hat A^{(p)}_n(s_{ij})$ has a unique kinematic structure when $p=0$ or $p=2$, given by 
\bea \label{eq:p0p2}
\hat{A}^{(0)}_n(s_{ij}) =1 \, , \qquad \hat{A}^{(2)}_n(s_{ij}) = \mathcal{O}_{n}^{(2)} \, , 
\eea
where $\mathcal{O}_{n}^{(2)}$ is defined in \eqref{eq:On2On3}. The soft limits relate  $\hat A^{(p)}_n(s_{ij})$ to $\hat A^{(p)}_{n-1}(s_{ij})$ trivially for these cases, and the coefficients are again related by a covariant derivative
\bea
F^{(p)}_{n-4} (\tau^0)  = 2 \, \cD F^{(p)}_{n-5} (\tau)\big |_{\tau=\tau^0}  \, .
\eea

\sm

The cases with $p=3$ are more interesting.  For all $n\geq 6$, there are two independent kinematic invariants  which are denoted $\mathcal{O}^{(3)}_{n,1}$ and $\mathcal{O}^{(3)}_{n,2}$. They satisfy the soft relations, 
  \begin{align} 
 {\mathcal{O}^{(3)}_{n,1}} \big |_{p_n\rightarrow 0} =   \mathcal{O}^{(3)}_{n-1,1}\, ,\qquad \quad 
  \mathcal{O}^{(3)}_{n,2} \big|_{p_n\rightarrow 0} =   \mathcal{O}^{(3)}_{n-1,2} \, .
\end{align}
The kinematic invariants for $n>6$ are uniquely determined using the expressions of $\mathcal{O}^{(3)}_{6,1}$ and $\mathcal{O}^{(3)}_{6,2}$ given in \eqref{sixinv1} and \eqref{sixinv2} and the above soft relations,  
 \begin{align}  
 \label{eq:On12}
 \mathcal{O}^{(3)}_{n,1} &= {1\over 32} \left(  (28-3n)  \sum_{i<j} s_{ij}^3 + 3 \sum_{i<j<k} s_{ijk}^3 \right) \, , \cr
  \mathcal{O}^{(3)}_{n,2} &=   (n-4) \sum_{i<j} s_{ij}^3 -\sum_{i<j<k} s_{ijk}^3  \, ,
   \end{align}
where $\mathcal{O}^{(3)}_{4,1}=s^3+t^3+u^3$ corresponds to $d^6R^4$. The maximal $U(1)$-violating  amplitudes are then given by
  \begin{align} 
A^{(3)}_{n,1} =\kappa\,  F^{(3)}_{n-4, 1} (\tau^0)\, \delta^{16}(Q_n) \,  \mathcal{O}^{(3)}_{n,1} \, , \qquad
A^{(3)}_{n,2} = \kappa\, F^{(3)}_{n-4, 2} (\tau^0)\,  \delta^{16}(Q_n) \,  \mathcal{O}^{(3)}_{n,2} \, ,
   \end{align}
where the coefficients are related by covariant derivatives, {i.e.} $F^{(3)}_{m, i} (\tau^0) = 2 \cD F^{(3)}_{m-1, i} (\tau^0)$ for $i=1,2$. 
   
\sm

In these expressions the coefficient  $F^{(3)}_{n-4, 1}(\tau^0)$ is determined by nested covariant derivatives acting on $F^{(3)}_{0}(\tau^0)$,  i.e., on the coefficient of $d^6R^4$.   Since these coefficients are associated with kinematic factors $\mathcal{O}^{(3)}_{n,1}$, their tree-level contributions are related and non-zero.    However,   $F^{(3)}_{n-4, 2}(\tau^0)$ with $n>6$ is determined in terms of nested derivatives acting on  the coefficient  $F^{(3)}_{2,2}(\tau^0)$.  These terms have no tree-level contributions.  As discussed in the previous section,  $F^{(3)}_{2,2}(\tau^0)$ is  constrained by supersymmetry and satisfies 
 \eqref{eq:DbE22}, which will also be seen to  emerge from the structure of super-amplitudes in the next section.

\section{Super-amplitude constraints on first-order differential equations}
\label{constraints}

We have seen how type IIB amplitudes with different numbers of particles are related by consideration of the soft $\Z $ limit.  This relates the $(n+1)$-particle  amplitude with  one soft $\Z $ state to the $n$-particle amplitude  with the soft $Z$ removed. In the case of maximal $U(1)$-violating amplitudes  this involves the relation between the coefficient modular forms of the form $F_{w+1} ^{(p)}(\tau) \sim \cD \, \F_w^{(p)}(\tau)$, that was  encountered in  sections \ref{effact}  and  \ref{eighthbps}  and applies to the coefficients  for any value of $p$.  This relationship applies to terms for which the kinematic  factors are related in the soft limit in the manner of \eqref{eq:softn2}. 

\sm

In order to show how the conjugate first order differential equations involving $\bar \cD$ are determined by  supersymmetry constraints,  we will extend the procedure devised in \cite{Wang:2015jna} for determining the constraints based on the structure of super-amplitudes. 
The key ingredients in this procedure are encapsulated in the following  statements:

{\it\begin{itemize}
\item 

Supersymmetric $F$-terms are contact interactions corresponding to $p \leq 3$ terms in the low-energy expansion of maximal $U(1)$-violating amplitudes.

\item
Supersymmetric contact terms of dimension $\leq 14$ are not allowed for non-maximal $U(1)$-violating processes. The absence of a supersymmetric contact term provides powerful constraints on the $F$-term effective interactions.

 \item
Interactions with dimension more than 14 are $D$-terms, whose couplings in general are not constrained by supersymmetry.  
 \end{itemize} } 

\sm

As an example, let us consider low-order terms in the low-energy  expansion of a six-particle amplitude (terms with $p=0,2,3$),  such as the amplitude with four gravitons, one $\Z $ field and one $\bar{\Z }$ field. It is straightforward to see that a supersymmetric contact term with $p\le 3$ (i.e. with a number of derivatives not greater than $14$)  does not exist for such an amplitude.  Indeed for this particular case, the corresponding super-amplitude contains $24$ $\eta$'s which enter into a supersymmetric invariant that can be expressed in  the following form, 
 \begin{align} \label{eq:contact}
\delta^{16}(Q_n) (\bar Q_n)^{16} (\eta_i)^8 (\eta_j)^8 (\eta_k)^8 \, ,
  \end{align}
since $(\bar Q_n)^{16}$ annihilates $16$ $\eta$'s (recalling that $\bar Q_n$ is defined in \eqref{totq}).   This has $16$ powers of momentum  whereas BPS terms ($p\le 3$)  have at most 14 powers.  Therefore, in order to describe a supersymmetric  term there must be intermediate poles (inverse momentum factors) so that supersymmetric  contact terms are not allowed. 

\sm

This  argument has an important and subtle loophole for $n=5$.   For instance $R^4\bar{\Z }$ also requires $24$ $\eta$'s.  But  since  $R^4\bar{\Z }$ is just the complex conjugate of $R^4{\Z }$ it  obviously does have a supersymmetric completion. It turns out that the following expression for the  $R^4\bar{\Z }$ amplitude, which appears to have a higher-order pole is actually a contact term
 \begin{align} 
\delta^{16}(Q_5) {  (\bar Q_5)^{16} (\eta_1)^8 (\eta_2)^8 (\eta_3)^8 \over (s_{45})^4 } \, . 
  \end{align}
To see that this  is non-singular as $s_{45} \to 0$ it is sufficient to  note that this expression is in fact invariant under permutations of the external states, although this is not manifest. 

\sm

The fact that it is not possible to write a supersymmetric contact term for a non-maximal $U(1)$-violating process  of dimension $\le 14$ implies that the low-energy expansion up to dimension 14 of the 
super-amplitude is uniquely determined by lower-point amplitudes via factorisation on intermediate poles as determined by tree-level unitarity.

\sm 

This strongly constrains the components of the effective action.  In particular, the contact terms in the component  action are related to the non-local factorisation diagrams.  In other words, the contact terms that enter the component action are not independent vertices since they cannot be supersymmetrised in isolation from the rest  of  the amplitude. This implies that there must be a linear relation between the coefficients of component contact terms and those of the factorisation terms.
 This approach using only on-shell data and tree-level unitarity is an efficient way of imposing supersymmetric constraints on the coefficient modular functions of $F$-terms.

\begin{figure}[h]
\begin{center}
\begin{tikzpicture}[scale=1]
\begin{scope}[xshift= -5.5  cm,yshift=0.0cm,very thick, every node/.style={sloped,allow upside down}]
  \filldraw [black]  (0,0) ellipse (.3 and .3);
\draw [ultra thick] (-1,1) -- (0,0);
\draw [ultra thick]  (1,1) --  (0,0);
\draw [ultra thick] (-1.0,-1.0) -- (0,0);
\draw [ultra thick]  (-1.4,0) --  (0,0);
\draw [ultra thick, decorate,decoration=snake]  (0.8,-1.3) --  (0,0);
\draw [ultra thick, decorate,decoration=snake]  (1.4,0) --  (0,0);
           \draw  (1.7, 0.1) node{$Z$};
           \draw  (1.0,-1.4) node{$\bar  Z$};
   \draw  (-2.8,0.0) node{$\bar \cD \F^{(p)}_{1}(\tau^0)$};
             \draw  (0.0,-1.9) node{$(a)$};
          \end{scope}
\begin{scope}[xshift= 1.0  cm,yshift=0.0cm,very thick, every node/.style={sloped,allow upside down}]
\filldraw [black]  (0,0) ellipse (.3 and .3);
\draw [ultra thick] (-1,1) -- (0,0);
\draw [ultra thick]  (1,1) --  (0,0);
\draw [ultra thick] (-1.0,-1.0) -- (0,0);
\draw [ultra thick]  (0.5,-0.4) --  (0,0);
\draw [ultra thick, decorate,decoration=snake]  (1.4,0) -- (0.5,-0.4) ;
\draw [ultra thick, decorate,decoration=snake]  (1.0,-1.0) -- (0.5,-0.4) ;
\draw [ultra thick, decorate,decoration=snake]  (1.0,-1.0) -- (1.6, -0.6) ;
\draw [ultra thick] (1.0,-1.0) -- (1.5,-1.4);
           \draw  (1.7,0.0) node{$\bar Z$};
                 \draw  (1.9, -0.5) node{$Z$};
        \draw  (-1.4, 0.0) node{$\F^{(p)}_{0}(\tau^0)$};
             \draw  (0.0,-1.8) node{$(b)$};
          \end{scope}

\begin{scope}[xshift= -5.5  cm,yshift=-4.0cm,very thick, every node/.style={sloped,allow upside down}]
\filldraw [black]  (0,0) ellipse (.3 and .3);
\draw [ultra thick] (-1,1) -- (-0.6,0.6);
\draw [ultra thick, decorate,decoration=snake] (-0.6,0.6) -- (0,0);
\draw [ultra thick, decorate,decoration=snake] (-0.6,0.6) -- (0.3,1.2);
\draw [ultra thick]  (1,1) --  (0,0);
\draw [ultra thick] (-1.0,-1.0) -- (0,0);
\draw [ultra thick]  (1.0,-1.0) --  (0.6,-0.6);
\draw [ultra thick, decorate,decoration=snake]  (0.6,-0.6) --  (0,0);
\draw [ultra thick, decorate,decoration=snake]  (1.4,-0.15) -- (0.6,-0.6);
           \draw  (1.7,0.0) node{$\bar Z$};
                      \draw  (0.4,1.4) node{$Z$};
        \draw  (-1.4, 0.0) node{$\F^{(p)}_{0}(\tau^0)$};
             \draw  (0.0,-1.8) node{$(c)$};
          \end{scope}

\begin{scope}[xshift= 1.0  cm,yshift=-4.0cm,very thick, every node/.style={sloped,allow upside down}]
\filldraw [black]  (0,0) ellipse (.3 and .3);
\draw [ultra thick] (-1,1) -- (0,0);
\draw [ultra thick]  (1,1) --  (0,0);
\draw [ultra thick] (-1.0,-1.0) -- (0,0);
\draw [ultra thick]  (0.5,-1.2) --  (0,0);
\draw [ultra thick]  (1.3, -0.3) --  (0,0);
\draw [ultra thick, decorate,decoration=snake]  (1.3,-0.3) -- (2.1, -1.0) ;
\draw [ultra thick, decorate,decoration=snake]  (1.3,-0.3) -- (2.1, 0.5) ;
                 \draw  (2.4, -1.1) node{$Z$};
               \draw  (2.4, 0.5) node{$\bar{Z}$};
        \draw  (-1.4, 0.0) node{$\F^{(p-1)}_{R^5}(\tau^0)$};
             \draw  (0.0,-1.8) node{$(d)$};
          \end{scope}
\end{tikzpicture}
\end{center}
\caption{  A diagrammatic interpretation of the pieces of the  first-order differential equation   relating  $\F_{1}^{(p)}(  \tau^0)$ to $\F_{0}^{(p)}(  \tau^0)$  for a dimension-$8$ ($p=0$) or dimension-$12$ ($p=2$)  contribution to the amplitude for $4$ gravitons together with a $\Z $ and a $\bar \Z $. (a) A contact term describing the emission of  a $\bar \Z $ from a $d^{2p}R^4 \Z $ five-particle contact term.  (b) and (c)  Two examples of factorisation contributions formed by attaching  $\Z $  and  $\bar  \Z $ to  the  external  legs  of the $d^{2p}R^4$-type contact terms with supergravity vertices.  (d) The factorisation contribution by attaching  a $Z$-$\bar Z$-graviton vertex to $d^{2p-2}R^5$ (it does not contribute when $p=0$).  }
\label{fig:sixfirst}
\end{figure}
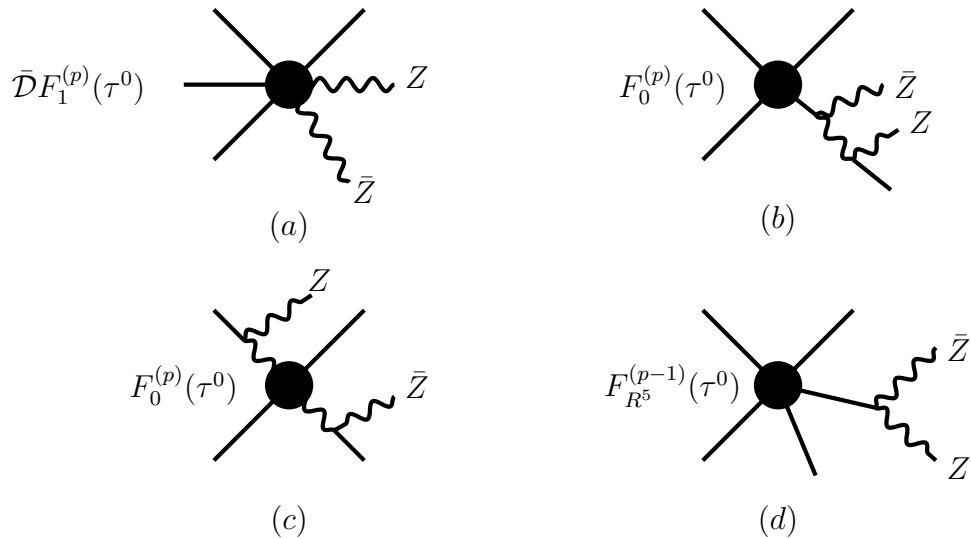

\sm 

Before applying the above idea to derive first-order differential equations satisfied by the modular forms that are the coefficients of the $F$-terms, we should emphasise again that the $D$-terms are in general not constrained.    The existence of supersymmetric contact terms such as \eqref{eq:contact} imply that  tree-level unitarity is not enough to determine the $D$-term contributions to super-amplitudes. In other words, one can always add contact terms with arbitrary coefficients to a given expression without modifying the factorisation conditions. Therefore, as is well-known, $D$-terms are not constrained by supersymmetry. 

\subsubsection*{Terms with $p=0$ and $p=2$}

These supersymmetry constraints  lead to particularly simple first-order equations in the $p=0$ and $p=2$ cases that are  illustrated by the processes depicted in figure~\ref{fig:sixfirst}.  This  shows  the contributions to the four-graviton-$\Z $-$\bar \Z $ amplitude. Such an ampliude is not maximal $U(1)$ violating  and has massless intermediate poles.  Figure~\ref{fig:sixfirst}(a) illustrates a contact  interaction in which the coefficient is expressed as $\bar \cD \F^{(p)}_{1}(\tau^0)$ where $\F^{(p)}_{1}(\tau^0)$ is the coefficient of a five-particle maximal $U(1)$-violating amplitude for four  gravitons and one $\Z $.  

\sm

The  absence of a supersymmetric  contact  term implies that there is a linear relation between the coefficients of each term that contributes to the component amplitude as shown in figure~\ref{fig:sixfirst}.  There are two classes of diagrams.  One is the contact interaction of figure~\ref{fig:sixfirst}(a), while the  others are factorisation contributions  that contain intermediate poles, such as the processes shown in figure~\ref{fig:sixfirst}(b), (c) and (d). In these factorisation diagrams $\Z $ and $\bar  \Z $ states are attached via supergravity interactions to external  legs  of  the four-graviton $d^{2p}R^4$  interaction, or (in the case of figure~\ref{fig:sixfirst}(d)) the five-graviton $d^{2p-2}R^5$ interaction. There are several other analogous diagrams to  take into account  that we have not  drawn. It would be complicated to calculate all of these pole contributions precisely.   However the contributions from  figures~\ref{fig:sixfirst}(b) and (c) are proportional to $\F^{(p)}_{0}(\tau^0)$.  In order to complete this discussion we will now demonstrate that  $\F_{R^5}^{(p-1)}$ (the coefficient of  $d^{2p-2}R^5$) in  figure~\ref{fig:sixfirst}(d),  is also proportional to $\F^{(p)}_{0}(\tau^0)$.

\sm

In order to determine properties of $\F_{R^5}^{(p-1)}$ we need to consider properties of the five-graviton amplitude, which is not a maximal $U(1)$-violating process. The diagrams that contribute to this interaction are the local vertex, $d^{2p-2}R^5$    (with coefficient $\F_{R^5}^{(p-1)}(\tau^0)$) and pole terms arising from attaching a three-graviton vertex to $d^{2p}R^4$. Since the interaction $d^{2p-2}R^5$ has dimension  $\le 14$ (when $p\le 3$),  our  previous argument implies that  the super-amplitude containing this process cannot have a supersymmetric contact term.    The absence of such a contact term implies that $d^{2p-2}R^5$ is related to $d^{2p}R^4$, which leads to a  linear relation between their coefficients
\bea \label{eq:dpR5}
\F^{(p-1)}_{R^5}(\tau^0) + a \,  \F^{(p)}_{0}(\tau^0) =0\, ,
\eea
that is  in agreement with \cite{Richards:2008jg}. 
Therefore,  $\F^{(p-1)}_{R^5}(\tau^0)$ is proportional to  $\F^{(p)}_{0}(\tau^0)$  (when $p \leq 3$). 
We note, in particular that the absence of a four-graviton interaction with $p=1$,  of the form $d^2R^4$ implies the absence of a $R^5$ contact interaction.

\sm

Returning to our consideration of the contributions in figure~\ref{fig:sixfirst} we now see that 
the uniqueness of the super-amplitude implies that there must be a linear relation between the coefficients of the various contributions, of the form 
\bea
\bar \cD \F^{(p)}_{1}(\tau^0)= c\, \F^{(p)}_{0}(\tau^0) \, , \qquad\qquad\qquad  p=0,2\,.
\label{dbarf}
\eea 
This is the structure of the relationship between the coefficients that  was discussed in section~\ref{effact} where the modular form coefficients were identified with  Eisenstein modular forms that satisfy \eqref{dbarzsn}.  In principle the value of $c$ should be determined by explicitly constructing the super-amplitude and  evaluating the  various supergravity insertions, but we have not done this.  An indirect way of fixing the value of $c$ is to note that    \eqref{dbarf}, together with the relation $\F^{(p)}_{1}(\tau^0) = 2 \cD\, \F^{(p)}_{0}(\tau^0)$, imply the well-known Laplace eigenvalue equation  $(4 \bar\cD \cD- 2c) F^{(p)}_{0}(\tau^0)=0$.  From our earlier discussion of such equations we know that when $p=0$ we must have $c=3/8$ (so  the  eigenvalue is $3/4$) and when $p=2$ we must have $c=15/8$ (so the  eigenvalue is $15/4$).   Equivalently, the value of $c$ 
is also be  fixed by inputting the string theory tree-level contribution to $\F^{(p)}_{0}(\tau^0)$  in \eqref{treeamp}.

\sm

\subsubsection*{Terms with $p=3$ and $w=1$}
\begin{figure}
\begin{center}
\begin{tikzpicture}[scale=1]
\begin{scope}[xshift= -4.0  cm,yshift=0.0cm,very thick, every node/.style={sloped,allow upside down}]
\filldraw [black]  (0,0) ellipse (.3 and .3);
\draw [ultra thick] (-1,1) -- (0,0);
\draw [ultra thick]  (1,1) --  (0,0);
\draw [ultra thick] (-1.0,-1.0) -- (0,0);
\draw [ultra thick]  (-1.4,0) --  (0,0);
\draw [ultra thick, decorate,decoration=snake]  (1.0,-1.0) --  (0,0);
\draw [ultra thick, decorate,decoration=snake]  (1.4,0) --  (0,0);
           \draw  (1.7, 0.0) node{$\Z $};
           \draw  (1.3,-1.0) node{$\bar \Z $};
   \draw  (-2.4,0.0) node{$\bar \cD\, \F^{(3)}_1(\tau^0)$};
             \draw  (0.0,-1.8) node{$(a)$};
          \end{scope}
\begin{scope}[xshift= 4.0  cm,yshift=0.0cm,very thick, every node/.style={sloped,allow upside down}]
\filldraw [black]  (0,0) ellipse (.3 and .3);
\draw [ultra thick] (-1,1) -- (0,0);
\draw [ultra thick]  (1,1) --  (0,0);
\draw [ultra thick] (-1.0,-1.0) -- (0,0);
\draw [ultra thick]  (0.5,-0.4) --  (0,0);
\draw [ultra thick, decorate,decoration=snake]  (1.2,0) -- (0.5,-0.4) ;
\draw [ultra thick, decorate,decoration=snake]  (1.0,-1.0) -- (0.5,-0.4) ;
\draw [ultra thick, decorate,decoration=snake]  (1.0,-1.0) -- (1.6, -0.6) ;
\draw [ultra thick] (1.0,-1.0) -- (1.5,-1.4);
           \draw  (1.7,0.0) node{$\bar \Z $};
                 \draw  (1.9, -0.5) node{$\Z $};
        \draw  (-1.4, 0.0) node{$\F^{(3)}_0(\tau^0)$};
             \draw  (0.0,-1.8) node{$(b)$};
          \end{scope}
\begin{scope}[xshift= -4.0  cm,yshift=-4.0cm,very thick, every node/.style={sloped,allow upside down}]
\filldraw [black]  (0,0) ellipse (.3 and .3);
\draw [ultra thick] (-1,1) -- (0,0);
\draw [ultra thick]  (1,1) --  (0,0);
\draw [ultra thick] (-1.0,-1.0) -- (0,0);
\draw [ultra thick]  (0.6,-0.6) --  (0,0);
\draw [ultra thick]  (1.2,0) -- (0.6,-0.6) ;
\draw [ultra thick]  (1.1,-1.1) -- (0.6,-0.6) ;
\draw [ultra thick, decorate,decoration=snake]  (1.1,-1.1) -- (1.6, -0.6) ;
\draw [ultra thick,   decorate,decoration=snake] (1.1,-1.1) -- (1.5,-1.4);
           \draw  (1.7,-1.5) node{$\bar \Z $};
                 \draw  (1.9, -0.5) node{$\Z $};
        \draw  (-1.4, 0.0) node{$\F^{(3)}_0(\tau^0)$};
             \draw  (0.0,-1.8) node{$(c)$};
          \end{scope}
\begin{scope}[xshift= 4.0  cm,yshift=-4.0cm,very thick, every node/.style={sloped,allow upside down}]
\filldraw [black]  (0,0) ellipse (.3 and .3);
\draw [ultra thick]  (1,1) --  (0,0);
\draw [ultra thick] (-1.0,-1.0) -- (0,0);
\draw [ultra thick]  (1.2,0) -- (0.6,-0.6) ;
\draw [ultra thick]  (0,1.2) -- (-0.6,0.6) ;
\draw [ultra thick, decorate,decoration=snake]  (1.0,-1.0) -- (0.0,-0.0) ;
\draw [ultra thick, decorate,decoration=snake]  (-1.0,1.0) -- (0.0,-0.0) ;
           \draw  (1.3,-1.1) node{$\bar \Z $};
                 \draw  (-1.3, 1.1) node{$\Z $};
        \draw  (-1.4, 0.0) node{$\F^{(3)}_0(\tau^0)$};
             \draw  (0.0,-1.8) node{$(d)$};
          \end{scope}
\begin{scope}[xshift= -4.0  cm,yshift=-8.0cm,very thick, every node/.style={sloped,allow upside down}]
\filldraw [black]  (0,0) ellipse (.3 and .3);
\draw [ultra thick] (-1,1) -- (0,0);
\draw [ultra thick]  (1,1) --  (0,0);
\draw [ultra thick] (-1.0,-1.0) -- (0,0);
\draw [ultra thick] (0.7,-0.34) -- (0,0);
\draw [ultra thick]  (-1.4,0) --  (0,0);
\draw [ultra thick, decorate,decoration=snake]  (1.0,-1.0) --  (0.7,-0.34) ;
\draw [ultra thick, decorate,decoration=snake]  (1.4,0) -- (0.7,-0.34) ;
           \draw  (1.7, 0.0) node{$\Z $};
           \draw  (1.3,-1.0) node{$\bar \Z $};
   \draw  (-2.4,0.0) node{$\F_{R^5}^{(2)}(\tau^0)$};
             \draw  (0.0,-1.8) node{$(e)$};
          \end{scope}
\begin{scope}[xshift= 2.5  cm,yshift=-8.0cm,very thick, every node/.style={sloped,allow upside down}]
\filldraw [black]  (0,0) ellipse (.3 and .3);
\draw [ultra thick, decorate,decoration=snake] (-1,1) -- (0,0);
\draw [ultra thick]  (1,1) --  (0,0);
\draw [ultra thick, decorate,decoration=snake] (-1.0,-1.0) -- (0,0);
\draw [ultra thick] (3,0) --  (0,0);
           \draw  (-1.3,-1.1) node{$\bar \Z $};
 \draw  (-1.3, 1.1) node{$\Z $};

\filldraw [black]  (0,0) ellipse (.3 and .3);
           \draw  (-1.5,0.0) node{$\F^{(0)}_0(\tau^0) $};
\filldraw [black]  (3,0) ellipse (.3 and .3);

\draw [ultra thick] (2,1) -- (3,0);
\draw [ultra thick]  (4,1) --  (3,0);]
\draw [ultra thick]  (4.0,-1.0) --  (3,0);
           \draw  (4.5,0.0) node{$\F^{(0)}_0(\tau^0)$};
               \draw  (1.5,-1.8) node{$(f)$};
\end{scope}
\end{tikzpicture}
\end{center}
\caption{   A subset of the many contributions to the first-order differential equation \eqref{barddef} in terms of a $\bar \Z $ insertion in a maximal $U(1)$-violating five-point function.  (a)  A contact contribution obtained by applying $\bar\cD$ to a five-point contact interaction, which comes from expanding $\F^{(3)}_1(\tau^0)\, d^6R^4\Z $.   (b)  A contribution in which a pair of supergravity $g Z\bar Z$ vertices is attached to an external line  on   $\F^{(3)}_0(\tau^0) \, d^6R^4$.   (c)  Another contribution with a supergravity $gg Z\bar Z$ tree attached to a   $d^6R^4$  contact interaction.  (d) A $\F_0^{(3)} (\tau^0) d^6 R^2 Z \bar Z$ contact term with two $g Z \bar Z$ vertices attached.  (e)   A contribution with a $g Z \bar Z$  vertex    attached  to an external graviton line on $\F^{(3)}_0(\tau^0) \, d^4R^5$.  (f) A  contribution to the inhomogeneous term from the product of  two $R^4$-type vertices with coefficients $\F^{(0)}_0(\tau^0)$.}
\label{fig:sixpt}
\end{figure}
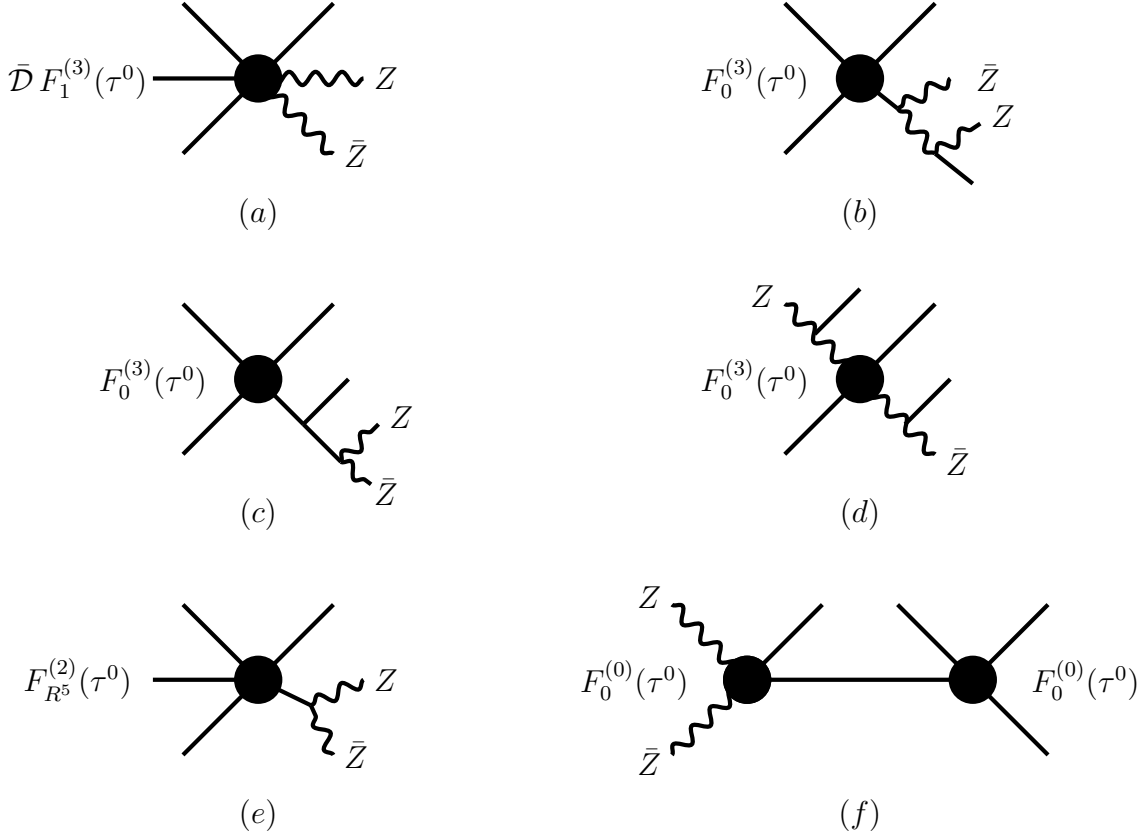

\sm

In considering the  first-order differential equations for the $p=3$  terms  (dimension-14 terms) in the low-energy  expansion we expect to meet the novel features of the coefficients that were described in section~\ref{eighthbps}.  This may again be seen by considering the absence of a contact term of the dimension-$14$ super-amplitude that contains the component amplitude of four gravitons, one $\Z $ field and one $\bar{\Z }$ field.   The contribution with $p=3$ to the low-energy expansion of this amplitude receives contributions that are schematically shown in figure~\ref{fig:sixpt}. In this case, the  terms such as figures~\ref{fig:sixpt}(b)-\ref{fig:sixpt}(e) are of the same form as those in the $p=0, 2$ cases.
However,  dimensional counting shows that the amplitude also has a contribution in which it factorises on an intermediate pole that separates two four-particle higher-dimensional interactions, as shown in figure~\ref{fig:sixpt}(f).
\sm

Again, the absence of a supersymmetric contact term implies that there must be  a relation among all three types of terms shown in the figure~\ref{fig:sixpt}
\bea
\bar \cD\, \EE_1(\tau^0) + a\, \EE_0 (\tau^0)+ b\, E_0(\threeh, \tau^0)E_0(\threeh, \tau^0)=0\, ,
\label{eonederiv}
\eea
which is the same form as (\ref{barddef}). Using $\EE_1 = 2\, \cD \EE_0$ (as follows from  the soft limit \eqref{eq:softn2}), the above equation leads to an inhomogeneous  Laplace equation for $\EE_0$ of the form \eqref{d6eq}. The constants $a$ and $b$  would be  determined if we were to evaluate all contributions to  this six-point amplitude, including those shown in figure~\ref{fig:sixpt}  and others,  which we have not done.  However, a shortcut is to input the known tree-level and one-loop terms of $\EE_0$, which leads to  $a=-6$, $b=1/2$, as given in (\ref{eq:DEE1}). 

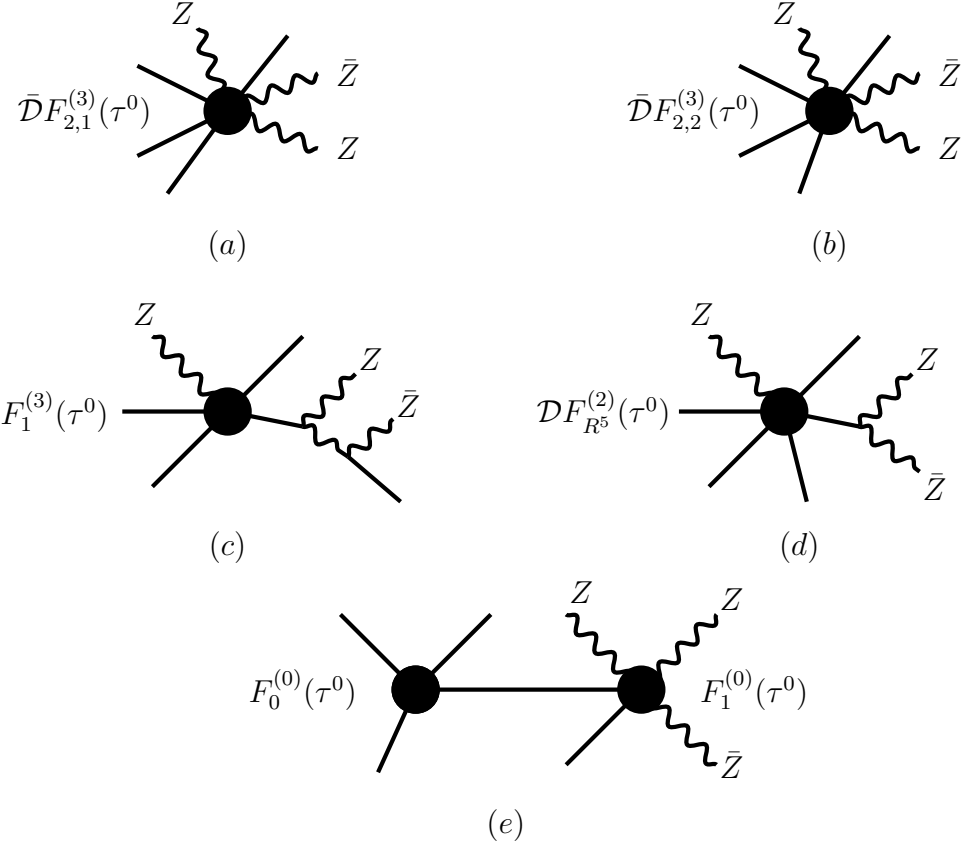
\begin{figure}
\begin{center}
\begin{tikzpicture}[scale=1]
\begin{scope}[xshift= -4.0  cm,yshift=0.0cm,very thick, every node/.style={sloped,allow upside down}]

\filldraw [black]  (0,0) ellipse (.3 and .3);
\draw [ultra thick] (-1.2,0.6) -- (0,0);
\draw [ultra thick, decorate,decoration=snake]  (-0.4,1.1) --  (0,0);
\draw [ultra thick] (-0.8, -1.1) -- (0,0);
\draw [ultra thick] (-1.2,-.6) -- (0,0);
\draw [ultra thick]  (0.8,1) --  (0,0);
\draw [ultra thick, decorate,decoration=snake]  (1.2,0.5) --  (0,0);
           \draw  (1.6,0.5) node{$\bar \Z $};  
\draw [ultra thick, decorate,decoration=snake]  (1.2,-0.5) --  (0,0);
            \draw  (1.6,-0.5) node{$\Z $};
     \draw  (-0.6,1.3) node{$\Z $};        
         \draw  (-1.9,0.0) node  {$\bar\cD\F^{(3)}_{2,1}(\tau^0)$};
             \draw  (0.0,-1.8) node{$(a)$};
          \end{scope}

\begin{scope}[xshift= 4.0  cm,yshift=0.0cm,very thick, every node/.style={sloped,allow upside down}]

\filldraw [black]  (0,0) ellipse (.3 and .3);
\draw [ultra thick] (-1.2,0.6) -- (0,0);
\draw [ultra thick, decorate,decoration=snake]  (-0.4,1.1) --  (0,0);
\draw [ultra thick] (-0.4, -1.1) -- (0,0);
\draw [ultra thick] (-1.2,-.6) -- (0,0);
\draw [ultra thick]  (0.8,1) --  (0,0);
\draw [ultra thick, decorate,decoration=snake]  (1.2,0.5) --  (0,0);
           \draw  (1.6,0.5) node{$\bar \Z $};  
\draw [ultra thick, decorate,decoration=snake]  (1.2,-0.5) --  (0,0);
            \draw  (1.6,-0.5) node{$\Z $};
     \draw  (-0.6,1.3) node{$\Z $};        
         \draw  (-1.8,0.0) node{$\bar\cD\F^{(3)}_{2,2}(\tau^0)$};
             \draw  (0.0,-1.8) node{$(b)$};
    
\end{scope}


\begin{scope}[xshift= -4.0  cm,yshift=-4.0cm,very thick, every node/.style={sloped,allow upside down}]

\filldraw [black]  (0,0) ellipse (.3 and .3);
\draw [ultra thick, decorate,decoration=snake] (-1,1) -- (0,0);
\draw [ultra thick] (-1.4,0.0) -- (0,0);
\draw [ultra thick]  (1,1) --  (0,0);
\draw [ultra thick] (-1.0,-1.0) -- (0,0);
\draw [ultra thick] (1,-0.2) --  (0,0);
\draw [ultra thick, decorate,decoration=snake]  (1,-0.2) --  (1.7, 0.5);
\draw [ultra thick, decorate,decoration=snake]  (1,-0.2) --  (1.6, -0.6);
\draw [ultra thick, decorate,decoration=snake]  (1.6, -0.6) --  (2.2, -0.1);
\draw [ultra thick] (1.6, -0.6)  --  (2.3, -1.2);
         \draw  (-2.3,0.0) node{$\F^{(3)}_{1}(\tau^0)  $};
   \draw  (-1.1,1.3) node{$\Z $};    
      \draw  (1.9, 0.7) node{$\Z $};  
            \draw (2.4, 0.1) node{${\bar \Z }$};  

 \draw  (-0,-1.8) node{$(c)$};
 
          \end{scope}

\begin{scope}[xshift= 3.4  cm,yshift=-4.0cm,very thick, every node/.style={sloped,allow upside down}]

\filldraw [black]  (0,0) ellipse (.3 and .3);
\draw [ultra thick, decorate,decoration=snake] (-1,1) -- (0,0);
\draw [ultra thick] (-1.4,0.0) -- (0,0);
\draw [ultra thick]  (1,1) --  (0,0);
\draw [ultra thick] (-1.0,-1.0) -- (0,0);
\draw [ultra thick] (1,-0.2) --  (0,0);
\draw [ultra thick] (0.3,-1.2) --  (0,0);
\draw [ultra thick, decorate,decoration=snake]  (1,-0.2) --  (1.7, 0.5);
\draw [ultra thick, decorate,decoration=snake]  (1,-0.2) --  (1.8, -0.8);
         \draw  (-2.4,0.0) node{$\cD F^{(2)}_{R^5}(\tau^0)$};
   \draw  (-1.1,1.3) node{$\Z $};    
      \draw  (1.9, 0.7) node{$\Z $};  
            \draw (2.0, -1) node{${\bar \Z }$};  
               \draw  (0.2,-1.8) node{$(d)$};

\end{scope}

\begin{scope}[xshift= -1.5  cm,yshift=-7.7cm,very thick, every node/.style={sloped,allow upside down}]

\filldraw [black]  (0,0) ellipse (.3 and .3);
\draw [ultra thick] (-1,1) -- (0,0);
\draw [ultra thick]  (1,1) --  (0,0);
\draw [ultra thick] (-0.5,-1.1) -- (0,0);
\draw [ultra thick] (3,0) --  (0,0);

\filldraw [black]  (0,0) ellipse (.3 and .3);
           \draw  (-1.5,0.0) node{$\F^{(0)}_0(\tau^0) $};
\filldraw [black]  (3,0) ellipse (.3 and .3);

\draw [ultra thick] (2.0,-1.0) -- (3,0);
\draw [ultra thick, decorate,decoration=snake] (2,1) -- (3,0);
\draw [ultra thick, decorate,decoration=snake]  (4,1) --  (3,0);]
           \draw  (4.2,-1.0) node{$\bar \Z $};
\draw [ultra thick, decorate,decoration=snake]  (4.0,-1.0) --  (3,0);
           \draw  (4.5,0.0) node{$\F_1^{(0)}(\tau^0) $};

 \draw  (4.2,1.2) node{$\Z $};
 \draw   (2.2,1.3) node{$\Z $};
   \draw  (1.2,-1.8) node{$(e)$};

\end{scope}
\end{tikzpicture}

\end{center}
\caption{Terms that contribute to the  $p=3$ contribution to the seven-particle amplitude of four gravitons, two $\Z $'s and one $\bar{\Z }$.  (a) The contact interaction obtained by expanding $\F_0^{(3)}(\tau)d^6 R^4$ to give $\bar{\cD}  \F^{(3)}_{2,1}(\tau) d^{6}_{(1)}R^4\Z ^2\bar{\Z }$.    (b) The other contact interaction obtained by expanding  $\F_{2,2}^{(3)}(\tau)\,d^6_{(2)}  R^4  \Z ^2$, to give the seven-point interaction $\bar \cD \F_{2,2}^{(3)}(\tau) \,d^6_{(2)}   R^4  \Z ^2\bar{\Z}$.  (c)  A  contribution arising from the $\Z $ and $\bar \Z $  joining to a graviton attached to a leg of  a  $p=3$, $n=5$  interaction. (d)  A  contribution arising from a $g \Z \bar \Z $ vertex attached to a graviton line of $d^4R^5Z$ (which has coefficient proportional to $\cD F^{(2)}_{R^5}(\tau^0)$). (e) A factorising contribution  with a pole linking a $p=0$, $n=4$  interaction with a  $p=0$,  $n=5$ interaction. }
\label{fig:E22}
\end{figure}

\sm

\subsubsection*{Terms with $p=3$ and $w=2$}
Let us now extend the argument to impose constraints on the six-particle $p=3$ terms with coefficients  $F^{(3)}_{2, 1} (\tau)$ and  $F^{(3)}_{2, 2} (\tau)$.   We will proceed  by  considering the example of the seven-particle amplitude with external states consisting of four gravitons, two $\Z $'s and one $\bar{\Z }$.   Again, a supersymmetric contact term cannot exist for such an amplitude, and therefore the super-amplitude is fully determined by the lower-point amplitudes via tree-level factorisitions. The terms which contribute to this amplitude are schematically shown in figure~\ref{fig:E22}.   In this case, there are two independent contact vertices shown in figure~\ref{fig:E22}(a) and figure~\ref{fig:E22}(b), and the examples of factorising contributions are shown in figure~\ref{fig:E22}(c), (d) and (e).  

\sm 

The contribution of~\ref{fig:E22}(d) represents the vertex $d^4R^5Z$, which is proportional to $\cD F^{(2)}_{R^5}(\tau^0)$. As we have argued previously in \eqref{eq:dpR5} that $F^{(2)}_{R^5}(\tau^0)$ is proportional $\F^{(3)}_0(\tau^0)$ therefore we have $\cD F^{(2)}_{R^5}(\tau^0) \sim \cD \F^{(3)}_0(\tau^0) \sim \F^{(3)}_1(\tau^0)$. Again, the amplitude also has a factorisation contribution that involves a four-particle and five-point higher-dimensional interactions, as shown in figure~\ref{fig:E22}(e). The absence of supersymmetric contact terms implies that the coefficient of each contact vertex is linearly related to the coefficients of the factorising terms, therefore we have following relations, 
\bea
\bar\cD\,\EE_{2,1}  (\tau^0)+ b_1 \, \EE_{1}  (\tau^0)+ b_2\, E_0(\threeh, \tau^0)  E_1(\threeh, \tau^0)&=& 0 \, , 
\label{ee21res}
\eea
and 
\bea
 \bar\cD\, \EE_{2,2} (\tau^0) + c_1 \, \EE_{1}  (\tau^0)+ c_2\, E_0(\threeh, \tau^0) E_1(\threeh, \tau^0) &=& 0 \, .
\label{ee22res}
\eea
As discussed earlier, $\EE_{2,1}$ (the coefficient of $\cO_{6,1}^{(3)}$) is related to $\EE_{0}$ by the action of covariant derivatives. With the normalisation of section \ref{sec: EE21}, we have $\EE_{2,1} =4 \,  \cD \cD \EE_0$. From this, we find
\bea
b_1 = -5 \, ,\qquad b_2 ={3\over 2}\, ,
\eea
as shown in \eqref{eq:DbE21} .

\sm 

The modular function $\EE_{2,2}$ is genuinely new and more interesting.  The fact that $\EE_{2,2}$ does not contain a tree-level term (which requires that $c_1=-2\, c_2$), results in 
 the expression (\ref{eq:DbE22})  for  $\EE_{2,2}$, 
\bea \label{eq:DbEE22-2}
\bar\cD\,\EE_{2,2}  (\tau^0)= c_1 \left(\EE_{1} (\tau^0) -  {1\over 2} E_0(\threeh, \tau^0)\, E_1(\threeh, \tau^0)\right)  \, .
\eea
We see that the modular  forms  $\EE_{2,1}(\tau)$ and $\EE_{2,2}(\tau)$ satisfy two distinct first-order differential relations that involve different linear combinations of $ \EE_{1}(\tau)$ and $E_0(\threeh, \tau)E_1(\threeh, \tau)$.  
Various features of  $\EE_{2,1}$ and $\EE_{2,2}$,   such as their perturbative expansions, were discussed in section~\ref{eighthbps}.
\sm

The constant $c_1$ is in principle determined by supersymmetry by  considering the seven-particle super-amplitude. This constant  could also be fixed by an explicit evaluation of the dimension-$14$ contribution to the low-energy  expansion  of the  six-particle one-loop string amplitude although we have not done this.  Finally,  once the six-particle terms are obtained, all the BPS maximal $U(1)$-violating interactions are completely fixed by soft limits as discussed in section \ref{softsec}. In particular, the coefficient of the $n$-particle kinematic  structure $\cO^{(3)}_{n,2}$ with $n>6$ is determined by acting with  covariant derivatives on $\EE_{2,2}(\tau)$.   So we conclude that all the BPS maximal $U(1)$-violating amplitudes are determined up to the constant $c_1$ that we have not evaluated. 
  
\sm 

\section{ Summary and discussion}
\label{discussion}

The aim  of  this paper has been to determine the first-order  differential equations that 
determine the moduli-dependent coefficients of BPS-protected terms in the low-energy expansion of type IIB superstring theory in a flat ten-dimensional Minkowski space  background. 
 The terms  in the  action \eqref{chargeq}  are $SL(2,\ZZ)$-invariant higher-derivative interactions that have the form  of moduli-independent $U(1)$-violating interactions multiplied by  $\tau$-dependent coefficients. These coefficients are modular forms, which transform by a phase under the action of $SL(2,\ZZ)$,  which compensates for the  $U(1)$-violation.
 The  BPS-protected interactions are the ones with $p=0$,  $p=2$ and $p=3$, which have dimension $\le 14$ (where classical Einstein gravity has dimension $2$).

\sm
 
The considerations of this paper followed two interrelated paths, investigating  the  higher-derivative effective action  \eqref{chargeq} in sections \ref{effact}  and \ref{eighthbps} and properties of $U(1)$-violating scattering amplitudes in sections~\ref{uoneamp}, \ref{sec:softB} and \ref{constraints}.\footnote{ There are a number of other conjectured generalisations to the coefficients of higher-dimension interactions, such as those of  \cite{Berkovits:1998ex, Basu:2008cf}, which are distinct from the consideration of this paper.  }  

\subsubsection*{Summary of the  effective interactions}

The lowest-order terms in the low-energy expansion  beyond  classical supergravity are those with $p=0$ (of order $R^4$)  and $p=2$ (of order  $d^4R^4$), which were the subject of earlier work.  These have  coefficients  proportional to Eisenstein modular  forms  with properties summarised in appendix~\ref{holosol},
\bea
\F_w^{(p)}(\tau)  =  c_w^{(p)}\, E_w(s,\tau)\,, \qquad \qquad\quad s=\frac{3+p}{2}\,, \qquad p=0,2\,,
\label{modp02}
\eea
where  $c_w^{(p)}$ are  numerical constants  that according to our convention, are determined  by  \eqref{eq:softn2}.
These functions are related  to each other by covariant  derivatives that raise  and lower the modular weights  as in \eqref{dzsn}  and  \eqref{dbarzsn},
\bea
E_{w+1}(s,\tau)=\frac{2}{s+w}\, \cD_wE_w(s,\tau)\,, \quad\qquad  E_{w-1}(s,\tau)=\frac{2}{s-w}\,\bar \cD_{-w}E_w(s,\tau)\,, 
\label{psval}
\eea
which imply  the Laplace eigenvalue equations 
\bea
\left(\Delta_{(-)}-s(s-1)+w(w-1)\right)\,  E_w(s,\tau)=0\,.
\label{lapone}
\eea

The structure  of  the $1/8$-BPS terms, for which $p=3$,  were  determined in section  \eqref{eighthbps} based on consistency with the coefficient of the $w=0$ case (the $d^6 R^4$ interaction).   In  these cases the coefficients are  modular forms given by 
\bea
\F_{w,i}^{(3)}(\tau)  =  c_{w,i}^{(3)}\, \EE_{w,i}(\tau)\,,
\label{modp02}
\eea
where $ \EE_{w,i}(\tau)$ satisfy  the following first-order differential equations\footnote{We have chosen the normalisation such that $c_{w,i}^{(3)}=1$ for all $w$. } 
\bea 
\label{firstp3}
\cD \EE_0 (\tau) &=& \frac{1}{2}\, \EE_1 (\tau)\,, \qquad\quad\quad \bar \cD \EE_1 (\tau) = 6\, \EE_0 (\tau)- \frac{1}{2}\left( E_0(\threeh,\tau)\right)^2\,,  \nn\\
 \cD \EE_1 (\tau) &=& \frac{1}{2}\, \EE_{2,1} (\tau)\,, \qquad\quad \quad \! \bar\cD \EE_{2,1}(\tau) = 5\, \EE_1 (\tau)- \frac{3}{2}  E_0(\threeh,\tau) E_1(\threeh,\tau) \,,  \\
&& \bar \cD \EE_{2,2} (\tau) = c_1 \left(\EE_1 (\tau)- \frac{1}{2}  E_0(\threeh,\tau) E_1(\threeh,\tau)\right) \,,  \nn
\eea
which imply the  following inhomogeneous Laplace  eigenvalue  equations
\bea \label{pthree}
\left(\Delta_{(-)}  -12\right)\, \EE_0(\tau) &=& - \left( E_0(\threeh,\tau)\right)^2 \,,\nn\\
\left(\Delta_{(-)}  - 12 \right)\, \EE_1(\tau) &=& -3\, E_0(\threeh,\tau) E_1(\threeh,\tau) \,,\\
\left(\Delta_{(-)}  -10\right)\, \EE_{2,1}(\tau) &=& -\frac{15}{2} \left( E_0(\threeh,\tau) E_2(\threeh,\tau)  + \frac{3}{5} \left(E_1(\threeh,\tau) \right)^2 \right) \,,\nn\\
\left(\Delta_{(-)}  -10\right)\, \EE_{2,2}(\tau) &=& -\frac{5c_1}{2} \left(E_0(\threeh,\tau) E_2(\threeh,\tau) -\left(E_1(\threeh,\tau) \right)^2\right) \nn \,.
\eea

\sm

We note,  in  particular, that the six-particle coefficient,  $\EE_{2,1}(\tau)$, has perturbative tree-level, one-loop and three-loop contributions. 
 It multiplies the  kinematic invariant $\cO^{(3)}_{6,1}$, which reduces to the  unique five-particle invariant  $\cO^{(3)}_{5,1}$ when any of the six  external momenta vanishes.  Combined with the first-order differential relation \eqref{firstp3}  this relates the  non-perturbative $p=3$ term in the low-energy expansion of the  six-particle amplitude to the $p=3$ term in the expansion of the five-particle amplitude.

\sm 

By contrast, the  coefficient  $\EE_{2,2}(\tau)$ has no perturbative tree-level coefficient, which is consistent with the  fact  that it multiplies the six-particle kinematic invariant $\cO^{(3)}_{6,2}$.  This invariant vanishes when any of the six particles has zero momentum, so it is  not related to a five-particle amplitude in the soft limit.

\subsubsection*{Summary of constraints from maximal $U(1)$-violating amplitudes}

Scattering amplitudes are evaluated in backgrounds with constant background, $\tau=\tau^0$ leading to maximal  $U(1)$-violating  amplitudes depend  on  the complex coupling in a manner that is  described by  appropriate modular forms.   
In section~\ref{uoneamp} we described  the  relation between the higher-derivative protected terms in the action and such amplitudes. 

\sm

Although ``naked'' factors of the modulus  $\tau$ cannot arise in $\cP_n(\{\Phi\})$, general maximal $U(1)$-violating scattering amplitudes have external complex scalar states, in addition to the fields in $\cP_n(\{\Phi\})$. These  are obtained by expanding the modular form  coefficients, $\F_{w,i}^{(p)}(\tau)$, in fluctuations of $\tau$.  It is essential to choose an appropriate parameterisation of these moduli  fluctuations in order to preserve manifest invariance under $SL(2,\ZZ)$ acting on $\tau^0$  as well as on the fluctuations.  This procedure, which is the normal coordinate expansion  for the $SL(2,\RR)/U(1)$ non-linear  sigma model,  leads to an  expansion in powers of the field $\Z $ defined  in \eqref{bdefone}.  The coefficients of the terms with the same power counting but with different numbers of $\Z $ fields are related by covariant derivatives. 
This procedure is consistent with the soft limits that relate higher-point amplitudes with lower-point ones, which were confirmed explicitly in \eqref{treeamp}  using coefficients of  the low-energy expansion of the $n=4, \, 5$ and $6$ superstring tree amplitudes (which were kindly  provided by  Oliver Schloterer~\cite{SchlottereNew}) as well as $n=4, \, 5$ one-loop results obtained from reference~\cite{Green:2013bza}. The explicit tree low-energy expansion of the tree amplitudes determine ratios of the coefficients $c_{w,i}^{(p)}$, as given in \eqref{cdefs}.

\sm

The constraints imposed by supersymmetry were analysed in section~\ref{constraints}  by extending the procedure in \cite{Wang:2015jna}.   This considers $n$-particle BPS-protected vertices together with some extra external $\Z $ and $\bar \Z $ states.  
Supersymmetry forbids contact interactions for such augmented amplitudes, which leads to conditions that constrain the modular coefficients of the BPS-protected vertices.  In this manner we  recover the conditions on the  coefficient modular forms
of sections \ref{effact}  and \ref{eighthbps}, which demonstrates directly  that the  first-order  equations  are indeed  a direct consequence of supersymmetry.

\subsection{Discussion}

\begin{itemize}

\item
We saw from \eqref{multideriv} that  a term in the large-$\tau_2$ expansion that contributes a negative integer power of $\tau_2$ is annihilated by a sufficient number of covariant derivatives. As  a result, we saw that  the  modular form $\F_{2,1}^{(3)} (\tau)\sim  \cD_1\cD_0 \F_0^{(3)}(\tau)$, which is   coefficient of 
 $d^6_{(1)}\, R^4 \Z ^2$,  has a vanishing two-loop term.  Similarly, it is easy to see that the coefficients $\F^{(3)}_{m,1} (\tau)$ with $m>3$  have vanishing two-loop and three-loop contributions.   Similarly, $\cD_{v+1} \dots \cD_2\F_{2,2}^{(3)} (\tau)$ not only has no tree-level term, but  the three-loop term also vanishes for all $v>2$. 

\sm 
 
 \item
 There has been a significant literature on the generalisation of the equations for the modular-invariant coefficients of the $w=0$ BPS-protected interactions (such as $d^{2p}R^4$ with $p\le 3$)  to type II superstring theory  compactified on a $d$-torus to $D=10-d$  dimensions.  The solutions are specific automorphic functions associated with the higher-rank duality groups in the $E_{d+1}$ series (see, for example, \cite{Green:2010kv, Pioline:2010kb} and references therein).
 It would be of interest to generalise these considerations to include processes in which the R-symmetry is broken, perhaps along the lines of \cite{Basu:2011he} and \cite{Wang:2015aua}.
 
  \sm

\item 

Finally, we note that one of the motivations for studying the constraints imposed by maximal supersymmetry
and $SL(2,\ZZ)$ duality in the ten-dimensional type IIB theory is to better understand  the holographic connection
with $SL(2,\ZZ)$  Montonen--Olive duality in four-dimensional ${\cal N} = 4$, $SU(N)$ supersymmetric
Yang--Mills theory. In particular, the pattern of $U(1)$-violation in type IIB superstring
amplitudes is the holographic image of the violation of the ``bonus'' $U(1)$ of  \cite{Intriligator:1998ig} in the gauge
theory.    In order to exhibit the $SL(2,\ZZ)$ duality  it is necessary to choose $g_{YM}$ and $N$ (rather than the 'tHooft coupling $\lambda=g_{YM}^2 N$ and $N$)  as independent parameters in the large-$N$ limit of  the  Yang--Mills theory.  This was discussed in \cite{Basu:2004dm} making use of properties of the operator product expansion of the composite gauge invariant Yang--Mills operator that is the holographic dual of $\tau$.  Plausibility arguments were given that in the large-$N$ fixed limit the dependence on $g_{YM}$ of certain BPS-protected correlation functions of gauge invariant operators in the 1/2-BPS Yang--Mills current supermultiplet is determined by the same $SL(2,\ZZ)$-covariant differential equations as those satisfied by the 1/2-BPS terms in the low-energy expansion of type IIB superstring amplitudes. The same limit also entered in \cite{Binder:2019jwn} in the context of the holographic connection between  four-dimensional maximally supersymmetric large-$N$ Yang--Mills and the  flat-space limit of the $AdS_5\times S^5$ type IIB superstring  theory.

\end{itemize}

\section{Acknowledgements}

We thank Nathan Berkovits, Paolo Di Vecchia, Carlos Mafra, Rodolfo Russo and Oliver Schlotterer  for very useful conversations.   CW is supported by a Royal Society University Research Fellowship No. UF160350 and is grateful to the Niels Bohr international Academy for its hospitality.   MBG has been partially supported by STFC consolidated grant ST/L000385/1 and by a Leverhulme Emeritus Fellowship.  He also thanks the  Niels Bohr international Academy for the support of a Simons Visiting Professorship. We thank the Galileo Galilei Institute for Theoretical Physics and INFN for hospitality and partial support during the workshop ``String Theory from a worldsheet perspective" where part of this work has been done.

\appendix

\section{Review of classical type IIB  supergravity} 
\label{app:SUGRA}

We here review some features of type IIB supergravity that are needed in the main body of the paper.  This is to some extent  based on
\cite{Schwarz:1983qr} and the appendices of \cite{Green:1998by}.\footnote{Whereas in \cite{Schwarz:1983qr}  the scalar fields were taken to parameterise the coset space $SU(1,1,\RR)/U(1)$, in \cite{Green:1998by} they were taken to paramterise $SL(2,\RR)/U(1)$.}.

\subsection{The field content}
The  fields of type IIB supergravity transform in  representations of $SL(2,\RR)\times SO(2)$, where $SL(2,\RR)$ is a global symmetry and $SO(2)\sim U(1)$  R-symmetry is a local symmetry.  The fermions are charged under  $SO(2)$ but are $SL(2,\RR)$ singlets.  With the exception of the scalar fields, the bosons are neutral with respect to the $SO(2)$ but transform in non-trivial representations of $SL(2,\RR)$.    

\sm

The scalar fields parameterise a $SL(2,\RR)$ matrix, which has three independent real components.   But this description is redundant since the local $SO(2)$ symmetry can be used to eliminate one scalar field, which restricts the scalar fields to the coset $SL(2,\RR)/U(1)$.
In order to understand the parameterisation of the fields it is useful to review properties of these scalar fields and their restriction to the coset.

\subsubsection*{The scalar fields and the   $SL(2,\RR)/U(1)$ coset}

A general $SL(2,\RR)$ matrix 
can be written in the  $N\times A \times K$ Iwazawa form,
\bea
\hat V(\tau_1,\tau_2,\phi) &=&
 \begin{pmatrix}
1 & \tau_1\\
0 &1
\end{pmatrix}
\begin{pmatrix}
\tau_2^{1/2} &  0 \\
0 &\tau_2^{-1/2}
\end{pmatrix}
\begin{pmatrix}
\cos \phi & -\sin \phi\\
\sin \phi &\cos \phi
\end{pmatrix} \\
&=& 
\frac{1}{\sqrt{\tau_2}}
 \begin{pmatrix}
\tau_2 \cos \phi + \tau_1 \sin\phi  &-\tau_2  \sin\phi+\tau_1 \cos \phi   \\
 \sin\phi &\cos \phi
\end{pmatrix}\,.
\label{izawa}
\eea
The indices on the matrix $\hat V^\alpha_{\ \ i}$ indicate that it transforms on the left by the global $SL(2,\RR)$ and on the right by the local $SO(2)$, i.e.,
\bea
\label{vtrans}
\hat V^\alpha_j \to (U^{-1})^\alpha_{\ \beta}
\, \hat V^\beta_i\, R_{ij} (\Sigma)
\,,
\eea
where 
\bea
U^\alpha_{\ \beta}= \begin{pmatrix}
a  & b \\
c &d
\end{pmatrix}\,, \qquad\quad  a,b,c,d\in \RR, \qquad\det U =1\,,
\label{Sdef}
\eea
is a $SL(2,\RR)$ matrix and $R_{ij}(\Sigma)$ is a rotation through an angle $\Sigma$. Note that
\bea
\hat V^\alpha_{\ \ i} \hat V_i^{\ \beta}:= M^{\alpha\beta}= \frac{1}{\tau_2}\begin{pmatrix}
\tau_1^2 + \tau_2^2 & \tau_1 \\
\tau_1 &1
\end{pmatrix}  \, ,
\label{Mdef}
\eea
is a $SL(2,\RR)$  matrix that is independent of $\phi$.

\sm 

The  local $SO(2)$ gauge symmetry  can be used to set $\phi=0$, which restricts the scalar fields to the two-dimensional coset  $SL(2,\RR)/SO(2)$.  In that case we have 
\bea
\hat V(\tau_1,\tau_2,0) =
 \frac{1}{\sqrt{\tau_2}}
 \begin{pmatrix}
\tau_2  & \tau_1 \\
0 &1
\end{pmatrix}\,.
\label{phigauge}
\eea
 
\sm 

We often use a complex $U(1)$  basis by adding and subtracting the columns in \eqref{izawa} and setting $\phi=0$, which takes $\hat V(\tau_1,\tau_2,0) \to V(\tau)$ defined by
\bea  \label{vdef}
V(\tau)=
\begin{pmatrix}
V^1_- & V^1_+ \\
                            V^2_- & V^2 _+
                            \end{pmatrix}
                            = {1\over \sqrt{-2i\tau_2}}
\begin{pmatrix}
\bar \tau & \tau \\
                1 & 1 
\end{pmatrix}\,,
                \eea
where $\tau=\tau_1+i \tau_2$ and $V_\pm=V_1\pm iV_2$.    In this basis the coset space is $SL(2,\RR)/U(1)$.
 
\sm 

After making the gauge choice $\phi=0$,  the action of the global  $SL(2,\RR)$ must be accompanied by a compensating gauge transformation together with a nonlinear redefinition of $\tau$ in order to ensure that $V(\tau)$ remains of the form \eqref{vdef}.
A general transformation \eqref{vtrans} that preserves the gauge combines a $SL(2,\RR)$ transformation with a compensating $U(1)$ transformation that leaves the form of $\hat V$ in \eqref{vdef} unchanged has the form 
\bea  
\label{vtwo}
(V_+^\alpha(\tau) ,V_-^\alpha (\tau))  \to (U^{-1})^\alpha_{\ \beta}
\left( V^\beta_+(\tau') e^{- i\Sigma},  V^\beta_- (\tau')e^{ i\Sigma}\right)\,
\eea
where 
\bea
\tau'= \frac{a\tau + b}{c \tau +d}\, ,
\label{tutranss}
\eea
and the compensating $U(1)$ transformation is given by 
\bea
e^{ i\Sigma} =\left( \frac{c \tau +d}{c\bar  \tau+d}\right)^\half\,.
\label{compensate}
\eea
 
 So we see that after restricting the scalar fields to the coset,  a $SL(2,\RR)$ transformation induces a $U(1)$ transformation that acts on the fermions, even though they were originally  $SL(2,\RR)$ singlets.
 
\sm 
 
\subsubsection*{The supergravity fields}

There are 128 physical bosonic states, of which $64$ come from the  Neveu--Schwarz/Neveu--Schwarz ($NSNS$)  and $64$ from the  Ramond--Ramond ($RR$) sector.  The fields of the  $NSNS$ sector consist of the graviton, which is a $U(1)$ and $SL(2,\ZZ)$ singlet; the second-rank antisymmetric potential $B_2$ with field strength $H=dB_2$, which is a $U(1)$ singlet and forms part of a $SL(2,\ZZ)$ doublet; the dilaton $\varphi$, which enters into the imaginary part of the complex modulus field, $\tau_2=e^{-\varphi}$.
 
\sm 

The fields of the $RR$ sector consist of the even-rank potentials, $C^{(p)}$ ($p=0,2,4$), with field strengths  $F^{(p+1)} = d C^{(p)}$.  The pseudoscalar  defines the real part of the complex modulus, $\tau_1=C^{(0)}$. The RR field strength $F^{(3)}$ is a $U(1)$ singlet that forms the other part of the $SL(2,\ZZ)$ doublet.  The five-form field strength $F^{(5)}$ is a $U(1)$ and $SL(2,\ZZ)$ singlet which satisfies a self-duality condition, $F^{(5)} =-* F^{(5)}$.
 
\sm 

The $128$ physical fermionic states are described by fermions in the $NSNS\times RR$ sector together with those of the $RR\times NSNS$ sector, which can be combined to form a complex chiral gravitino, $\psi_{\mu} = \psi_{1\, \mu}+ i\psi_{2\,\mu}$,  and a complex spin-half dilatino, $\Lambda=\Lambda_1+i\Lambda_2$ of the opposite chirality (where the subscripts $1$ and $2$ indicate the $NS\times R$ sector and the $R\times NS$ sector, respectively).  These fermion fields are invariant under $SL(2,\RR)$, while $\psi$ carries a $U(1)$ charge $q_{\psi} =  -1/2$ and $\Lambda$ carries charge $q_\Lambda=  - 3/2$.

 \subsubsection*{The $q=\pm2$ scalar fields}

The two-derivative  supergravity action can be conveniently expressed in a covariant form by appropriate  parameterisation of the  scalar fields.  The $SL(2,\RR)$ singlet expressions 
\bea 
\label{pdef}
P_\mu = - \epsilon_{\alpha\beta}V_+^\alpha \partial_\mu V^\beta_+ = i \frac{\partial_\mu \tau}{2 \tau_2} \, e^{-2\pi i \phi},\qquad \bar P_\mu = - \epsilon_{\alpha\beta}V_-^\alpha \partial_\mu V^\beta_- = - i \frac{\partial_\mu \bar \tau}{2 \tau_2} \, e^{2\pi i \phi},
\eea
manifestly transform with $U(1)$ charges $q_{P}= -2$ and $q_{\bar P}=2$, respectively.  Upon fixing the gauge $\phi=0$  they transform  int the following manner  under the $U(1)$ transformations induced from the $SL(2,\RR)$ transformations
\bea
P_\mu\to \left(\frac{c\bar \tau+d}{c \tau+d} \right) \, P_\mu\,,\quad\qquad  \bar P_\mu\to  \left(\frac{c \tau+d}{c\bar \tau+d} \right)  \bar P_\mu\,.
\label{ptrans}
\eea

 \subsubsection*{The $U(1)$ connection and covariant derivatives}
 
Space-time derivatives need to be augmented with a $U(1)$ gauge connection in order to express the action in a $SL(2,\RR)$-invariant manner.  The $SL(2,\RR)$ singlet expression,
\bea
\label{qdef}
Q_\mu =-\frac{i}{2} \epsilon_{\alpha\beta} \, (V^\alpha_+
\partial_\mu V^\beta_-  -  V^\beta_-
\partial_\mu V^\alpha_+ )= \frac{\partial_\mu \tau_1}{2\tau_2}-\partial_\mu \phi\,,
\eea
is the composite $U(1)$ gauge connection that  transforms as $Q \to Q -\partial_\mu \Sigma$ under the local transformation \eqref{vtwo}.
 Thus,   we define the covariant space-time derivative acting on charge-$q =-2w$ fields
\bea
\D_w :=   \partial_\mu + i \, q\,Q_\mu \,.
\label{covdu}
\eea
 
\sm 

It is easy to verify that under the induced $U(1)$ transformation that accompanies a $SL(2,\RR)$ transformation in the gauge $\phi=0$,  the transformation of $Q_\mu$  is given by
\bea
Q_\mu=\frac{\partial_\mu \tau_1}{2\tau_2} &\to& \frac{\partial_\mu \tau}{4\tau_2}\left(\frac{c\bar \tau+d}{c \tau+d} \right)  +\frac{\partial_\mu\bar \tau}{4\tau_2}  \left(\frac{c \tau+d}{c\bar \tau+d} \right)     \nn\\
 & =& Q_\mu + \partial_\mu\Sigma  \,,
\label{qtrans}
\eea 
where $\Sigma$ was defined in \eqref{compensate}.
 
\sm 

To verify that $\D_w$ is indeed a $U(1)$-covariant derivative note that under a $SL(2,\ZZ)$ transformation \eqref{tautrans} a charge-$q=-2w$ field transforms as
\bea
\Phi_q \to \frac{(c\bar \tau+d)^w}{(c \tau+d)^w}\, \Phi_q= \Phi_q\, e^{iq\Sigma}\,,
\label{qtrans}
\ee
so we have 
\bea
\partial_\mu\, \Phi_q  &\to&    \left(   \frac{c \,w\,\partial_\mu \bar \tau}{c\bar \tau+d}   -  \frac{c\, w\,\partial_\mu \tau}{c\tau + d }  \right)
 \frac{(c \bar \tau+d)^w}{(c  \tau+d)^w}\, \Phi_q +  \frac{(c\bar \tau+d)^w}{(c  \tau+d)^w}\,\partial_\mu \Phi_q\nn\\
&=&  \frac{(c \bar\tau+d)^w}{(c\tau+d)^w}   \,\left(\partial_\mu - i q \partial_\mu \Sigma  \right)\Phi_q
\, .
\label{derivphi}
\ee
Therefore
\bea
(\partial_\mu + i q Q_\mu)\, \Phi_q  \to  \frac{(c\bar \tau+d)^w}{(c\tau+d)^w}   \,\left(\partial_\mu + i q Q_\mu \right)\Phi_q 
\, .
\label{derivphi}
\ee
 
\sm 

\subsection{Terms in the type IIB supergravity action}

\noindent{\it The scalar field action}

The scalar field kinetic  term has the form  (in Einstein frame)

\bea
S_\tau=-\frac{1}{\kappa^2} \int d^{10} x \,e\, \frac {\partial_\mu \tau  \partial^\mu\bar\tau}{2\tau_2^2} =-\frac{2}{\kappa^2} \int d^{10} x \,e\, 
P_\mu\,  \bar P^\mu\,.
\label{scalact}
\eea 
  
\sm 

\noindent{\it The  other bosonic fields}

The two antisymmetric second-rank potentials,  $B_{\mu\nu}$ and
$C^{(2)}_{\mu\nu}$, have   field strengths $H=dB_2$   and $F^{(3)}=dC^{(2)}$.
that form an  $SL(2,\RR)$ doublet, $F^\alpha$.  In discussing the $SL(2,\ZZ)$ properties of the theory is very natural to package
them into the $SL(2,\RR)$ singlet fields,
\bea
\label{gdef}
G = - \epsilon_{\alpha\beta} V_+^\alpha F^\beta ,
\qquad \bar G =-\epsilon_{\alpha\beta} V_-^\alpha F^\beta\,,
\eea
which carry $U(1)$ charges $q_G =-1$ and $q_{\bar G}= +1$, respectively.
The kinetic term involving $G$ in the action is given by 
\bea
S_G=  -\frac{1}{\kappa^2} \int d^{10} x \,e\,\frac{1}{2} G\, \bar G\,.  
\label{Gact}
\eea
 
\sm 

The antisymmetric fourth-rank potential, $C^{(4)}$,
with self-dual  field strength $F_5 = d C^{(4)}$,  has an equation of motion that is expressed by the self-duality
condition $F_5 = * F_5$, which cannot be obtained
from a globally well-defined Lagrangian.

\sm 

\noindent{\it The fermion field action }
 
The covariant Dirac action for the  dillatino has the form
\bea
\label{lambdact}
S_\Lambda = \frac{i}{\kappa^2} \int d^{10} x\, e\, \bar\Lambda \gamma^\mu\, (\partial_\mu + \frac{3}{2}i\, Q_\mu) \Lambda\,.
\eea
Similarly, in a fixed gauge   $\gamma_\mu \psi^\mu=0$ the Rarita--Schwinger equation for $\psi_\mu$ reduces to $\partial_\mu \psi^\mu=0$ and $\gamma\cdot D \psi_\mu=0$ and the action for  the  Rarita--Schwinger field can be  written as 
\bea
\label{psiact}
S_\psi= \frac{ i}{\kappa^2} \int d^{10} x\, e\, \bar\psi^\nu\,  \gamma^\mu\, (\partial_\mu - \frac{1}{2}i\, Q_\mu) \, \psi_\nu\,.
\eea

\sm

\noindent {\it Interaction terms}

Although we do not need the explicit supergravity  interaction terms in this paper we note that  they are invariant under $SL(2,\ZZ)$ and they conserve the local $U(1)$, which means that the phase  $\phi$ cancels out of the action.   For example, the complex  scalar field interacting with the fermions has the form 
\bea
\label{lag2}
S_{\Lambda\psi*}^{P} = \frac{i}{\kappa^2} \int d^{10} x\, e\,   \bar \Lambda
\gamma^\mu\gamma^\omega \bar \psi_\mu
 P_\omega+ c.c. \, .
 \eea
 
  
\subsection{The $SU(1,1)$ parameterisation of the complex scalar field fluctuations}
 \label{app:B-field}

For much of this paper we use moduli fields that parameterise the coset space $SL(2,\RR)/U(1)$, which is the upper half $\tau$ plane.   This is well suited to making the discrete identifications that are implied by invariance under the T transformation, $\tau\to \tau +1$, and the S transformation, $\tau \to -1/\tau$, which restrict $\tau$ to a fundamental domain of $SL(2,\ZZ)$.  
\sm 

However, as is common in coset space nonlinear sigma models, in discussing amplitudes with external scalar fields it is important that we define the fluctuating fields in a parameterisation that transforms covariantly under the symmetry.   Therefore we want to consider fluctuations of the bosonic fields  around a constant background $\tau=\tau^0$,  that transform covariantly under the $U(1)$ induced by $SL(2,\ZZ)$ transformations.  This is realised by the field redefinition of \eqref{bdefone}
\bea
\label{teautoB}
\Z= \frac{\tau - \tau^0}{\tau-\bar\tau^0}\,,
\eea
which is a $SL(2,\CC)$ transformation that maps the upper-half $\tau$ plane to the unit disk in the $\Z $ plane. The origin of the disk is the mapping of the point $\tau=\tau^0$ and its boundary is the real axis of the $\tau$ plane. It is easy to see that transforming $\tau$ and $\tau^0$ by  $SL(2,\ZZ)$ gives  the linear transformation 
\bea
\Z  \to \frac{c \tau^0+d}{c \bar\tau^0 + d}\,\Z \, .
\label{slact}
\eea
The advantage of describing the background in the $SL(2,\RR)$ parameterisation is that the duality transformations lie in the arithmetic subgroup $SL(2,\ZZ) \in SL(2,\RR)$ which is obtained by making discrete identifications of $\tau$ that restrict it to a single fundamental domain. This restriction is very unnatural in the $SU(1,1)$ parameterisation.
 
\sm

The definition of $\Z $ given in  \eqref{teautoB} leads to the expression
\bea
\tau_2 = \tau^0_2 \, \frac{1-\bar \Z  \Z }{(1-\Z )(1-\bar \Z )}\,,
\label{taub2}
\eea
and the field $P_\mu$  in \eqref{pdef}  becomes
\bea
P_\mu = i \frac{\partial_\mu \tau}{2 \tau_2} \, e^{-2\pi i \phi} = \frac{\partial_\mu \Z }{1-\bar \Z \Z }\, \left(\frac{1- \bar \Z }{1-\Z }\right) e^{-2 \pi i \phi}\,.
\label{predef}
\eea 
In our analysis we are setting $\phi=0$ in order to describe the coset in terms of $\tau$.

\sm

Likewise the expression for the connection becomes
\bea
Q_\mu = \frac{\partial_\mu \tau_1}{2\tau_2}  =
\frac{i}{2}\,  \frac{\Z \partial_\mu \bar \Z - \bar \Z  \partial_\mu \Z }{1-\bar \Z \Z } + \frac{i}{2}\,\partial_\mu \log \left(\frac{1- \bar \Z }{1-\Z }\right)  \,.
\label{newq}
\eea

\sm
 
 It is very simple to transform terms in the action, such as  the scalar kinetic term $S_\tau$ in \eqref{scalact}, or the interaction term $S_{\Lambda\psi*}^{P}$ \eqref{lag2}, from functions of $\tau$ to functions of $\Z $.
 
\sm 

Although we want to stay in the gauge  $\phi=0$, we note that in order for the transformation to reproduce the form of the $SU(1,1)/U(1)$ coset with $W^\alpha_{\ \pm}$  it would be necessary to change the $U(1)$ gauge so that 
\bea
e^{2\pi i \phi}= \left(    \frac{1- \bar \Z }{1-\Z }  \right) \,,
\label{newphi}
\eea
in which case \eqref{predef} and  \eqref{newq}  are the same as those in  \cite{Schwarz:1983qr}.

\section{Properties of modular forms and Eisenstein series}
\label{holosol}

We will here discuss some properties of the modular functions and modular forms that arise in the text.  
The simplest examples are the Laplace equations \eqref{laplaceone}  for $w=0$ and  $s\in \CC$,  
\bea
\left(\Delta - s(s-1) \right) f^{(0,0)}(\tau) =0\,,
\label{eiseneq}
\eea
where $\Delta = 4 \tau_2^2(\partial_\tau \, \partial_{\bar \tau})$ and $f^{(0,0)}(\tau)$  is  a $SL(2,\ZZ)$ modular function (so $w=0$)  that satisfies the boundary condition  $\lim_{\tau_2\to \infty} f^{(0,-0)}(\tau)< \tau_2^a$, where $a$ is a real number.  This condition of power boundedness follows from  string perturbation theory, where the most singular term has the tree-level behaviour.  The unique solution to this  Laplace  eigenvalue  equation with these boundary  conditions  is the non-holomorphic Eisenstein series, which has the form
\bea
\label{eisendef}
E(s,\tau) = \sum_{(m,n) \neq\,(0,0)}\frac{\tau_2^s}{|m+n\tau|^{2s}}     =   \sum_{N\in\ZZ} \cF_N(s,\tau_2) \, e^{2\pi i N \tau_1}\,,
\eea
where the zero mode consists of two power behaved terms,
\bea
\cF_0(s,\tau_2)  = 2\zeta(2s)\,  \tau_2^s \  +  \ \frac{2\sqrt \pi \,\Gamma(s-\frac{1}{2}) \zeta(2s-1)}{\Gamma(s)}\, \tau_2^{1-s} \,,
\label{eisenzero}
   \eea
and the non-zero modes  are proportional to $K$-Bessel functions,
\bea
\cF_N(s,\tau_2)  =   \frac{4\,\pi^s}{\Gamma(s)}\,  |N|^{s-\half} \, \sigma_{1-2s}(|N|)
\sqrt{\tau_2}\,K(s-\half, 2\pi |N|\tau_2) \,, \  \ \ N\neq 0\,,
\label{nonzeroeisen}
\eea
where the divisor sum is defined by  
\bea \label{eq:divisor-sum}
\sigma_p(N)=\sum_{d>0, {d|N}}  d^{2p}\, , \quad {\rm for} \quad N>0 \, ,
\eea 
and $\sigma_{-p}(N) = N^{-p}\, \sigma_p(N)$. 

\sm

The lowest order example of such a modular invariant coefficient is $\F^{(0)}_0(\tau) = E(\threeh,\tau)$, the coefficient of the $R^4$ interaction, which is the $p=0$ (i.e  dimension-8) term in the low-energy expansion  of the four-graviton amplitude.  This has a zero mode that contains two power-behaved terms given by \eqref{eisenzero} with $s=3/2$.  Taking into account the power of $\tau_2^{1/2}$ in transforming to the string frame in \eqref{chargeq}, these powers are $\tau_2^2$  and $\tau_2^0$, which correspond to tree-level and one-loop perturbative superstring contributions.  The $p=2$ term of order $d^4 R^4$ has a coefficient $E(\fiveh,\tau)$ that has  tree-level and two-loop perturbative contributions.

\sm

We are generally interested in modular forms with weights $(w,- w)$, or $U(1)$ charge $q=2w$.
Using the definitions of covariant derivatives in (\ref{covderdef})  we have,
\be
\label{dzsn}
\cD_w\, E_{w} (s,\tau)= {s+w\over 2}\, E_{w+1}(s,\tau)\, ,
\ee
and
\be
\label{dbarzsn}
\bar \cD_{-w}\, E_{w}(s,\tau) = {s-w\over 2}\, E_{w-1}(s,\tau)\, .
\ee
Note, in particular, that with this normalisation
\be
E_{w} (s,\tau)= \frac{2^w \Gamma(s) } { \Gamma(s+w)}\, \cD_{w-1} \cdots \cD_0\, E_{0}(s,\tau)\, ,
\label{zwdef}
\ee
where $E_{0}(s,\tau) := E (s,\tau)$.  It is straightforward to show that 
\be
\label{zswdef}
E_{w}(s,\tau) =\sum_{(m,n)\ne (0,0)} \left({m+n\bar\tau\over m+n\tau}\right)^w \, {\tau_2^s \over |m + n \tau|^{2s}} \, .
\ee

Iterating these equations gives the Laplace equations
\be
\label{laplaceplusn}
\Delta_{(+)}^{(w)} E_{w}(s,\tau) := 4\bar \cD_{-w-1} \cD_w E_{w} (s,\tau)= (s+w)(s-w-1) \, E_{w}(s,\tau)\, ,
\ee
\be
\label{laplaceminusn}
\Delta_{(-)}^{(w)}  E_{w} (s,\tau):=4\cD_{w-1} \bar \cD_{-w} E_{w} (s,\tau)= (s-w)(s+w-1) \, E_{w}(s,\tau)\, .
\ee
Note that the two laplacians acting on weight-$(w,-w)$ modular forms satisfy
\be
\label{difflap}
\Delta^{(w)}_{(+)} - \Delta^{(w)}_{(-)} = -2w\,, \qquad\quad   \Delta^{(0)}_{(+)}=  \Delta^{(0)}_{(-)}=\Delta \, .
\ee
Hence we see that the non-holomorphic  modular form ${f^{(w,-w)}}$ satisfying  the $SL(2,\ZZ)$-covariant Laplace eigenvalue equation \eqref{laplaceone}  has the solution
\bea
f_s^{(w,-w)}(\tau) := E_{w}(s,\tau)  \,.
\label{modw}
\eea
In the case $s=3/2$ that is relevant for the coefficients of the  $O(\left(\alpha')^{-1}\right)$ terms, this has a Fourier expansion of the form

\bea
\label{expadef}
E_{w}(\threeh,\tau)  = 
2\zeta(3)\, \tau_2^{{3\over 2}} + \frac{4 \zeta(2)}{1-4w^2} \, \tau_2^{-{1\over
2}} +  
\sum_{N =1}^\infty \left( \cF_{N,4-w} (\threeh,\tau_2)e^{2\pi i N\tau_1} + 
\cF_{N,4+w}  (\threeh,\tau_2)e^{-2\pi i N \tau_1} \right)\,.\nn\\
\eea
The first two terms in (\ref{expadef}) have the interpretation of contributions that should arise in string perturbation theory at   tree-level and one loop, while the instanton and anti-instanton terms are contained in
\bea
\label{znpddef}
\cF_{N,4+w}  (\threeh,\tau_2)
 = (8\pi)^{\half} \, \sigma_{-2}(N)\,   (2\pi N)^{\half} \,  \sum_{k=w}^\infty    \frac{a_{4+w, k}}   {(2\pi N \tau_2)^k}\,
e^{-2\pi  N\tau_2}
\eea
where
\begin{equation}
\label{ckrdef}{a_{n,k} = {(-1)^n \over 2^k (k-n+4)!}   \frac {\Gamma(\threeh) }{ 
\Gamma(n-\fiveh)} \frac{\Gamma (k -\half)}{ \Gamma(- k -\half)}}\,.
\end{equation} %
The instanton sum in (\ref{znpddef}) 
begins with the power $\tau_2^w$ for  D-instantons  (which have phases $e^{2\pi i N \tau_1}$)   
while the series of corrections to the anti D-instanton  (with phases $e^{-2\pi i N \tau_1}$)    starts with the power $\tau_2^{-w}$.  These powers are consistent with the requirement of saturating the fermionic zero modes that are present in the D-instanton background.

\section{Linearised supersymmetry and higher derivative terms}
\label{summact}

The supersymmetries of the ten-dimensional type IIB  theory are associated with two sixteen-component chiral fermionic $SO(9,1)$ spinors, $\theta_1$ and $\theta_2$, which have the same chirality.  It is convenient to combine these into a complex  supercharge $\theta=\theta_1+i\theta_2$,  and its complex conjugate, $\bar \theta$.  The linearised expressions for the effective interactions that preserve half of the $32$ supersymmetries can be simply obtained by packaging the physical fields or their field strengths into a constrained
superfield $\Phi(x^\mu - i\bar \theta\gamma^\mu \theta,\theta)$ where $\theta^A$  ($A=1, \dots,16$)
is  a complex Grassmann coordinate that transforms as a Weyl spinor of
$SO(9,1)$.  This superfield satisfies the 
holomorphic condition \cite{Howe:1983sra},
\begin{equation}\label{holocon}{ \bar D_{\theta} \Phi=0,} 
\end{equation} %
and is further constrained by imposing the condition,
\begin{equation}\label{onshell}{\DD^4  \Phi = \bar  \DD^{4}  \bar \Phi,}
\end{equation} %
where
\begin{equation}
\label{covderiv}{
\DD_A = {\partial \over \partial \theta^A} +2i (\gamma^\mu  \bar\theta)_A
\partial_\mu, \qquad   \bar\DD_A = - {\partial\over \partial  \bar \theta^{A}}}
\end{equation} %
are the holomorphic and anti-holomorphic covariant
derivatives that anticommute with
the rigid supersymmetries 
\begin{equation}\label{susys}{
Q_A ={\partial \over \partial \theta^A},
\qquad   \bar Q_A = - {\partial \over \partial \bar
\theta^{A}} +  2i
(\bar \theta \gamma^\mu  )_A \partial_\mu .}
\end{equation} %
The  constraints (\ref{holocon}) and (\ref{onshell}) ensure that the field $\Phi$ has an expansion in powers of $\theta$ (but not $
\bar\theta$), that terminates after
the
$\theta^8$ term  and  the 256 component fields satisfy the linearised field equations and Bianchi identities.
\begin{eqnarray}
\Phi & = & \tau^0_2 + \tau^0_2\,\Delta \nn\\
& = & \tau^0_2  + \tau^0_2( \hat {\tau}  + \theta \Lambda +  \theta^2 G+  \theta^3 \partial \psi
+    \theta^4( 
\cR_{\mu\sigma\nu\tau}+  \partial
 F_5)+
 \cdots  + \theta^8 \partial^4 \hat{\bar\tau})\nn \\
&:= &\tau^0_2+ \tau^0_2 \sum_{r =0}^8 \theta^r  \Phi^{(r)} ,
\label{expphi}
\end{eqnarray} %
where  we have suppressed all details of the  spinor  and tensor indices. The quantity  $\tau^0_2\, \Delta$ is the linearised fluctuation 
around   a constant  purely imaginary  flat background, $\tau^0_2 =g_s^{-1}$. 
The  fields 
 $G$ and $\bar G$ are complex combinations of the $RR$ and $NSNS$  field
strengths.
The $\theta^4$ terms are   the
Weyl curvature,  $R$, 
and the $RR$ five-form field strength, $ F_5$.     The fermionic field $\Lambda$ 
is the complex dilatino and $\psi$ is the complex gravitino.  
The terms  indicated by $\cdots$ in (\ref{expphi}) 
fill in the remaining members of the ten-dimensional $N=2$ chiral
supermultiplet, comprising (in symbolic notation)   $\partial
\psi$,  $ \partial ^2 \bar G$ and  $\partial^3 \bar\Lambda$.  The complex conjugate superfield $\bar\Phi$ is a function of $\bar \theta$ and has a similar expansion with the component fields interchanged with their complex conjugates.
 
\sm 

The $U(1)$ R-symmetry  charge $q_r$  of any   component 
$\Phi^{(r)}$  is  correlated with the powers of $\theta$. Assigning a
charge $-1/2$ to  $\theta$ and an overall charge $-2$ to the superfield leads
to the charge for the field with $r$ powers of $\theta$,
\begin{equation}\label{uchar}{q_r=- 2 + {r\over 2}}\,.
\end{equation}
Thus,   
$q_{\hat\tau} = -2$; $q_{\Lambda} =- 3/2$; $q_G=-1$; $q_{\psi} = -1/2$;  $q_R = q_{F_5} =0$.
 
\sm 

Although the linearised theory cannot capture the full structure of the terms in
the effective action it can be used to relate various terms in
the limit of weak coupling, $\tau^0_2 = g_s^{-1}\to \infty$ (where $g=e^{\phi_0}$ is the string 
coupling constant).  The linearised approximations to the complete 
interactions are those that
arise by integrating a function of $\phi$ over the sixteen components of $\theta$,
\begin{equation}\label{actdef}{
S_{linear} = \int d^{10}x  d^{16} \theta \, e\,  \H[\Phi] +\ {\rm c.c.},}
\end{equation} %
which is manifestly invariant under the rigid supersymmetry
transformations, (\ref{susys}).
The various  component interactions contained in (\ref{actdef}) are obtained
\ from the $\theta^{16}$ term in the expansion,
\begin{equation}\label{expands}{
\H[\Phi] = \H(\tau^0_2) + \Delta {\partial \over \partial \tau^0_2}
\H(\tau^0_2) + \frac{1}{2} \Delta^2 \left( {\partial \over \partial \tau^0_2}\right)^2  \H(\tau^0_2)   +
\cdots.}
\end{equation} %
Using the expression for  $\Delta$ in (\ref{expphi}) and substituting into (\ref{actdef})
leads to all the possible interactions at order $1/\alpha'$,  
\begin{eqnarray}\label{nonpert} 
S_{linear} & = & 
\int d^{10}x\, e\, \left(
h^{(12,-12)} \Lambda^{16} + h^{(11,-11)}  G \Lambda^{14} + \ldots
\right .\nn \\
& &  \left.
+ h^{(8,-8)}  G^8 +\ldots + h^{(0,0)} \, R^4 + \ldots +
h^{(-12,12)} \Lambda^{* \, 16}\right),
\end{eqnarray} %
where $h^{(w,-w)}$ are functions of $\tau^0_2$.
 
\sm 

The superscripts that label the coefficients $h^{(w,-w)}$ are related
to the violation of the $U(1)$ charge.  Thus, the linearised form of 
the general term in (\ref{nonpert}) contains a product of $p$ fields, 
\begin{equation}\label{genterm}{
\int d^{10}x\,  e\, h^{(w,-w)} \prod_{k=1}^p \Phi^{(r_k)},}
\end{equation} %
which violates the $U(1)$ charge by the units of
\begin{equation}\label{uviol} {\sum_{k=1}^p q_{r_k} =-2w= 8-2p} \, ,
\end{equation} %
where we have used (\ref{uchar}) and the fact that the total power of
$\theta$ must be $\sum_k r_k = 16$.  For example the $R^4$ term ($w=0$) 
conserves the $U(1)$ charge while the 
$\Lambda^{16}$ term ($w=12$) violates the $U(1)$ charge by $-24$  and
there are many other terms that violate the charge by any even number. 
 
\sm 

In the linearised approximation,  $g_s\to 0$ ($\tau^0_2 \to  \infty$),  
the coefficients $h^{(w,-w)}$ are constants that
are related to each other by use of the Taylor expansion, (\ref{expands}).
For example, the $R^4$ term has coefficient $\partial_{\tau^0_2}^4 \H$
while the $\Lambda^{16}$ term has coefficient $\partial_{\tau^0_2}^{16} \H$ so
that, at the linearised level,
\begin{equation}\label{linrels}{h^{(12,-12)} \sim \left( \tau^0_2 {\partial \over
\partial \tau^0_2}\right)^{12} h^{(0,0)},}
\end{equation} %
where for the moment we are not concerned about the overall constant.
In writing this we have used the fact that
the linearised approximation  is valid only if  the inhomogeneous term
in the modular covariant derivative, $\cD$  is negligible, which requires that 
\begin{equation}\label{validl}{2\tau^0_2 \partial_{\tau^0_2} h^{(w, - w)}  \gg w  h^{(w, - w)}}
\end{equation} %
since only in this case does the modular covariant derivative reduce to the ordinary derivative.   
This inequality is obviously not satisfied by terms in the expansion of $h^{(w,- w)}$ that are
powers of $\tau^0_2$. However, when acting on a factor  such as 
$(\tau^0_2)^ne^{ - 2\pi |N|\tau^0_2}$ (where $n$ is any constant) 
which is  characteristic of a  charge-$N$ D-instanton, the
inhomogeneous term may be neglected in the limit  $\tau^0_2 \to \infty$ and the  covariant
derivative linearises. Therefore, a linearised superspace
expression such as  (\ref{actdef}) should  contain the exact  leading
multi-instanton contributions to the    $R^4$ and related   terms.
These  leading  instanton terms arise by  substituting  the expression
\begin{equation}\label{leadinst}{F_N (\phi)=  c_N e^{2\pi i |N| \phi}}
\end{equation} %
into  (\ref{actdef}). 
 
\sm 

In the nonlinear theory 
the $SL(2,\ZZ)$ symmetry of the type IIB theory requires that the   $h^{(w,-w)}(\tau)$ are modular forms with
holomorphic and anti-holomorphic weights as indicated in the
superscripts.  
The relative coefficients of the interactions of different $U(1)$ charge could, in principle, be determined by supersymmetry, but we have not determined them in that manner.


\begin{thebibliography}{99}


\bibitem{Gaberdiel:1998ui}
  M.~R.~Gaberdiel and M.~B.~Green,
  ``An SL(2, Z) anomaly in IIB supergravity and its F theory interpretation,''
  JHEP {\bf 9811} (1998) 026
doi:10.1088/1126-6708/1998/11/026
  [hep-th/9810153].



\bibitem{Gross:1986iv}
  D.~J.~Gross and E.~Witten,
  ``Superstring Modifications of Einstein's Equations,''
  Nucl.\ Phys.\ B {\bf 277} (1986) 1.
doi:10.1016/0550-3213(86)90429-3
  
\bibitem{Grisaru:1986px}
  M.~T.~Grisaru, A.~E.~M.~van de Ven and D.~Zanon,
  ``Four Loop beta Function for the N=1 and N=2 Supersymmetric Nonlinear Sigma Model in Two-Dimensions,''
  Phys.\ Lett.\ B {\bf 173} (1986) 423.
doi:10.1016/0370-2693(86)90408-9
  

\bibitem{Green:1997tv}
  M.~B.~Green and M.~Gutperle,
  ``Effects of D instantons,''
  Nucl.\ Phys.\ B {\bf 498} (1997) 195
doi:10.1016/S0550-3213(97)00269-1
  [hep-th/9701093].
  
\bibitem{Green:1997as}
  M.~B.~Green, M.~Gutperle and P.~Vanhove,
``One loop in eleven-dimensions,''
  Phys.\ Lett.\ B {\bf 409} (1997) 177
doi:10.1016/S0370-2693(97)00931-3
  [hep-th/9706175].
  
\bibitem{Green:1998by}
  M.~B.~Green and S.~Sethi,
  ``Supersymmetry constraints on type IIB supergravity,''
  Phys.\ Rev.\ D {\bf 59} (1999) 046006
doi:10.1103/PhysRevD.59.046006
  [hep-th/9808061].
  
  

\bibitem{Boels:2012zr}
  R.~H.~Boels,
  ``Maximal R-symmetry violating amplitudes in type IIB superstring theory,''
  Phys.\ Rev.\ Lett.\  {\bf 109} (2012) 081602
doi:10.1103/PhysRevLett.109.081602
  [arXiv:1204.4208 [hep-th]].


\bibitem{Howe:1983sra}
  P.~S.~Howe and P.~C.~West,
  ``The Complete N=2, D=10 Supergravity,''
  Nucl.\ Phys.\ B {\bf 238} (1984) 181.
doi:10.1016/0550-3213(84)90472-3

\bibitem{Sinha:2002zr}
  A.~Sinha,
  ``The G(hat)**4 lambda**16 term in IIB supergravity,''
  JHEP {\bf 0208} (2002) 017
doi:10.1088/1126-6708/2002/08/017
  [hep-th/0207070].
  


\bibitem{Green:2005ba}
  M.~B.~Green and P.~Vanhove,
  ``Duality and higher derivative terms in M theory,''
  JHEP {\bf 0601} (2006) 093
doi:10.1088/1126-6708/2006/01/093
  [hep-th/0510027].
 
  
\bibitem{Wang:2015jna}
  Y.~Wang and X.~Yin,
  ``Constraining Higher Derivative Supergravity with Scattering Amplitudes,''
  Phys.\ Rev.\ D {\bf 92} (2015) no.4,  041701
doi:10.1103/PhysRevD.92.041701
  [arXiv:1502.03810 [hep-th]].
  
\bibitem{Elvang:2010jv}
  H.~Elvang, D.~Z.~Freedman and M.~Kiermaier,
  ``A simple approach to counterterms in N=8 supergravity,''
  JHEP {\bf 1011} (2010) 016
  doi:10.1007/JHEP11(2010)016
  [arXiv:1003.5018 [hep-th]].
  
\bibitem{Chen:2015hpa} 
  W.~M.~Chen, Y.~t.~Huang and C.~Wen,
  ``Exact coefficients for higher dimensional operators with sixteen supersymmetries,''
  JHEP {\bf 1509}, 098 (2015)
doi:10.1007/JHEP09(2015)098
  [arXiv:1505.07093 [hep-th]].
  
\bibitem{Wang:2015aua} 
  Y.~Wang and X.~Yin,
 ``Supervertices and Non-renormalization Conditions in Maximal Supergravity Theories,''
  arXiv:1505.05861 [hep-th].
  
\bibitem{Lin:2015dsa} 
  Y.~H.~Lin, S.~H.~Shao, Y.~Wang and X.~Yin,
  ``Supersymmetry Constraints and String Theory on K3,''
  JHEP {\bf 1512}, 142 (2015)
doi:10.1007/JHEP12(2015)142
  [arXiv:1508.07305 [hep-th]].
  
\bibitem{Bianchi:2016viy} 
  M.~Bianchi, A.~L.~Guerrieri, Y.~t.~Huang, C.~J.~Lee and C.~Wen,
  ``Exploring soft constraints on effective actions,''
  JHEP {\bf 1610}, 036 (2016)
doi:10.1007/JHEP10(2016)036
  [arXiv:1605.08697 [hep-th]].
  
  \bibitem{Green:1999pu}
  M.~B.~Green, H.~h.~Kwon and P.~Vanhove,
  `Two loops in eleven-dimensions,''
  Phys.\ Rev.\ D {\bf 61} (2000) 104010
doi:10.1103/PhysRevD.61.104010
  [hep-th/9910055].
  
\bibitem{DHoker:2005jhf}
  E.~D'Hoker, M.~Gutperle and D.~H.~Phong,
  ``Two-loop superstrings and S-duality,''
  Nucl.\ Phys.\ B {\bf 722} (2005) 81
doi:10.1016/j.nuclphysb.2005.06.010
  [hep-th/0503180].

\bibitem{DHoker:2013fcx}
  E.~D'Hoker and M.~B.~Green,
  ``Zhang-Kawazumi Invariants and Superstring Amplitudes,''
  Journal of Number Theory, Vol 144 (2014) page 111
  [arXiv:1308.4597 [hep-th]].

\bibitem{Gomez:2013sla}
  H.~Gomez and C.~R.~Mafra,
  ``The closed-string 3-loop amplitude and S-duality,''
  JHEP {\bf 1310} (2013) 217
doi:10.1007/JHEP10(2013)217
  [arXiv:1308.6567 [hep-th]].
   

\bibitem{Schwarz:1983qr}
  J.~H.~Schwarz,
  ``Covariant Field Equations of Chiral N=2 D=10 Supergravity,''
  Nucl.\ Phys.\ B {\bf 226} (1983) 269.
doi:10.1016/0550-3213(83)90192-X
    
\bibitem{Green:2003an}
  M.~B.~Green and C.~Stahn,
  ``D3-branes on the Coulomb branch and instantons,''
  JHEP {\bf 0309} (2003) 052
doi:10.1088/1126-6708/2003/09/052
  [hep-th/0308061].

\bibitem{Rajaraman:2005ag}
  A.~Rajaraman,
  ``On the supersymmetric completion of the R**4 term in M-theory,''
  Phys.\ Rev.\ D {\bf 74} (2006) 085018
doi:10.1103/PhysRevD.72.125008
  [hep-th/0512333].
  
\bibitem{Green:2014yxa}
  M.~B.~Green, S.~D.~Miller and P.~Vanhove,
  ``$SL(2, \mathbb{Z})$-invariance and D-instanton contributions to the $D^6 R^4$ interaction,''
  Commun.\ Num.\ Theor.\ Phys.\  {\bf 09} (2015) 307
doi:10.4310/CNTP.2015.v9.n2.a3
  [arXiv:1404.2192 [hep-th]].
  
\bibitem{Green:2013bza}
  M.~B.~Green, C.~R.~Mafra and O.~Schlotterer,
 ``Multiparticle one-loop amplitudes and S-duality in closed superstring theory,''
  JHEP {\bf 1310} (2013) 188
  doi:10.1007/JHEP10(2013)188
  [arXiv:1307.3534 [hep-th]].
  
\bibitem{CaronHuot:2010rj}
  S.~Caron-Huot and D.~O'Connell,
  ``Spinor Helicity and Dual Conformal Symmetry in Ten Dimensions,''
  JHEP {\bf 1108} (2011) 014
doi:10.1007/JHEP08(2011)014
  [arXiv:1010.5487 [hep-th]].
  
  
\bibitem{Boels:2012ie}
  R.~H.~Boels and D.~O'Connell,
  ``Simple super-amplitudes in higher dimensions,''
  JHEP {\bf 1206} (2012) 163
doi:10.1007/JHEP06(2012)163
  [arXiv:1201.2653 [hep-th]].
  

  
\bibitem{Mafra:2011nv} 
  C.~R.~Mafra, O.~Schlotterer and S.~Stieberger,
  ``Complete N-Point Superstring Disk Amplitude I. Pure Spinor Computation,''
  Nucl.\ Phys.\ B {\bf 873}, 419 (2013)
doi:10.1016/j.nuclphysb.2013.04.023
  [arXiv:1106.2645 [hep-th]].
  
\bibitem{Mafra:2011nw} 
  C.~R.~Mafra, O.~Schlotterer and S.~Stieberger,
  ``Complete N-Point Superstring Disk Amplitude II. Amplitude and Hypergeometric Function Structure,''
  Nucl.\ Phys.\ B {\bf 873}, 461 (2013)
 doi:10.1016/j.nuclphysb.2013.04.022
  [arXiv:1106.2646 [hep-th]].
  
\bibitem{Schlotterer:2012ny} 
  O.~Schlotterer and S.~Stieberger,
  ``Motivic Multiple Zeta Values and Superstring Amplitudes,''
  J.\ Phys.\ A {\bf 46}, 475401 (2013)
doi:10.1088/1751-8113/46/47/475401
  [arXiv:1205.1516 [hep-th]].
  
  \bibitem{SchlottereNew}
O.~Schlotterer, Private communication.

\bibitem{ArkaniHamed:2008gz} 
  N.~Arkani-Hamed, F.~Cachazo and J.~Kaplan,
  ``What is the Simplest Quantum Field Theory?,''
  JHEP {\bf 1009}, 016 (2010)
doi:10.1007/JHEP09(2010)016
  [arXiv:0808.1446 [hep-th]].
  
\bibitem{Carrasco:2013ypa} 
  J.~J.~M.~Carrasco, R.~Kallosh, R.~Roiban and A.~A.~Tseytlin,
  ``On the U(1) duality anomaly and the S-matrix of N=4 supergravity,''
  JHEP {\bf 1307}, 029 (2013)
doi:10.1007/JHEP07(2013)029
  [arXiv:1303.6219 [hep-th]].
  
\bibitem{Huang:2015sla} 
  Y.~t.~Huang and C.~Wen,
  ``Soft theorems from anomalous symmetries,''
  JHEP {\bf 1512}, 143 (2015)
doi:10.1007/JHEP12(2015)143
  [arXiv:1509.07840 [hep-th]].
  
\bibitem{Ademollo:1975pf} 
  M.~Ademollo, A.~D'Adda, R.~D'Auria, F.~Gliozzi, E.~Napolitano, S.~Sciuto and P.~Di Vecchia,
``Soft Dilations and Scale Renormalization in Dual Theories,''
  Nucl.\ Phys.\ B {\bf 94}, 221 (1975).
  doi:10.1016/0550-3213(75)90491-5
  
\bibitem{Shapiro:1975cz} 
  J.~A.~Shapiro,
  ``On the Renormalization of Dual Models,''
  Phys.\ Rev.\ D {\bf 11}, 2937 (1975).
  doi:10.1103/PhysRevD.11.2937
  
\bibitem{DiVecchia:2015jaq} 
  P.~Di Vecchia, R.~Marotta, M.~Mojaza and J.~Nohle,
  ``New soft theorems for the gravity dilaton and the Nambu-Goldstone dilaton at subsubleading order,''
  Phys.\ Rev.\ D {\bf 93}, no. 8, 085015 (2016)
 doi:10.1103/PhysRevD.93.085015
  [arXiv:1512.03316 [hep-th]].

  
\bibitem{DiVecchia:2016szw} 
  P.~Di Vecchia, R.~Marotta and M.~Mojaza,
``Soft behavior of a closed massless state in superstring and universality in the soft behavior of the dilaton,''
  JHEP {\bf 1612}, 020 (2016)
  doi:10.1007/JHEP12(2016)020
  [arXiv:1610.03481 [hep-th]].
  
\bibitem{DiVecchia:2018dob} 
  P.~Vecchia, R.~Marotta and M.~Mojaza,
  ``Multiloop Soft Theorem for Gravitons and Dilatons in the Bosonic String,''
  JHEP {\bf 1901}, 038 (2019)
 doi:10.1007/JHEP01(2019)038
  [arXiv:1808.04845 [hep-th]].
  
\bibitem{Richards:2008jg}
  D.~M.~Richards,
  ``The One-Loop Five-Graviton Amplitude and the Effective Action,''
  JHEP {\bf 0810} (2008) 042
  doi:10.1088/1126-6708/2008/10/042
  [arXiv:0807.2421 [hep-th]].


\bibitem{Berkovits:1998ex}
  N.~Berkovits and C.~Vafa,
  ``Type IIB R**4 H**(4g-4) conjectures,''
  Nucl.\ Phys.\ B {\bf 533} (1998) 181
doi:10.1016/S0550-3213(98)00475-1
  [hep-th/9803145].
  
\bibitem{Basu:2008cf}
  A.~Basu and S.~Sethi,
``Recursion Relations from Space-time Supersymmetry,''
  JHEP {\bf 0809} (2008) 081
doi:10.1088/1126-6708/2008/09/081
  [arXiv:0808.1250 [hep-th]].


    
\bibitem{Green:2010kv}
  M.~B.~Green, S.~D.~Miller, J.~G.~Russo and P.~Vanhove,
  ``Eisenstein series for higher-rank groups and string theory amplitudes,''
  Commun.\ Num.\ Theor.\ Phys.\  {\bf 4} (2010) 551
doi:10.4310/CNTP.2010.v4.n3.a2
  [arXiv:1004.0163 [hep-th]].

  
\bibitem{Pioline:2010kb}
  B.~Pioline,
  ``R**4 couplings and automorphic unipotent representations,''
  JHEP {\bf 1003} (2010) 116
 doi:10.1007/JHEP03(2010)116
  [arXiv:1001.3647 [hep-th]].
  
\bibitem{Basu:2011he}
  A.~Basu,
  ``Supersymmetry constraints on the $R^4$ multiplet in type IIB on $T^2$,''
  Class.\ Quant.\ Grav.\  {\bf 28} (2011) 225018
doi:10.1088/0264-9381/28/22/225018
  [arXiv:1107.3353 [hep-th]].
  
  
\bibitem{Intriligator:1998ig}
  K.~A.~Intriligator,
  ``Bonus symmetries of N=4 superYang-Mills correlation functions via AdS duality,''
  Nucl.\ Phys.\ B {\bf 551} (1999) 575
doi:10.1016/S0550-3213(99)00242-4
  [hep-th/9811047].


\bibitem{Basu:2004dm}
  A.~Basu, M.~B.~Green and S.~Sethi,
  ``A Curious truncation of N=4 Yang-Mills,''
  Phys.\ Rev.\ Lett.\  {\bf 93} (2004) 261601
doi:10.1103/PhysRevLett.93.261601
  [hep-th/0406267].
  
\bibitem{Binder:2019jwn} 
  D.~J.~Binder, S.~M.~Chester, S.~S.~Pufu and Y.~Wang,
  ``$\mathcal{N}=4$ Super-Yang-Mills Correlators at Strong Coupling from String Theory and Localization,''
  arXiv:1902.06263 [hep-th].
  

  


\end{thebibliography}
\end{document}